\documentclass[12pt, a4paper]{article}
\usepackage{graphicx}
\usepackage{amssymb}
\usepackage{amsmath}
\usepackage{bm}
\usepackage{color}
\usepackage{cancel}
\usepackage{colortbl}
\usepackage{theorem}
\usepackage{subcaption}
\usepackage{braket}
\usepackage{cases}
\usepackage[utf8]{inputenc}

\usepackage[sort&compress,numbers, merge]{natbib}

\definecolor{Orange}{cmyk}{0,0.61,0.87,0}
\definecolor{JungleGreen}{cmyk}{0.99,0,0.52,0}
\definecolor{OliveGreen}{cmyk}{0.64,0,0.95,0.40}
\definecolor{Brown}{cmyk}{0,0.81,1,0.60}
\definecolor{RoyalBlue}{cmyk}{0.71,0.53,0,0.12}
\definecolor{Gray}{cmyk}{0,0,0,0.40}
\definecolor{LightPink}{cmyk}{0.0,0.25,0,0}
\definecolor{LLightPink}{cmyk}{0.0,0.10,0,0}
\definecolor{LightBlue}{cmyk}{0.25,0,0,0}
\definecolor{LightGray}{cmyk}{0,0,0,0.2}
\definecolor{LightGreen}{cmyk}{0.3,0,0.3,0}

\makeatletter
\newcommand{\overleftrightsmallarrow}{\mathpalette{\overarrowsmall@\leftrightarrowfill@}} 
\newcommand{\overarrowsmall@}[3]{%
  \vbox{%
    \ialign{%
      ##\crcr
      #1{\smaller@style{#2}}\crcr
      \noalign{\nointerlineskip}%
      $\m@th\hfil#2#3\hfil$\crcr
    }%
  }%
}
\def\smaller@style#1{%
  \ifx#1\displaystyle\scriptstyle\else
    \ifx#1\textstyle\scriptstyle\else
      \scriptscriptstyle
    \fi
  \fi
}
\makeatother

\usepackage[colorlinks=true, linkcolor=OliveGreen, citecolor=RoyalBlue,
urlcolor=black]{hyperref}

\begin{document}

\begin{titlepage}

\begin{flushright}
{\tt 
IPMU18-0085\\
TU-1071 \\
UT-18-16
}
\end{flushright}

\vskip 1.35cm
\begin{center}

{\large
{\bf
Singlet Dirac Fermion Dark Matter with Mediators at Loop
}
}

\vskip 1.2cm

Junji Hisano$^{a,b,c}$,
Ryo Nagai$^{d}$,
and
Natsumi Nagata$^e$

\vskip 0.4cm

{\it $^a$Kobayashi-Maskawa Institute for the Origin of Particles and the
 Universe, \\ Nagoya University, Nagoya 464-8602, Japan}\\[3pt]
{\it $^b$Department of Physics,
Nagoya University, Nagoya 464-8602, Japan}\\[3pt]
{\it $^{c}$Kavli IPMU (WPI), UTIAS, University of Tokyo, Kashiwa, Chiba
 277-8583, Japan}\\ [3pt] 
{\it $^{d}$Department of Physics, Tohoku University, Sendai, Miyagi
 980-8578, Japan}\\ [3pt] 
{\it $^e$Department of Physics, University of Tokyo, Bunkyo-ku, Tokyo
 113-0033, Japan} 

\date{\today}

\vskip 1.5cm

\begin{abstract}
 We study the phenomenology of singlet Dirac fermion dark matter in the
 simplified models where the dark matter interacts with the Standard
 Model particles at loop-level with the help of either colored or
 non-colored mediators. We especially focus on the implications
 of non-zero CP phases in the dark sector, which induce the electric
 dipole moments of the Dirac fermion dark matter as well as those of
 electron and nucleon. It is then found that the dark matter direct
 detection searches and the measurements of the electric dipole
 moments are able to test the singlet Dirac fermion dark matter
 scenario in the forthcoming experiments. 

\end{abstract}

\end{center}
\end{titlepage}

%%%%%%%%%%%%%%%%%%%%%%%%%%%%%%%%%%%%%%%%%%
\section{Introduction}
%%%%%%%%%%%%%%%%%%%%%%%%%%%%%%%%%%%%%%%%%%

Weakly-interacting massive particles (WIMPs) offer the most promising
framework to explain dark matter (DM) in the Universe, as their
thermal relic abundance naturally agrees with the observed DM
density. A variety of possibilities for WIMP DM candidates have been
proposed and studied so far \cite{Arcadi:2017kky}; for instance, a real or complex scalar particle that is
singlet under the Standard Model (SM) gauge interactions and is 
stabilized by, \textit{e.g.}, a $\mathbb{Z}_2$ symmetry is known to be
the simplest example for WIMP DM \cite{Silveira:1985rk, *McDonald:1993ex,
*Burgess:2000yq, *Davoudiasl:2004be}. This scalar DM
has a quartic coupling with the SM Higgs field at the renormalizable
level, which allows the DM particles to annihilate into SM
particles. This DM interacts with nucleons through this quartic coupling,
which enables us to test this DM model in future DM direct detection
experiments.

Fermionic DM candidates are, on the other hand, classified into either a
Majorana or Dirac fermion. It turns out that these two classes of DM candidates
have different phenomenological properties, as Majorana fermions have
neither vector nor dipole interactions, while Dirac fermions are able
to have both of those interactions. In addition, pair annihilation cross
sections of Majorana fermions into light fermions are suppressed
as the $s$-wave annihilation requires chirality flip, while those of
Dirac fermions are free from such suppression. Thus, for Dirac fermion
DM, heavier mass regions may be favored in terms of the thermal relic
abundance compared to Majorana fermion DM, which allows a simple
explanation for the null results in the existing DM searches such as the
LHC and DM direct detection experiments.

In this paper, we focus on the singlet Dirac fermion DM.  If DM is a
Dirac fermion, there should exist a U(1) symmetry, either global or
gauged, under which the DM has a non-zero charge so that the nature of
Dirac fermion is maintained.  This possibility is theoretically
interesting since it explains the stability of the DM as in the case
of proton, which is cosmologically stable because of an accidental
global U(1) symmetry (baryon number) in the SM.  Such a setup is also
motivated by the so-called asymmetric DM scenario \cite{Hut:1979xw,
*Nussinov:1985xr, *Barr:1990ca, *Barr:1991qn, *Kaplan:1991ah,
*Dodelson:1991iv, *Kuzmin:1996he, *Hooper:2004dc, *Kitano:2004sv,
*Farrar:2005zd, *Kaplan:2009ag, *Nagata:2016knk}, in which asymmetry
in the DM particle and antiparticle number densities accounts for the
observed DM density.

When DM is a fermion which is singlet under the SM gauge interactions, additional
fields called mediators are required in order for the DM to
interact with the SM sector. This is because there is no
renormalizable intearction of the DM with the SM fields. A variety
of options for introducing such mediators have been considered so
far. A simple way is to add a singlet scalar particle which couples to
both the fermionic DM and the SM Higgs boson \cite{Kim:2008pp,
  *Kanemura:2011vm, *Lindner:2011it, *Kanemura:2011mw, *Baek:2011aa, *Ko:2017uyb}.
Another way is to assume that DM couples to a neutral gauge boson which
interacts with the SM fields as well \cite{Okada:2010wd, Cheung:2007ut,
*Mambrini:2011dw, *An:2012va, *Barger:2012ey, *Chu:2013jja,
*Arcadi:2013qia, *Lebedev:2014bba, *Mambrini:2015sia}. Moreover, if there are extra
scalars that have the same quantum numbers as those of the SM
fermions, DM may couple to the SM fermions directly
with the help of these scalar particles \cite{Agrawal:2011ze,
*Weiner:2012gm,
*Bai:2013iqa, *DiFranzo:2013vra, *Kopp:2014tsa, *Bai:2014osa,
*Chang:2014tea, *Agrawal:2014ufa, *Agrawal:2014una, *Hamze:2014wca,
*Yu:2014mfa, *Kilic:2015vka, *Primulando:2015lfa, *Ko:2016zxg, *Chao:2016lqd,
Ibarra:2015fqa, Baker:2018uox}. If one introduces an SU(2)$_L$ doublet
fermion with hypercharge $\pm 1/2$, we may also couple the singlet
fermion DM to the SM Higgs field.

In all of the above cases, the fermionic DM couples to the SM sector
at tree level. On the other hand, in this paper, we study the cases
where the Dirac fermion DM does not couple to the SM sector at tree
level, but does couple to it at loop level. To that end, we consider
simplified Dirac fermion DM models, where extra fermions and scalars
with non-zero SM gauge charges are coupled with the Dirac fermion
DM.\footnote{See also Ref.~\cite{Herrero-Garcia:2018koq}
for a relevant work, where the Dirac fermion DM is supposed to interact
with the SM sector at loop level through non-colored mediators. In this
paper, we also consider colored mediators, which give rise to rich
phenomenology as we see below.  } 
As mentioned above, the Dirac fermion DM may have vector and tensor
couplings, which considerably affect the direct detection
rate of DM. However, the significance of these couplings depends on
models. For instance, if the Dirac fermion DM had a direct coupling with
light quarks via a mediator, this interaction would induce vector-current
four-Fermi interactions with light quarks at tree level after the
mediator is integrated out, but such interactions cause a large DM-nucleus
scattering cross section and thus have already been severely constrained
by the DM direct detection experiments. If, on the other hand, the Dirac
fermion DM is coupled to 
the SM sector at loop level, these vector interactions are
suppressed by a loop factor and the direct detection bound may be
evaded. It turns out that in this case the tensor
couplings may give significant contributions to the DM direct
detection even if they are also suppressed by one-loop factor. The
Dirac fermion DM may have a magnetic dipole moment (MDM)---and an electric
dipole moment (EDM) as well if the DM-mediator interactions contain CP
phases. These dipole moments open up possibilities to test this DM
candidate in DM direct detection experiments \cite{Banks:2010eh}.

As it turns out, the detectability of the DM through the dipole moments
is considerably affected by the size of the EDM, which depends on the CP
phases in the DM-mediator couplings. These CP phases, as well as the CP
phases in the mediator-Higgs couplings, also induce CP-odd quantities in
the SM sector at loop level, and thus we may probe them through the
measurements of electron and nucleon EDMs. In particular, if the Dirac
fermion DM is accompanied with colored mediators, they generate the
dimension-six Weinberg operator \cite{Weinberg:1989dx}, which then
induces nucleon EDMs. We thus expect a close correlation between the DM
and nucleon EDMs, which has significant implications for the testability
of this DM scenario in the future DM direct detection and EDM experiments. 

This paper is organized as follows. In the next section, we show our
simplified Dirac fermion DM models.  We then briefly discuss the thermal relic
abundance of DM in the models in Sec.~\ref{sec:relic}. The DM direct
detection and the electron and nucleon EDMs generated in these
model are discussed in Sec.~\ref{sec:dd} and Sec.~\ref{sec:edms},
respectively.  Then,
in Sec.~\ref{sec:results}, we show the current constraints on our 
simplified Dirac fermion DM models, and discuss their testability in future
experiments. Section~\ref{sec:concl} is devoted to conclusion and
discussion.

%%%%%%%%%%%%%%%%%%%%%%%%%%%%%%%%%%%%%%%%
\section{Simplified Dirac Fermion DM Models}
\label{sec:models}
%%%%%%%%%%%%%%%%%%%%%%%%%%%%%%%%%%%%%%%%

%%%%%%%%%%%%%%%%%%%%%%%%%%%%%%%%%%%%%%%%%%%%
\subsection{Field content and Lagrangian}
%%%%%%%%%%%%%%%%%%%%%%%%%%%%%%%%%%%%%%%%%%%%

To begin with, we show our simplified Dirac fermion DM models which we discuss
in this paper. The Dirac fermion DM $\chi$ is composed of two Weyl fermions,
$\xi_\chi$ and $\eta_\chi$, which are assumed to be singlet under the
SM gauge interactions.\footnote{Weyl fermions are
assumed to be left-handed throughout this paper.  }  We then introduce a global U(1)
symmetry, $\text{U}(1)_D$, and assume that these fermions have the
$\text{U}(1)_D$ charge $+1$ and $-1$, respectively. The SM particles
are supposed to be singlet under the U(1)$_D$ symmetry.

%%%%%%%%%%%%%%%%%%%%% TABLE %%%%%%%%%%%%%%%%%%%%%%%%%%%%%%%%%%%%%%%%%%
\begin{table}[t]
 \begin{center}
\caption{Quantum numbers of DM and mediators. 
A new global symmetry $\text{U}(1)_D$ is for stability of the DM.}
\label{tab:model}
\vspace{5pt}
\begin{tabular}{cccccc}
\hline
\hline
Field & Spin & SU(3)$_C$ & SU(2)$_L$ & U(1)$_Y$ & $\text{U}(1)_D$ \\
\hline
${\xi_\chi}$ & $1/2$ & {\bf 1} & {\bf 1} & $0$ & $+1$ \\
${\eta_\chi}$ & $1/2$ & {\bf 1} & {\bf 1} & $0$ & $-1$ \\
\rowcolor{LLightPink}
${\xi_Q}$ & $1/2$ & {\bf 3} & {\bf 2} & $\frac{1}{6}$ & $0$ \\
\rowcolor{LLightPink}
${\eta_Q}$ & $1/2$ & $\overline{\bf 3}$ & {\bf 2} & $-\frac{1}{6}$ &
		     $0$ \\
\rowcolor{LLightPink}
$\widetilde{Q}$ & $0$ & {\bf 3} & {\bf 2} & $\frac{1}{6}$ & $+1$ \\
\rowcolor{LLightPink}
${\xi_{\bar{u}}}$ & $1/2$ & $\overline{\bf 3}$ & {\bf 1} & $-\frac{2}{3}$ & $0$ \\
\rowcolor{LLightPink}
${\eta_{\bar{u}}}$ & $1/2$ & ${\bf 3}$ & {\bf 1} & $\frac{2}{3}$ &
		     $0$ \\
\rowcolor{LLightPink}
$\widetilde{\bar{u}}$ & $0$ & $\overline{\bf 3}$ & {\bf 1} & $-\frac{2}{3}$ & $-1$ \\
\rowcolor{LLightPink}
${\xi_{\bar{d}}}$ & $1/2$ & $\overline{\bf 3}$ & {\bf 1} & $\frac{1}{3}$ & $0$ \\
\rowcolor{LLightPink}
${\eta_{\bar{d}}}$ & $1/2$ & ${\bf 3}$ & {\bf 1} & $-\frac{1}{3}$ &
		     $0$ \\
\rowcolor{LLightPink}
$\widetilde{\bar{d}}$ & $0$ & $\overline{\bf 3}$ & {\bf 1} & $\frac{1}{3}$ & $-1$ \\
\rowcolor{LightGreen}
${\xi_L}$ & $1/2$ & {\bf 1} & {\bf 2} & $-\frac{1}{2}$ & $0$ \\
\rowcolor{LightGreen}
${\eta_L}$ & $1/2$ & ${\bf 1}$ & {\bf 2} & $\frac{1}{2}$ &
		     $0$ \\
\rowcolor{LightGreen}
$\widetilde{L}$ & $0$ & {\bf 1} & {\bf 2} & $-\frac{1}{2}$ & $+1$ \\
\rowcolor{LightGreen}
${\xi_{\bar{e}}}$ & $1/2$ & {\bf 1} & {\bf 1} & $1$ & $0$ \\
\rowcolor{LightGreen}
${\eta_{\bar{e}}}$ & $1/2$ & ${\bf 1}$ & {\bf 1} & $-1$ &
		     $0$ \\
\rowcolor{LightGreen}
$\widetilde{\bar{e}}$ & $0$ & {\bf 1} & {\bf 1} & $1$ & $-1$ \\

\hline
\hline
\end{tabular}
 \end{center}
\end{table}
%%%%%%%%%%%%%%%%%%%%%%%%%%%%%%%%%%%%%%%%%%%%%%%%%%%%%%%%%%%%%%%%%%%%%%%

In addition, we introduce vector-like fermions and complex
scalars as mediators, in order to couple the Dirac fermion DM to the SM sector
at loop level. These additional particles have SM gauge
interactions, which then induce the couplings of the DM with the SM
gauge bosons through quantum corrections. To avoid stable
charged/colored fermions, we assume the extra vector-like fermions to
have the same quantum numbers as those of the SM fermions and to mix
with the SM fermions by a small amount.  Thus, these extra fermions
should be singlet under U(1)$_D$. For the DM to couple with the extra
fermions, we need to assume that the extra scalar particles also have
the same quantum numbers as those of the SM fermions and are charged
under U(1)$_D$. These U(1)$_D$-charged scalars are supposed to be
heavier than the Dirac fermion DM to insure the stability of DM. 

We list sets
of such extra scalars and fermions and show their quantum numbers in
Table~\ref{tab:model}. Here, all of the fermionic fields are
introduced in a vector-like manner so that they form Dirac
fields. This simultaneously assures that these new fermions do not
cause gauge anomaly. In addition, since the Dirac fermion DM is the
only fermion that has non-zero U(1)$_D$ charge, this U(1)$_D$ symmetry
is also anomaly-free.\footnote{For this reason, we may also consider
the gauged U(1)$_D$ symmetry instead of the global symmetry. In this
case, the Dirac fermion DM has the U(1)$_D$ gauge interaction, which may
affect its phenomenological properties significantly. 
Although this is an interesting possibility, we do not consider this case in the
following discussion.  } 

In what follows, we consider two models where different sets of
fields are added to the SM besides the Dirac fermion DM $\chi$; Model I
contains $\xi_Q$, $\eta_Q$, $\widetilde{Q}$, $\xi_{\bar{u}}$,
$\eta_{\bar{u}}$, $\widetilde{\bar{u}}$, $\xi_{\bar{d}}$,
$\eta_{\bar{d}}$, and $\widetilde{\bar{d}}$, while Model II includes
$\xi_L$, $\eta_L$, $\widetilde{L}$, $\xi_{\bar{e}}$, $\eta_{\bar{e}}$,
and $\widetilde{\bar{e}}$. These sets of particles are shaded in pink
and green in Table~\ref{tab:model}, respectively. These two models are
qualitatively different, since in Model I the DM interacts with colored particles
while it does not in Model II.  We also denote the SU(2)$_L$ component
fields of $\xi_Q$, $\eta_Q$, $\xi_L$, and $\eta_L$ by
\begin{equation}
 \xi_Q =
\begin{pmatrix}
 \xi_u \\ \xi_d
\end{pmatrix}
~,~~~~
 \eta_Q =
\begin{pmatrix}
 \eta_u \\ \eta_d
\end{pmatrix}
~,~~~~
 \xi_L =
\begin{pmatrix}
 \xi_\nu \\ \xi_e
\end{pmatrix}
~,~~~~
 \eta_L =
\begin{pmatrix}
 \eta_\nu \\ \eta_e
\end{pmatrix}
~,
\end{equation}
respectively, while for $\widetilde{Q}$ and $\widetilde{L}$
\begin{equation}
 \widetilde{Q} =
\begin{pmatrix}
 \widetilde{u} \\ \widetilde{d}
\end{pmatrix}
~, ~~~~~~
 \widetilde{L} =
\begin{pmatrix}
 \widetilde{\nu} \\ \widetilde{e}
\end{pmatrix}
~.
\end{equation}
The mass terms of these fields are given by
\begin{align}
 {\cal L}_{\rm mass} 
= 
&- 
\Bigl[
\mu_\chi \xi_\chi \eta_\chi 
+\sum_{f} \mu_f \xi_f \eta_f +{\rm h.c.}
\Bigr]
- \sum_{f} \widetilde{m}^2_{f} |\widetilde{f}|^2 
 ~,
\end{align}
where the sum is taken over the extra matters in each
model. $\widetilde{m}^2_{f}$ are real, while $\mu_\chi$ and
$\mu_f$ are in general complex quantities. 

In both of the models, the Dirac fermion DM has the following interactions: 
\begin{align}
 {\cal L}_{\chi f \tilde{f}} &= 
 a_Q\, \xi_\chi \xi_Q \widetilde{Q}^*
+b_Q\, \eta_\chi \eta_Q \widetilde{Q}
+a_{\bar{u}}\, \xi_{\chi} \eta_{\bar{u}} \widetilde{\bar{u}}
+b_{\bar{u}}\, \eta_\chi \xi_{\bar{u}} \widetilde{\bar{u}}^*
+a_{\bar{d}}\, \xi_{\chi} \eta_{\bar{d}} \widetilde{\bar{d}}
+b_{\bar{d}}\, \eta_\chi \xi_{\bar{d}} \widetilde{\bar{d}}^*
\nonumber \\
&+a_L\, \xi_\chi \xi_L \widetilde{L}^*
+b_L\, \eta_\chi \eta_L \widetilde{L}
+a_{\bar{e}}\, \xi_{\chi} \eta_{\bar{e}} \widetilde{\bar{e}}
+b_{\bar{e}}\, \eta_\chi \xi_{\bar{e}} \widetilde{\bar{e}}^*
+{\rm h.c.}~.
\label{eq:lagdmint}
\end{align}
These interactions induce the annihilation of the DM into the extra
vector-like fermions via $t$-channel exchange of the extra scalars if the
vector-like fermions are lighter than the DM.

The vector-like fermions may have Yukawa couplings with the SM Higgs
field $H$ as they are allowed by the gauge invariance in the models,
\begin{align}
 {\cal L}_{\rm Yukawa}
=
&- \kappa_{\bar{u}} \xi_{\bar{u}} \epsilon_{\alpha\beta}(\xi_Q)_\alpha (H)_\beta
- \kappa_{\bar{u}}^\prime \eta_{\bar{u}}  (\eta_Q)^\alpha (\widetilde{H})_\alpha
\nonumber \\
&-\kappa_{\bar{d}} \xi_{\bar{d}} (\xi_Q)_\alpha  (H^\dagger)^\alpha 
- \kappa^\prime_{\bar{d}} \eta_{\bar{d}}(\eta_Q)^\alpha  (H)_\alpha 
\nonumber \\
&-\kappa_{\bar{e}} \xi_{\bar{e}} (\xi_L)_\alpha  (H^\dagger)^\alpha 
- \kappa^\prime_{\bar{e}} \eta_{\bar{e}}(\eta_L)^\alpha  (H)_\alpha 
+{\rm h.c.}~,
\label{eq:yukawa}
\end{align}
where $\alpha, \beta$ are SU(2)$_L$ indices, $\epsilon_{\alpha\beta}$
is the anti-symmetric tensor with $\epsilon_{12} = -\epsilon_{21}
=+1$, and $\widetilde{H} \equiv i\tau_2 H^\dagger$. Note that, in this
paper, we have defined $\eta_Q$ and $\eta_L$ such that they transform
as anti-fundamental representations of SU(2)$_L$; {\it i.e.}, $i\tau_2
\eta_Q$ and $i\tau_2 \eta_L$ are fundamental representations of
SU(2)$_L$, where $\tau_a$ ($a=1,2,3$) are the Pauli matrices.
As mentioned above, we assume
that the extra vector-like fermions have small but non-zero Yukawa
couplings with the SM fermions so that these vector-like fermions decay 
into the SM fermions;\footnote{The upper limits on the mixing angles between 
the SM fermions and the vector-like fermions imposed by flavor experiments 
are ${\cal O}(10^{-4})$ in the
most stringent cases (see, {\it e.g.}, Refs.~\cite{Botella:2012ju, Alok:2014yua, 
Bobeth:2016llm, Morozumi:2018cnc}); with this size of mixing angles, 
the vector-like fermions decay almost promptly.} we do not show these terms explicitly here and
hereafter. The new scalar fields may also have trilinear terms similar to the
terms in Eq.~\eqref{eq:yukawa}: 
\begin{align}
 {\cal L}_{\rm tri}
=- A_{\bar{u}} \widetilde{\bar{u}} \epsilon_{\alpha\beta} (\widetilde{Q})_\alpha 
(H)_\beta 
-A_{\bar{d}} \widetilde{\bar{d}} (\widetilde{Q})_\alpha  (H^\dagger)^\alpha 
-A_{\bar{e}} \widetilde{\bar{e}} (\widetilde{L})_\alpha  (H^\dagger)^\alpha 
+{\rm h.c.}~.
\label{tri}
\end{align}
In addition, there may be quartic couplings 
\begin{align}
 {\cal L}_{\rm quart} 
= - \sum_{f} \lambda_f |\widetilde{f}|^2 |H|^2
- \lambda^\prime_Q \widetilde{Q}^\dagger \tau_a \widetilde{Q} 
H^\dagger \tau_a H
- \lambda^\prime_L \widetilde{L}^\dagger \tau_a \widetilde{L} 
H^\dagger \tau_a H
+\dots ~,
\label{quart}
\end{align}
where dots indicate other quartic terms that contain only the new
scalar fields. Such terms are irrelevant to the following analysis,
and thus we neglect them in what follows. The extra-scalar interactions
in Eqs.~\eqref{tri} and \eqref{quart} give rise to the mass terms of
the extra scalars as we see in the next subsection.

%%%%%%%%%%%%%%%%%%%%%%%%%%%%%%%%%%%%
\subsection{Mass eigenstates}
%%%%%%%%%%%%%%%%%%%%%%%%%%%%%%%%%%%%

After the Higgs field develops a vacuum expectation value (VEV),\footnote{We
take $v$ to be real without loss of generality.}
\begin{equation}
 \langle H \rangle =\frac{1}{\sqrt{2}} 
\begin{pmatrix}
 0 \\v
\end{pmatrix}
~,
\end{equation}
with $v\simeq 246$~GeV, the extra fermions and scalars mix among each
other. For the fermionic part, the mass terms are
\begin{align}
 {\cal L}_{\rm mass}^{({\rm ferm})}
 &= - 
\sum_{f=u,d,e}
\left(\eta_f,~\xi_{\bar{f}}\right)
{\cal M}_f 
\begin{pmatrix}
 \xi_f \\ \eta_{\bar{f}}
\end{pmatrix}
- \mu_L \xi_\nu \eta_\nu 
+{\rm h.c.} ~,
\end{align}
where
\begin{align}
 {\cal M}_u =
\begin{pmatrix}
 \mu_Q & \frac{v}{\sqrt{2}} \kappa_{\bar{u}}^\prime \\
 \frac{v}{\sqrt{2}} \kappa_{\bar{u}} & \mu_{\bar{u}}
\end{pmatrix}
~,~~~~~
 {\cal M}_d =
\begin{pmatrix}
 \mu_Q & \frac{v}{\sqrt{2}} \kappa_{\bar{d}}^\prime \\
 \frac{v}{\sqrt{2}} \kappa_{\bar{d}} & \mu_{\bar{d}}
\end{pmatrix}
~,~~~~~
 {\cal M}_e =
\begin{pmatrix}
 \mu_L & \frac{v}{\sqrt{2}} \kappa_{\bar{e}}^\prime \\
 \frac{v}{\sqrt{2}} \kappa_{\bar{e}} & \mu_{\bar{e}}
\end{pmatrix}
~.
\end{align}
Each mass matrix ${\cal M}_f$ ($f = u,d,e$) is diagonalized by means of
biunitary transformation as
\begin{equation}
 V_{fR}^\dagger {\cal M}_f V_{fL}^{} =
\begin{pmatrix}
 m_{f_1} & 0 \\ 0 & m_{f_2}
\end{pmatrix}
~,
\end{equation}
where $m_{f_i}$ $(i =1,2)$ are real and non-negative, 
and we denote the corresponding Dirac fermions in the mass eigenbasis by
$\psi_{f_i}$. The relations between the mass eigenstates and the weak
eigenstates are given by
\begin{align}
 \begin{pmatrix}
  \xi_f \\ \eta_{\bar{f}} 
 \end{pmatrix}
=
V_{fL}
\begin{pmatrix}
 \psi_{f_1 L} \\  \psi_{f_2 L} 
\end{pmatrix}
~,~~~~~~
\begin{pmatrix}
 \eta_f^\dagger \\ \xi^\dagger_{\bar{f}}
\end{pmatrix}
= V_{fR} 
\begin{pmatrix}
 \psi_{f_1 R} \\  \psi_{f_2 R} 
\end{pmatrix}
~,
\end{align}
where $L$ ($R$) represents the left-handed (right-handed) components of
$\psi_{f_i}$. 
The Dirac mass term for $\xi_\nu$ and $\eta_\nu$ is taken to be
real and non-negative via an appropriate phase rotation, where 
the Dirac field is given by 
\begin{equation}
 \psi_\nu =
\begin{pmatrix}
 e^{\frac{i}{2} \theta_{L}} \xi_\nu \\
 e^{-\frac{i}{2} \theta_{L}} \eta^\dagger_\nu
\end{pmatrix}
~,
\end{equation}
with $\theta_L \equiv {\rm arg}(\mu_L)$, and its mass is given by $m_\nu
= |\mu_L|$.  
Similarly, by defining the Dirac fermion DM field 
\begin{equation}
 \chi \equiv 
\begin{pmatrix}
 e^{\frac{i}{2} \theta_{\chi}} \eta_\chi \\
 e^{-\frac{i}{2} \theta_{\chi}} \xi^\dagger_\chi
\end{pmatrix}
~,
\end{equation}
with $\theta_\chi \equiv {\rm arg}(\mu_\chi)$, we have the DM mass term
of the form
$- m_{\chi} \overline{\chi} \chi$ with $m_{\chi} \equiv
|\mu_\chi|$. 

As we see, the DM fermions $\xi_\chi$ and $\eta_\chi$ form a Dirac
fermion with an identical mass $m_\chi$. This remains unchanged even if
radiative corrections are included. This is because the interaction
terms presented in the previous subsection preserve the U(1)$_D$
symmetry, which is not spontaneously broken as the SM Higgs field $H$ is
not charged under the U(1)$_D$ symmetry. Since $\xi_\chi$ and $\eta_\chi$
are charged under U(1)$_D$, Majorana mass terms for these fields, with
which these two fields split into two Majorana fermions, are not
generated as long as the U(1)$_D$ symmetry is preserved.  

For the scalar fields, on the other hand, the mass terms are written as
\begin{align}
 {\cal L}_{\rm mass}^{({\rm sca})} 
&= -\sum_{f = u, d, e} \left(\widetilde{f}^*,~
\widetilde{\bar{f}}\right) \widetilde{\cal M}_f^2 
\begin{pmatrix}
 \widetilde{f} \\ \widetilde{\bar{f}}^*
\end{pmatrix}
-\widetilde{m}_\nu^2  \left|
\widetilde{\nu}
\right|^2 ~,
\end{align}
where 
\begin{align}
 \widetilde{\cal M}_u^2 
&=
\begin{pmatrix}
 \widetilde{m}_Q^2 +\frac{\lambda_Q-\lambda_Q^\prime}{2}v^2 &
 \frac{v}{\sqrt{2}} A_{\bar{u}}^* \\ 
 \frac{v}{\sqrt{2}} A_{\bar{u}} & \widetilde{m}^2_{\bar{u}}
 +\frac{\lambda_{\bar{u}}}{2} v^2 
\end{pmatrix}
~,~~~~~
 \widetilde{\cal M}_d^2 
=
\begin{pmatrix}
 \widetilde{m}_Q^2 +\frac{\lambda_Q + \lambda^\prime_Q}{2}v^2 
& \frac{v}{\sqrt{2}} A_{\bar{d}}^* \\
 \frac{v}{\sqrt{2}} A_{\bar{d}} & \widetilde{m}^2_{\bar{d}}
 +\frac{\lambda_{\bar{d}}}{2} v^2 
\end{pmatrix}
~, \nonumber \\[3pt]
 \widetilde{\cal M}_e^2 
&=
\begin{pmatrix}
 \widetilde{m}_L^2 +\frac{\lambda_L + \lambda^\prime_L}{2}v^2 
& \frac{v}{\sqrt{2}} A_{\bar{e}}^* \\
 \frac{v}{\sqrt{2}} A_{\bar{e}} & \widetilde{m}^2_{\bar{e}}
 +\frac{\lambda_{\bar{e}}}{2} v^2 
\end{pmatrix}
~,
~~~~~
\widetilde{m}_\nu^2 = \widetilde{m}_L^2 +\frac{\lambda_L-\lambda^\prime_L}{2}
 v^2 ~.
\end{align}
The mass matrices $\widetilde{\cal M}^2_f$ $(f=u,d,e)$ are Hermitian, and thus
diagonalized by unitary matrices $\widetilde{V}_f$ as
\begin{equation}
 \widetilde{V}_f^\dagger \widetilde{\cal M}_f^2 \widetilde{V}_f^{} 
=
\begin{pmatrix}
 \widetilde{m}_{f_1}^2 & 0 \\ 0&  \widetilde{m}_{f_2}^2
\end{pmatrix}
~,
\end{equation}
with the mass eigenstates given by
\begin{equation}
 \begin{pmatrix}
   \widetilde{f} \\ \widetilde{\bar{f}}^*
 \end{pmatrix}
= \widetilde{V}_f 
\begin{pmatrix}
 \widetilde{f}_1 \\ \widetilde{f}_2
\end{pmatrix}
~.
\end{equation}
The interactions of the new particles in the mass eigenbasis are
presented in Appendix~\ref{sec:interactions}.

%%%%%%%%%%%%%%%%%%%%%%%%%
\subsection{CP phases}
\label{sec:cp}
%%%%%%%%%%%%%%%%%%%%%%%%%

Many of the couplings newly introduced in our simplified Dirac fermion
DM models are in general complex. Field redefinition cannot remove all
of the CP phases, and thus the remaining CP phases are physical. These
physical CP phases induce the EDM of the Dirac fermion DM as well as
those of electron and nucleon. As it turns out, the EDM of the DM
significantly affects the DM direct detection rate.

For later use, let us summarize the physical CP phases and express them
in a rephasing invariant manner: 
\begin{equation}
\varphi_{\mu_f} \equiv \text{arg}(\mu_\chi \mu_f a_f^* b_f^*) ~,
\end{equation}
for $f=Q,\bar{u},\bar{d},L,\bar{e}$, and 
\begin{equation}
\varphi_{\kappa_{\bar{f}}} \equiv \text{arg}(\mu_f \mu_{\bar{f}}
 \kappa_{\bar{f}}^* \kappa_{\bar{f}}^{\prime *}) 
~, \qquad
\varphi_{A_{\bar{f}}} \equiv \text{arg}(A_{\bar{f}} \mu_f b_{\bar{f}} b_f^*
\kappa_{\bar{f}}^*) ~, 
\end{equation}
for $f=u,d,e$. 
The combinations $A_{\bar{f}} \mu_f^* a_f a_{\bar{f}}^*\kappa_{\bar{f}}^\prime
$ are also rephasing invariant 
though the phases are given by linear combinations of the above phases. 

%

%%%%%%%%%%%%%%%%%%%%%%%%%%%%%%%%%%%%%%%
\section{Thermal Relic Abundance}
\label{sec:relic}
%%%%%%%%%%%%%%%%%%%%%%%%%%%%%%%%%%%%%%

The DM couplings in Eq.~\eqref{eq:lagdmint} keep the Dirac fermion DM $\chi$
in thermal equilibrium with the SM particles in the early
Universe. Once the temperature of the Universe falls down below the DM
mass, its number density exponentially decreases, and eventually the
annihilation processes freeze out, and the number of the DM particles
in a unit comoving volume becomes constant. The resultant DM relic
density is thus determined by only the pair annihilation cross section
of the DM particles at the decoupling temperature, unless other
particles with nonzero U(1)$_D$ charges are degenerate with the DM
particles in mass so that significant amount of these particles is left
in the thermal bath at the decoupling temperature. 

In the present model, the DM particles annihilate through
the couplings in Eq.~\eqref{eq:lagdmint}; they annihilate into the
vector-like fermions through the $t$-channel scalar exchange processes
if the vector-like fermions are lighter than the Dirac fermion DM. If the
vector-like fermions are heavier than the DM, on the other hand, the
tree-level annihilation processes into the vector-like fermions are
kinematically forbidden and they annihilate into the SM particles at
loop level; therefore, in this case the annihilation cross section
is much smaller than the former case. To obtain a sizable annihilation
cross section, we assume throughout this work that the vector-like
fermions are lighter than the Dirac fermion DM, though we may find a
parameter region in which the correct relic abundance is obtained even
when the DM particle is lighter than the vector-like fermions.

The Dirac fermion and its antiparticle annihilate efficiently
via $s$-wave annihilation processes. Expanding in powers of the
relative velocity between these particles, $v_{\rm rel}$, and keeping
only the leading contribution, we obtain the cross section of the
annihilation process $\chi \overline{\chi} \to \psi_{f_i} \overline{\psi}_{f_j}$,
$\sigma_{\rm ann}^{(i,j)}$, as
\begin{align} 
 \sigma_{\rm ann}^{(i,j)} &v_{\rm rel} \simeq 
\frac{N_c m_{\chi}^2}{32\pi}
\biggl[
1-\frac{m_{f_i}^2 + m_{f_j}^2}{2 m_{\chi}^2} + \frac{(m_{f_i}^2 -
 m_{f_j}^2)^2}{16 m_{\chi}^4}
\biggr]^{\frac{1}{2}}
\nonumber \\[3pt]
&\times \sum_{k,l}
\frac{1}{\bigl[\widetilde{m}_{f_k}^2 +m_{\chi}^2
 -\frac{1}{2}(m_{f_i}^2 + m_{f_j}^2)\bigr]\bigl[\widetilde{m}_{f_l}^2 +m_{\chi}^2
 -\frac{1}{2}(m_{f_i}^2 + m_{f_j}^2)\bigr]}
\nonumber \\[3pt]
&\times
\biggl[\left(
C^{*ik}_{f\chi L} C^{il}_{f\chi L} 
+C^{*ik}_{f\chi R} C^{il}_{f\chi R} 
\right)
\biggl(1+ \frac{m_{f_i}^2 - m_{f_j}^2}{4 m_{\chi}^2} \biggr)
+\left(
C^{*ik}_{f\chi R} C^{il}_{f\chi L} 
+C^{*ik}_{f\chi L} C^{il}_{f\chi R} 
\right)\frac{m_{f_i}}{m_{\chi}}
\biggr]\nonumber \\[3pt]
&\times
\biggl[\left(
C^{*jl}_{f\chi L} C^{jk}_{f\chi L} 
+C^{*jl}_{f\chi R} C^{jk}_{f\chi R} 
\right)
\biggl(1- \frac{m_{f_i}^2 - m_{f_j}^2}{4 m_{\chi}^2} \biggr)
+\left(
C^{*jl}_{f\chi R} C^{jk}_{f\chi L} 
+C^{*jl}_{f\chi L} C^{jk}_{f\chi R} 
\right)\frac{m_{f_j}}{m_{\chi}}
\biggr]~,
\end{align}
where $N_c = 3$ (1) for vector-like quark (lepton) final states. 
The couplings $C^{ik}_{f\chi L/R}$ are given in
Appendix~\ref{app:dmcoups}. 
Notice that this $s$-wave contribution remains sizable even if the
DM couplings are purely chiral (either $C^{ik}_{f\chi R}$ or
$C^{ik}_{f\chi L}$ vanishes) and the final state fermions are massless
($m_{f_i} = 0$), contrary to the case of Majorana fermion DM.  

To roughly estimate the typical size of the annihilation cross section,
we take $a_f = a_{\bar{f}} = a$, $b_f = b_{\bar{f}} = \kappa_{\bar{f}} =
\kappa_{\bar{f}}^\prime = A_{\bar{f}} = \theta_\chi = 0$, $m_{f_i} = m_f$, and
$\widetilde{m}_{f_i} = \widetilde{m}_f$. We then have
\begin{align}
  \sigma_{\rm ann} v_{\rm rel} &\simeq \frac{N_c |a|^4}{32\pi}\biggl[
1-\frac{m_{f}^2}{m_{\chi}^2}
\biggr]^{\frac{1}{2}}
\frac{m_{\chi}^2}{\left[\widetilde{m}_f^2 + m_{\chi}^2
 -m_f^2\right]^2}
\nonumber \\[3pt]
&\simeq N_c |a|^4 \biggl[
1-\frac{m_{f}^2}{m_{\chi}^2}
\biggr]^{\frac{1}{2}} 
\biggl(\frac{m_{\chi}}{1~{\rm TeV}}\biggr)^2
\biggl(\frac{1~{\rm TeV}^2}{\widetilde{m}_f^2 + m_{\chi}^2
 -m_f^2}\biggr)^{-2} \times 10^{-25}~{\rm cm}^3/{\rm s} ~,
\label{eq:appanncros}
\end{align}
for each vector-like fermion. It is known that the thermal relic abundance of
Dirac fermion DM reproduces the observed value of the DM density,
$\Omega_{\rm DM}h^2\simeq 0.12$ \cite{Aghanim:2018eyx}, if its thermal-averaged
annihilation cross section is $\simeq 2 \times (2-3) \times 10^{-26}
~{\rm cm}^3/s$, where the factor of two is included as both DM
fermion and anti-fermion contribute to the relic density. This simple
estimation shows that the thermal
relic abundance of the singlet Dirac fermion DM agrees to the correct DM density
if the DM, vector-like fermions, and new scalars lie around the TeV
scale and the relevant couplings are ${\cal O}(1)$, which we confirm
below. 

In principle, one could require the condition that the thermal relic
abundance of the Dirac fermion DM should reproduce $\Omega_{\rm DM}h^2\simeq 0.12$
\cite{Aghanim:2018eyx} so that a parameter in this model is fixed as a
function of the other parameters, and make predictions for other
observables with this constraint imposed. It turns out, however, that
this restriction is not so much instructive; first and foremost, the
present model has many free parameters and this constraint does not
improve the predictivity of this model so much. Second, 
if the scalar particles are degenerate with the DM in
mass (with $\lesssim 10\%$ mass degeneracy), the thermal relic
abundance significantly decreases because of the coannihilation
effects \cite{Griest:1990kh}, while other observables are rather
insensitive to this degeneracy. In particular, when the colored scalars
are degenerate in mass with the DM, the Sommerfeld effect
\cite{Hisano:2003ec,Hisano:2004ds,Hisano:2006nn} and bound-state effect
\cite{Ellis:2015vaa,Liew:2016hqo} due to QCD interactions significantly
reduce the relic
abundance. This again spoils a tight correlation between the DM relic
density and other observables. Moreover, the Dirac fermion DM accommodates
various possibilities that affect its relic abundance, such as the
asymmetric DM scenario \cite{Hut:1979xw, *Nussinov:1985xr,
  *Barr:1990ca, *Barr:1991qn, *Kaplan:1991ah, *Dodelson:1991iv,
  *Kuzmin:1996he, *Hooper:2004dc, *Kitano:2004sv, *Farrar:2005zd,
  *Kaplan:2009ag, *Nagata:2016knk}, the presence of extra light U(1)
gauge bosons and/or singlet scalar fields, and so on. In most of the
cases, these new possibilities have little effect on the detectability
of the DM, while the regions favored by thermal relic
abundance are considerably changed.  For these reasons, in the
following analyses, we do not strictly impose the relic abundance
condition; we regard all variables in our model as free parameters and
show the parameter region favored by the thermal relic abundance just
for reference. We however note in passing that we assume the standard cosmological
history in our calculations of the DM relic abundance---deviations from 
the standard picture, such as the late-time entropy production after the
DM freeze-out, may also affect the DM relic abundance significantly.

%%%%%%%%%%%%%%%%%%%%%%%%%%%%%%%%%%%%%%%
\section{Direct Detection}
\label{sec:dd}
%%%%%%%%%%%%%%%%%%%%%%%%%%%%%
%%%%%%%%%%

In our simplified Dirac fermion DM models, the singlet Dirac fermion DM
couples with the SM particles through the radiative
corrections. Thus, the DM candidate may be tested in
direct DM searches. In this section, we calculate the relevant
effective interactions of the DM with SM fields, and then evaluate the
expected event rates in DM direct detection experiments.

%%%%%%%%%%%%%%%%%%%%%%%%%%%%%%%%%%%%%%%%%%%%%%%%%%%%%%%
\subsection{Effective interactions}
%%%%%%%%%%%%%%%%%%%%%%%%%%%%%%%%%%%%%%%

%%%figure%%%
\begin{figure}[t]
 \centering
  \includegraphics[height=40mm]{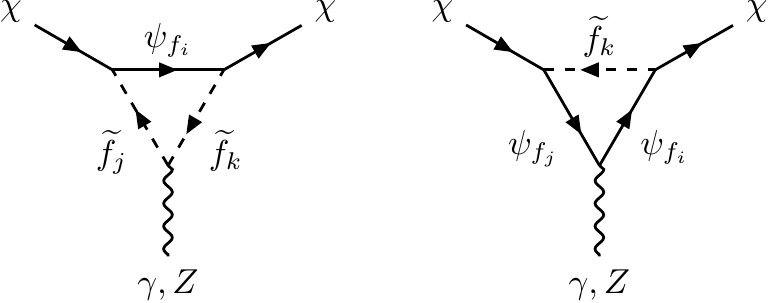}
 \caption{Feynman diagrams for effective DM-$\gamma/Z$ interactions.}
 \label{fig:DM-gamma}
\end{figure}
%%%figure%%%

To begin with, let us consider the effective couplings of DM with photon
($\gamma$). The Dirac fermion DM couples with photon at one-loop level
through the Feynman diagrams depicted in Fig.~\ref{fig:DM-gamma}. 
The relevant DM-$\gamma$ interactions for the
direct
detection experiments are given by
\begin{align}
\mathcal{L}_{\text{eff--}\gamma}
&=
\frac{1}{2}C^\gamma_M\overline{\chi}\sigma^{\mu\nu}\chi F_{\mu\nu}
-
\frac{i}{2}C^\gamma_E\overline{\chi}\sigma^{\mu\nu}\gamma_5\chi F_{\mu\nu}
+
C^\gamma_R\overline{\chi}\gamma^\mu\chi \partial^\nu F_{\mu\nu}~,
\label{eq:Effint_photon}
\end{align}
with $F_{\mu\nu}$ being the field strength of the photon field, $A_\mu$: 
$F_{\mu\nu}\equiv\partial_\mu A_\nu-\partial_\nu A_\mu$. Here
$C^\gamma_M$, $C^\gamma_E$, and $C^\gamma_R$ denote the Wilson
coefficients for the DM magnetic dipole moment (DM-MDM), the
DM electric dipole moment (DM-EDM), and the DM
charge radius (DM-CR), respectively. These Wilson coefficients are
obtained through the matching of our models onto the effective theory
with these interactions; the concrete expressions for these Wilson
coefficients are summarized in Appendix
\ref{app:DM-photon}. 
The DM-EDM is induced in the presence of non-zero CP-phases in the
DM-mediator couplings. 
We note in passing that we have ignored the DM anapole moment, described
by the effective operator $\bar{\chi}\gamma^\mu\gamma_5\chi\partial^\nu
F_{\mu\nu}$, as this contribution is always subdominant due to the
velocity suppression.

As we see in Sec.~\ref{sec:models}, there are many free parameters in
both Model I and II. To simplify the analysis, in what follows, we focus
on the following five cases and study their phenomenologies:
\begin{itemize}
\item Model I-A: Model I with  $a_Q=a_{\bar{u}}=a_{\bar{d}}\equiv a$ and
	     $b_Q=b_{\bar{u}}=b_{\bar{d}}\equiv b$. 
\item Model I-B: Model I with $a_Q=a_{\bar{d}}=b_Q=b_{\bar{d}}=0$,
	     $a_{\bar{u}}\equiv a$, and $b_{\bar{u}} \equiv b$.
\item Model I-C: Model I with $a_Q=a_{\bar{u}}=b_Q=b_{\bar{u}}=0$,
	     $a_{\bar{d}}\equiv a$, and $b_{\bar{d}}\equiv b$.
\item Model II-A: Model II with $a_L=a_{\bar{e}}\equiv a$ and $b_L=b_{\bar{e}}\equiv b$.
\item Model II-B: Model II with $b_L=a_L=0$, $a_{\bar{e}}\equiv a$,
	     and $b_{\bar{e}}\equiv b$.
\end{itemize}
The rest of the parameters are set to be certain appropriate values in each
analysis.

%%%figure%%%
%%%%%%%%%%
%%%figure%%%
\begin{figure}[t]
 \centering
  \includegraphics[width=50mm]{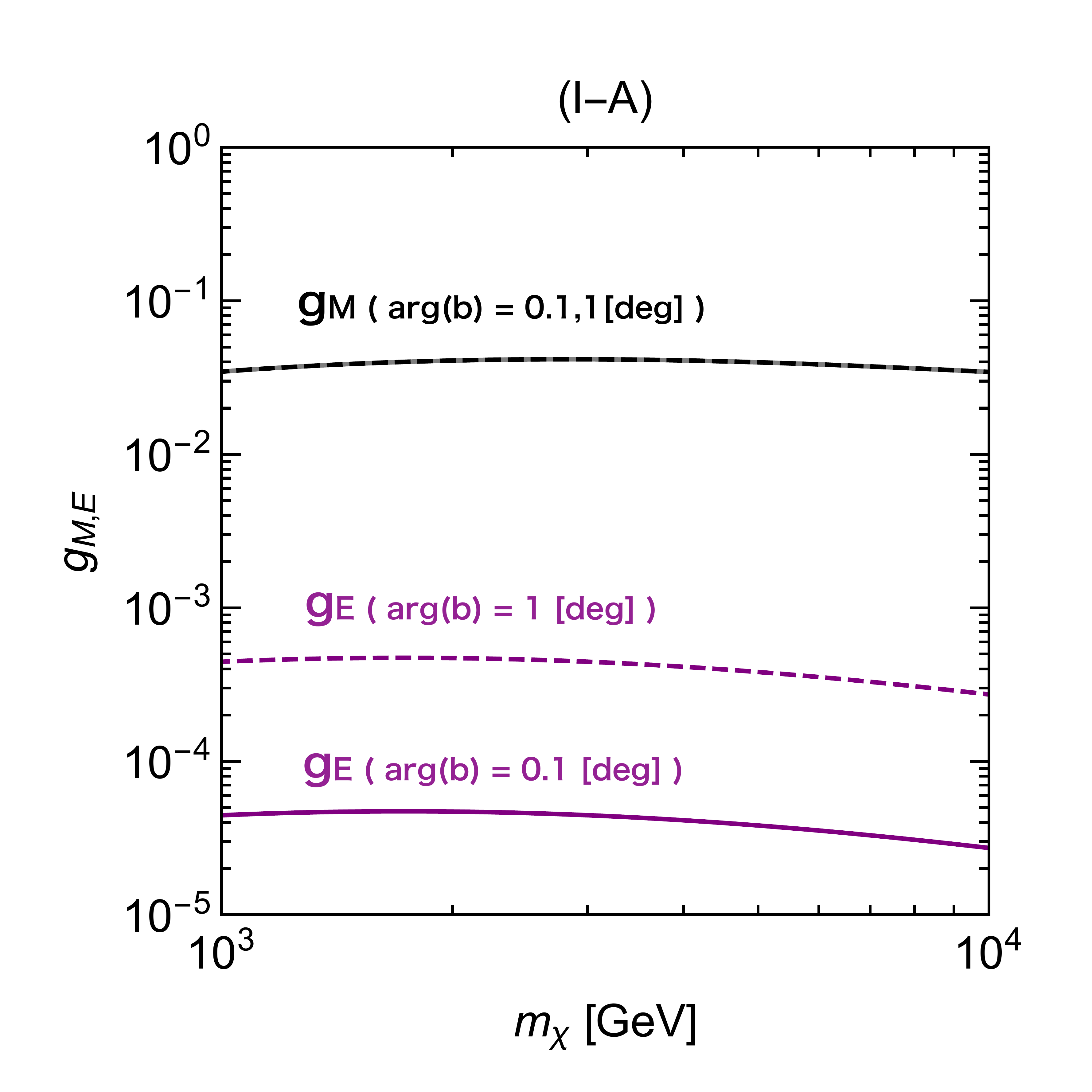}~~
  \includegraphics[width=50mm]{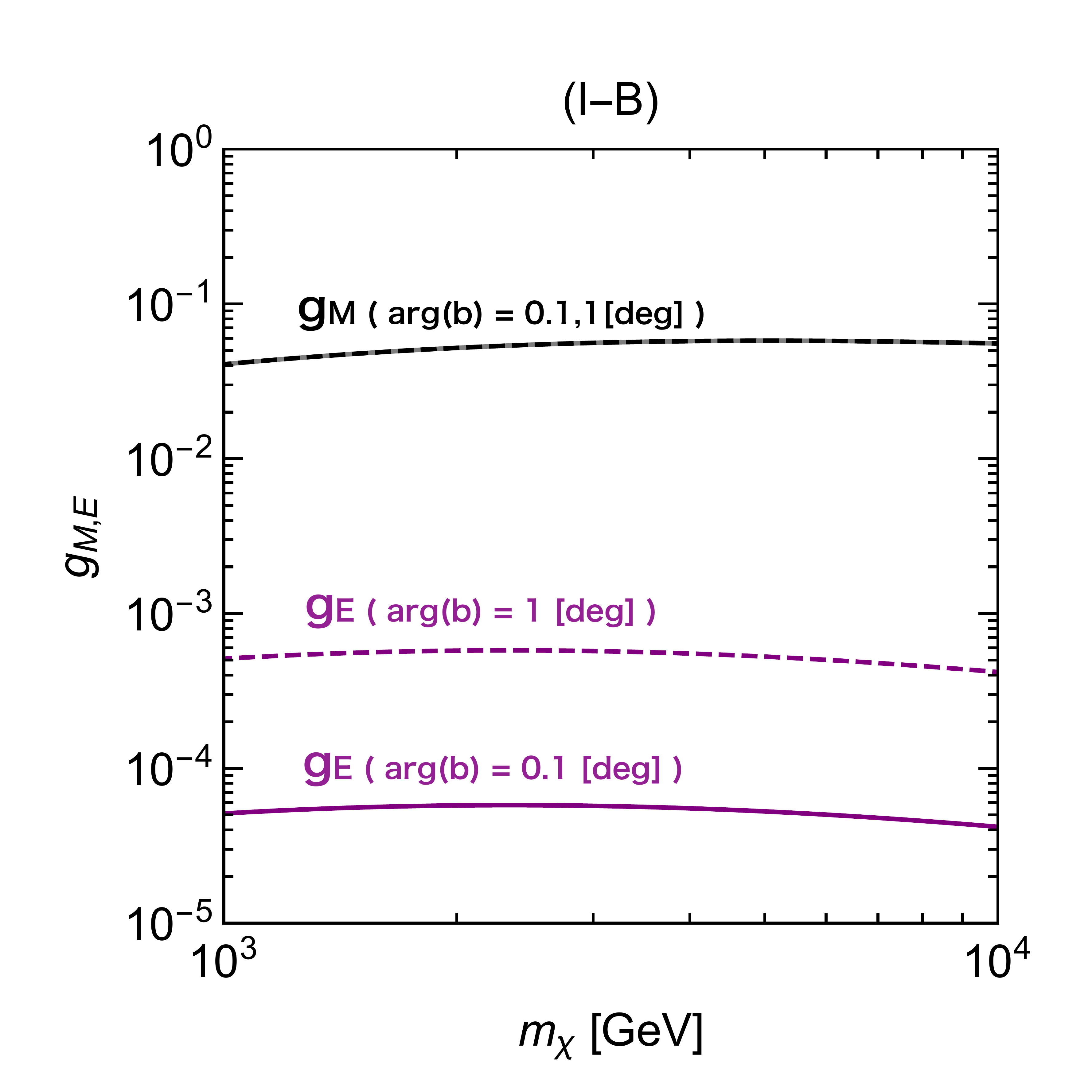}~~
  \includegraphics[width=50mm]{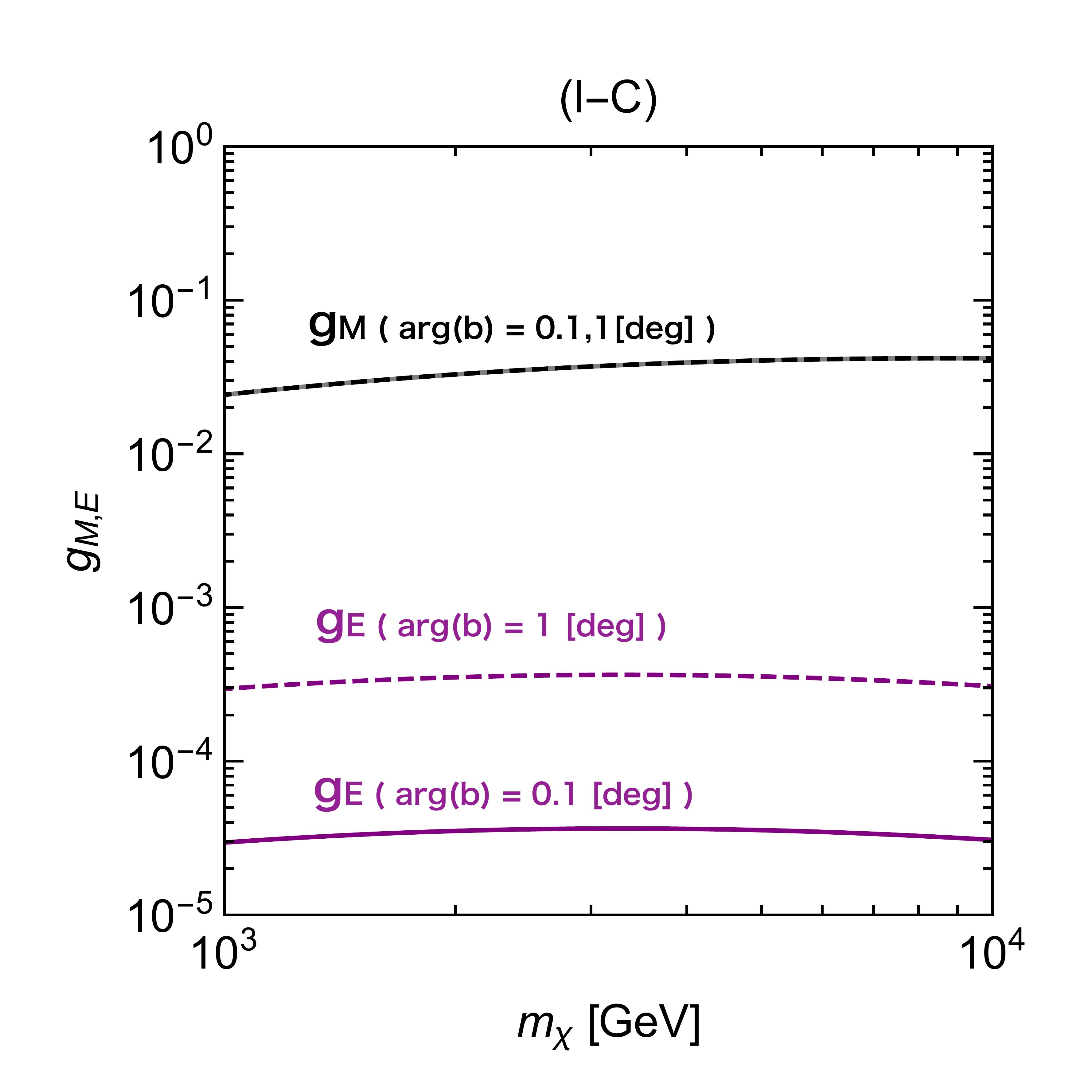}
  \\
  \includegraphics[width=50mm]{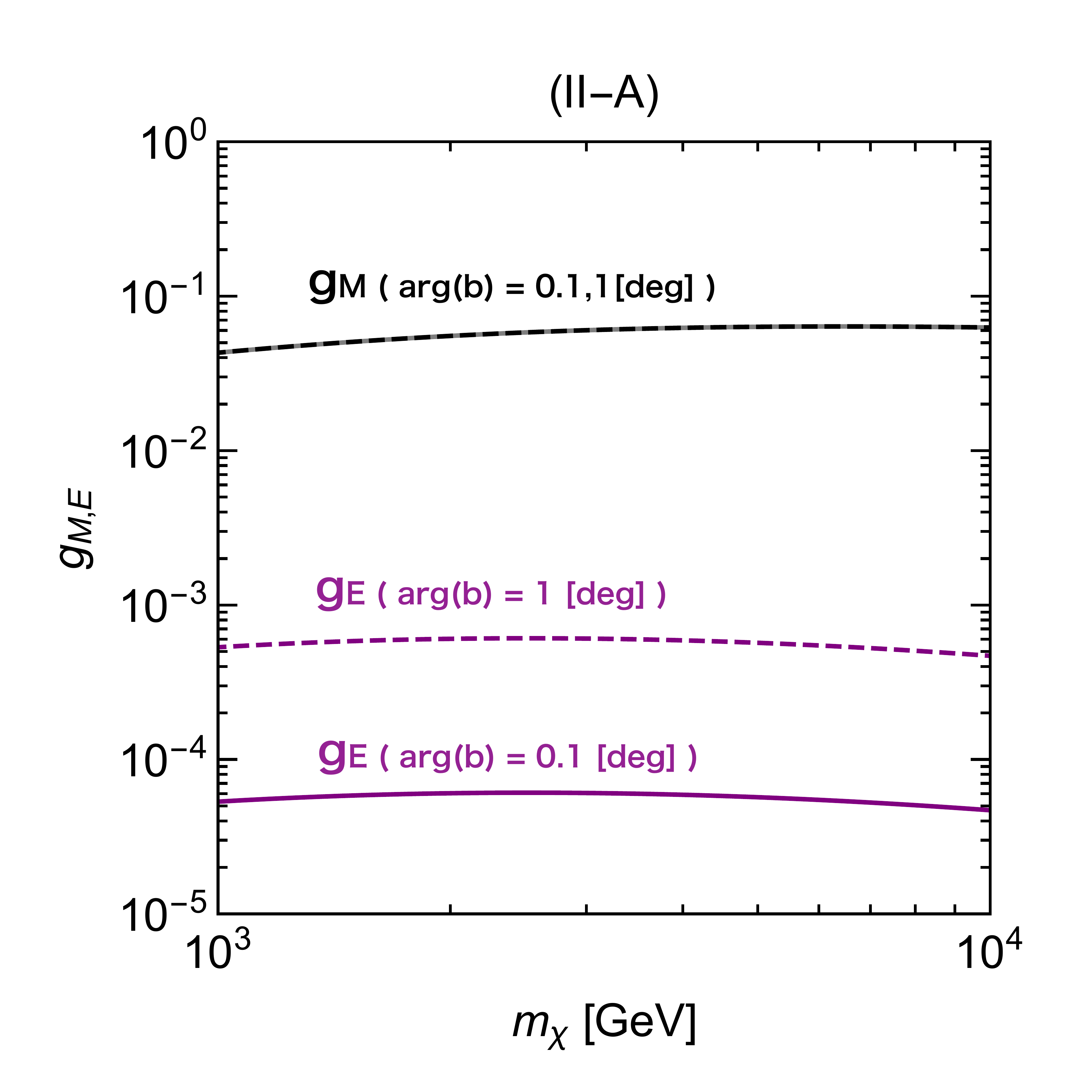}~~
  \includegraphics[width=50mm]{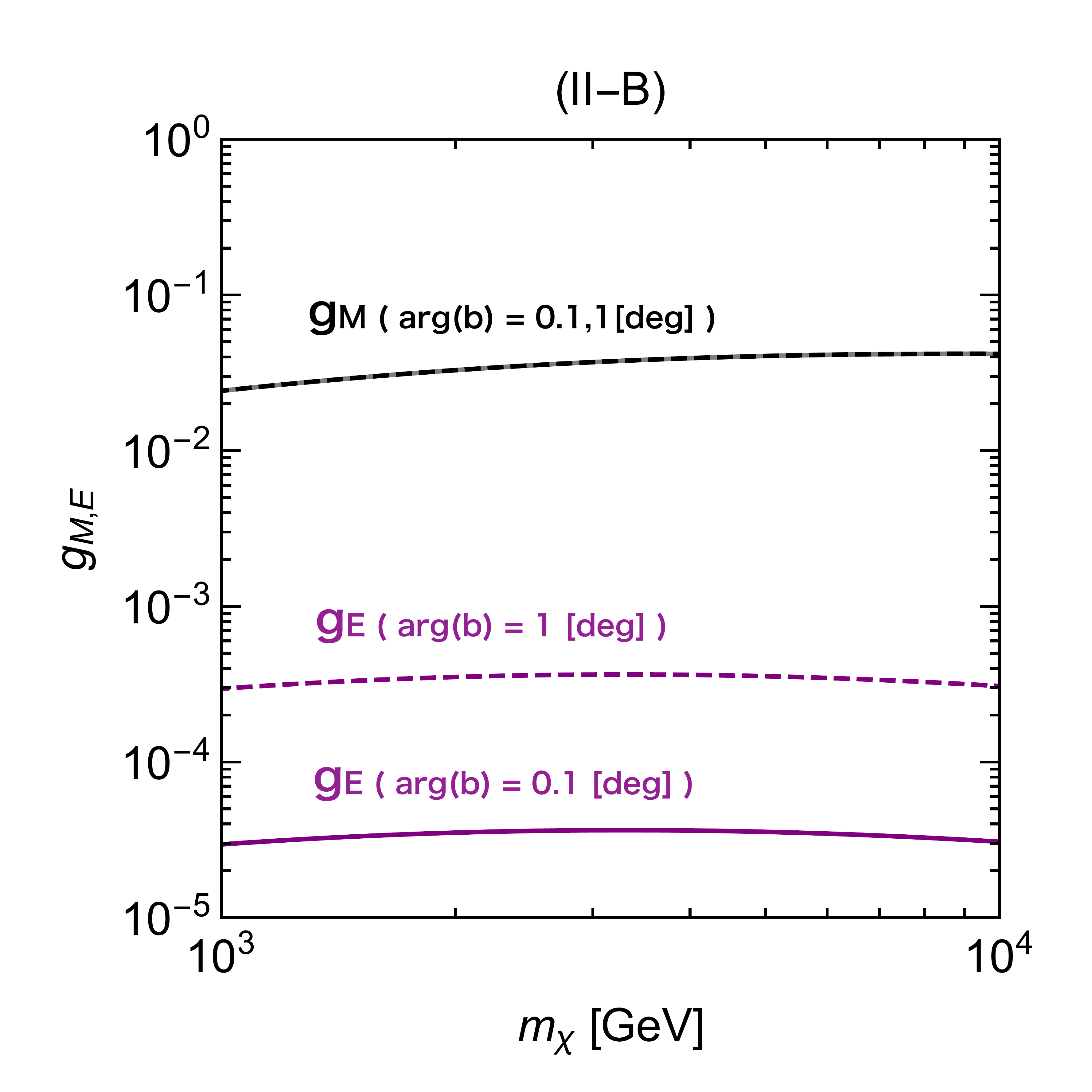}
 \caption{DM-MDM and DM-EDM, $g_M$ and $g_E$, as functions of DM mass $m_\chi$ in the black and
 purple lines, respectively. We take $a=|b|=1$, 
$\lambda^\prime_Q=\lambda^\prime_L=0$,
$\lambda_f=\kappa_{\bar{f}}=\kappa^\prime_{\bar{f}}=0.5$, 
$\mu_Q= \mu_L = 800$~GeV, $\mu_{\bar{u}} = 750$~GeV, $\mu_{\bar{d}} =
\mu_{\bar{e}} = 700$~GeV, while 
$\widetilde{m}_Q = \widetilde{m}_L = 1.2 M$,
$\widetilde{m}_{\bar{u}} = 1.1 M$,
$\widetilde{m}_{\bar{d}} = \widetilde{m}_{\bar{e}}= M$, and
$A_{\bar{f}} = 2 M$ $(M = 1.1 m_\chi)$. The solid and dashed lines correspond
to $\text{arg}(b) = 0.1$ and 1~degrees, respectively.}
 \label{fig:gME}
\end{figure}
%%%figure%%%

In Fig.~\ref{fig:gME}, we show the values of the DM-MDM
and DM-EDM as functions of the DM mass $m_\chi$ in the black and purple
lines, respectively, where the Wilson coefficients are normalized such
that  
\begin{equation}
 C_M^\gamma \equiv \frac{eg_M}{4m_\chi} ~, \qquad
 C_E^\gamma \equiv \frac{eg_E}{4m_\chi} ~.
\label{eq:gmgedef}
\end{equation}
In all of these plots we take $a=|b|=1$, 
$\lambda^\prime_Q=\lambda^\prime_L=0$,
$\lambda_f=\kappa_{\bar{f}}=\kappa^\prime_{\bar{f}}=0.5$, 
$\mu_Q= \mu_L = 800$~GeV, $\mu_{\bar{u}} = 750$~GeV, $\mu_{\bar{d}} =
\mu_{\bar{e}} = 700$~GeV, while
$\widetilde{m}_Q = \widetilde{m}_L = 1.2 M$,
$\widetilde{m}_{\bar{u}} = 1.1 M$,
$\widetilde{m}_{\bar{d}} = \widetilde{m}_{\bar{e}}=M$, and
$A_{\bar{f}} = 2 M$
$(M= 1.1 m_\chi)$.\footnote{
Let us give some comments on the present LHC bounds on the masses of the
extra scalars and fermions. For vector-like quarks, the limits strongly
depend on their decay modes. If they decay into third-generation quarks,
the present limits on the masses are as strong as $\gtrsim 1.3$~TeV
\cite{ATLAS-CONF-2018-032, Sirunyan:2018omb}. If, on the other hand,
they can decay into only the light quarks, the limits can be lower than
500~GeV, depending on their decay channels \cite{Aad:2015tba,
Sirunyan:2017lzl}. For colored scalars, we refer to squark searches; for
a DM mass of $\gtrsim 1$~TeV, colored scalars evade the LHC limits as
long as their masses are larger than the DM mass \cite{Aaboud:2017vwy,
Sirunyan:2017cwe}. The bounds on non-colored vector-like fermions and
scalars are much weaker than those on colored particles. Taking account
of these limits, in our analysis, we set the vector-like fermion masses
to be $\gtrsim 700$~GeV and the scalar masses to be $ > m_\chi \geq
1$~TeV.  
} 
The solid and dashed lines correspond
to $\text{arg}(b) = 0.1$ and 1~degrees, respectively. These plots show that
both DM-MDM and EDM described by $g_M$ and $g_E$, respectively, have
little dependence on the DM mass, as we fix the ratios between the DM
mass and the scalar masses. In addition, the DM-MDM is almost
independent of the CP phases in the DM-mediator couplings, while the
DM-EDM strongly depends on these phases as expected.

%%%%%%%%%%
%%%figure%%%
\begin{figure}[t]
 \centering
  \includegraphics[height=40mm]{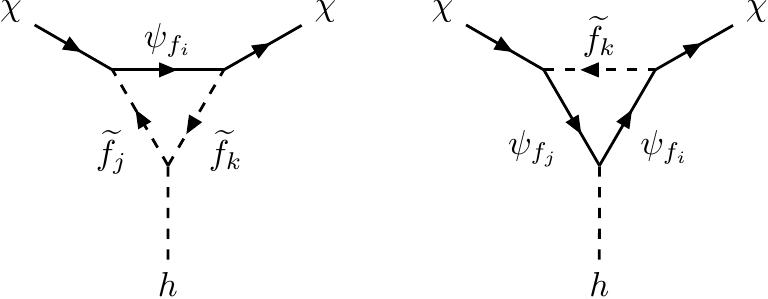}
 \caption{Feynman diagrams for effective DM-Higgs interactions.}
 \label{fig:DM-Higgs}
\end{figure}
%%%figure%%%

%%%%%%%%%%%%%% FIGURE %%%%%%%%%%%%%%%%%%%%%%%%%%%%%%%%%%%%
\begin{figure}[t]
  \centering
  \subcaptionbox{}{
  \includegraphics[width=0.25\columnwidth]{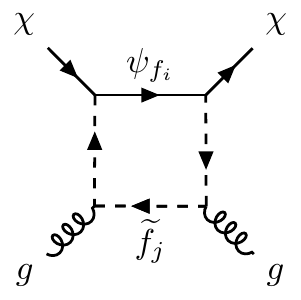}}
  \subcaptionbox{}{
  \includegraphics[width=0.27\columnwidth]{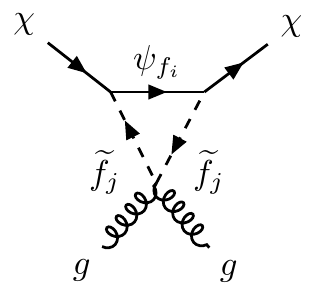}} \\
  \subcaptionbox{}{
  \includegraphics[width=0.3\columnwidth]{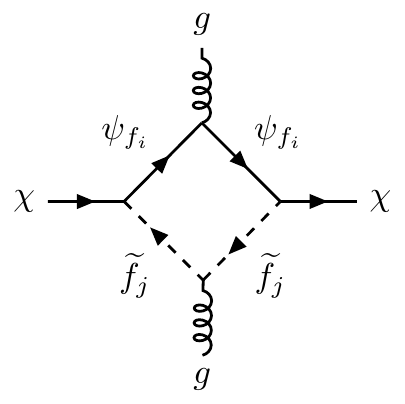}}
  \subcaptionbox{}{
  \includegraphics[width=0.25\columnwidth]{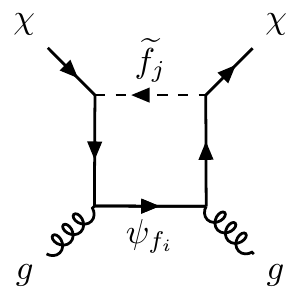}}
\caption{Feynman diagrams for effective DM-gluon interactions.} 
  \label{fig:dmg}
\end{figure}
%%%%%%%%%%%%%%%%%%%%%%%%%%%%%%%%%%%%%%%%%%%%%%%%%%%%%%%%%%

At one-loop level, the DM-$Z$ boson, DM-Higgs boson, and DM-gluon
interactions are also induced through the diagrams shown in
Fig.~\ref{fig:DM-gamma},  Fig.~\ref{fig:DM-Higgs}, and
Fig.~\ref{fig:dmg}, respectively. 
Explicit formulae for these effective interactions are given in
Appendix~\ref{app:DM-z}, \ref{app:DM-h}, and \ref{app:DM-g},
respectively. 
In what follows, we focus on the
spin-independent interactions since the limits on these interactions
from the direct detection experiments are much stronger than those on
the spin-dependent interactions. We also neglect the effective
interactions that are suppressed in the non-relativistic limit.

\begin{figure}[t]
 \begin{center}
  \includegraphics[width=50mm]{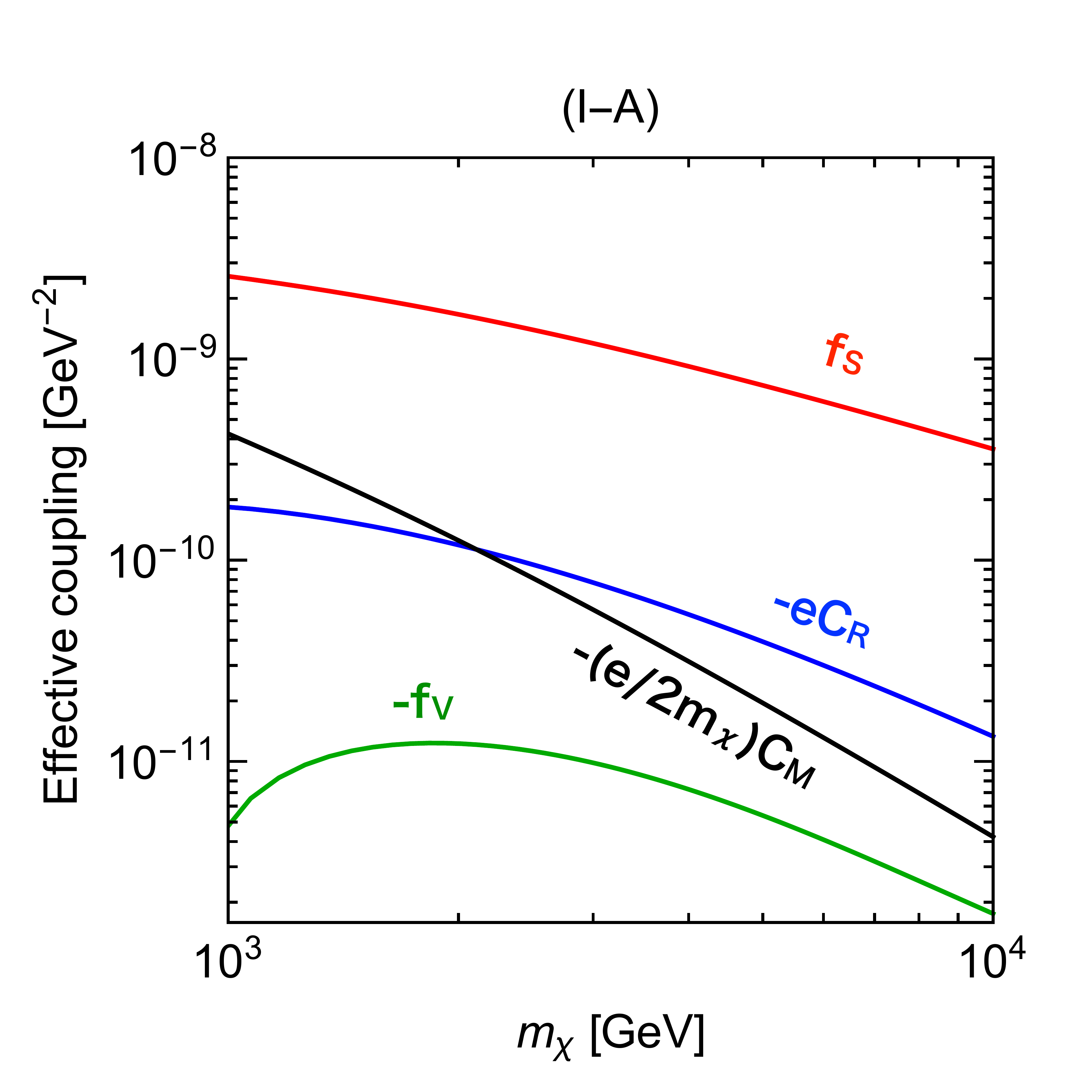}~~
  \includegraphics[width=50mm]{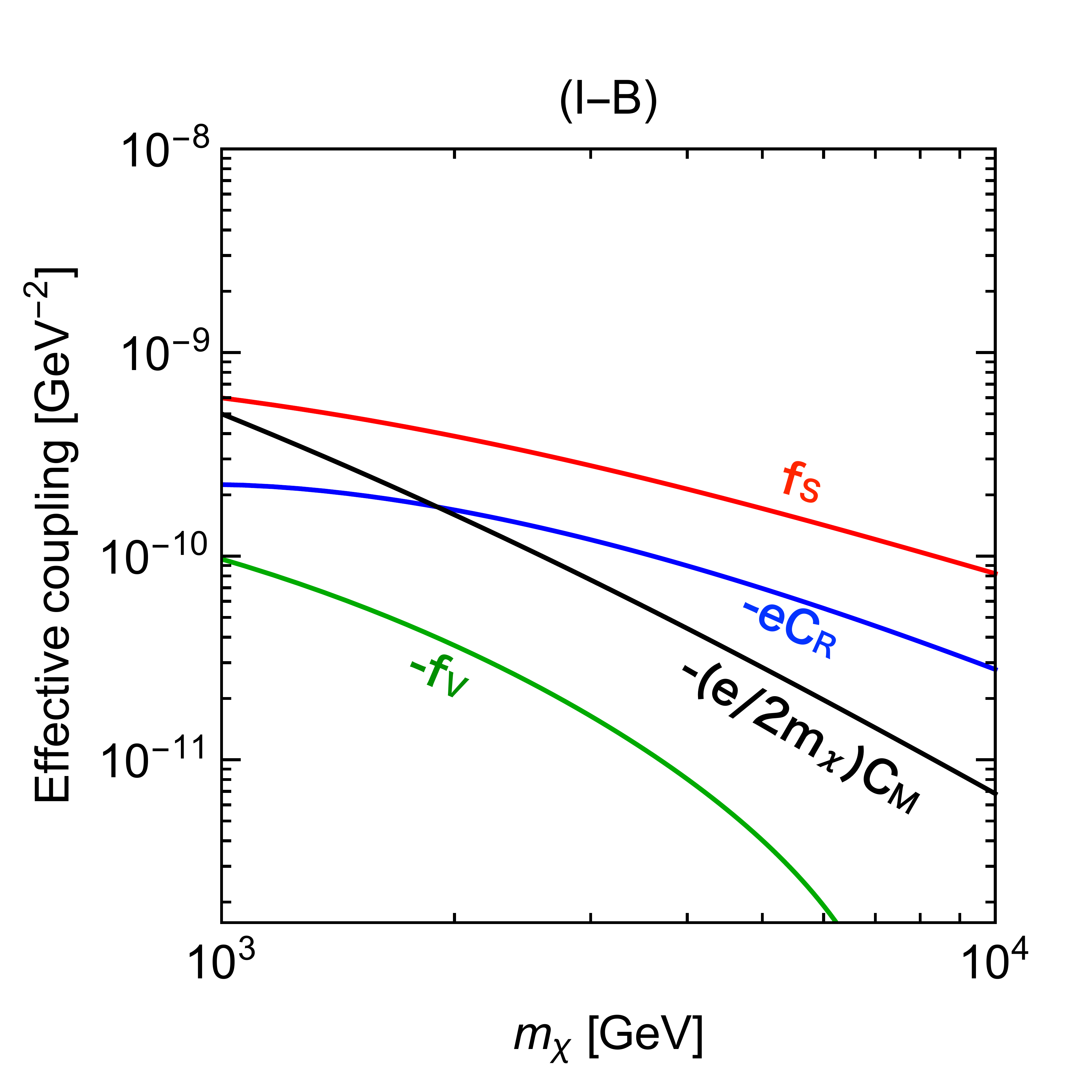}~~
  \includegraphics[width=50mm]{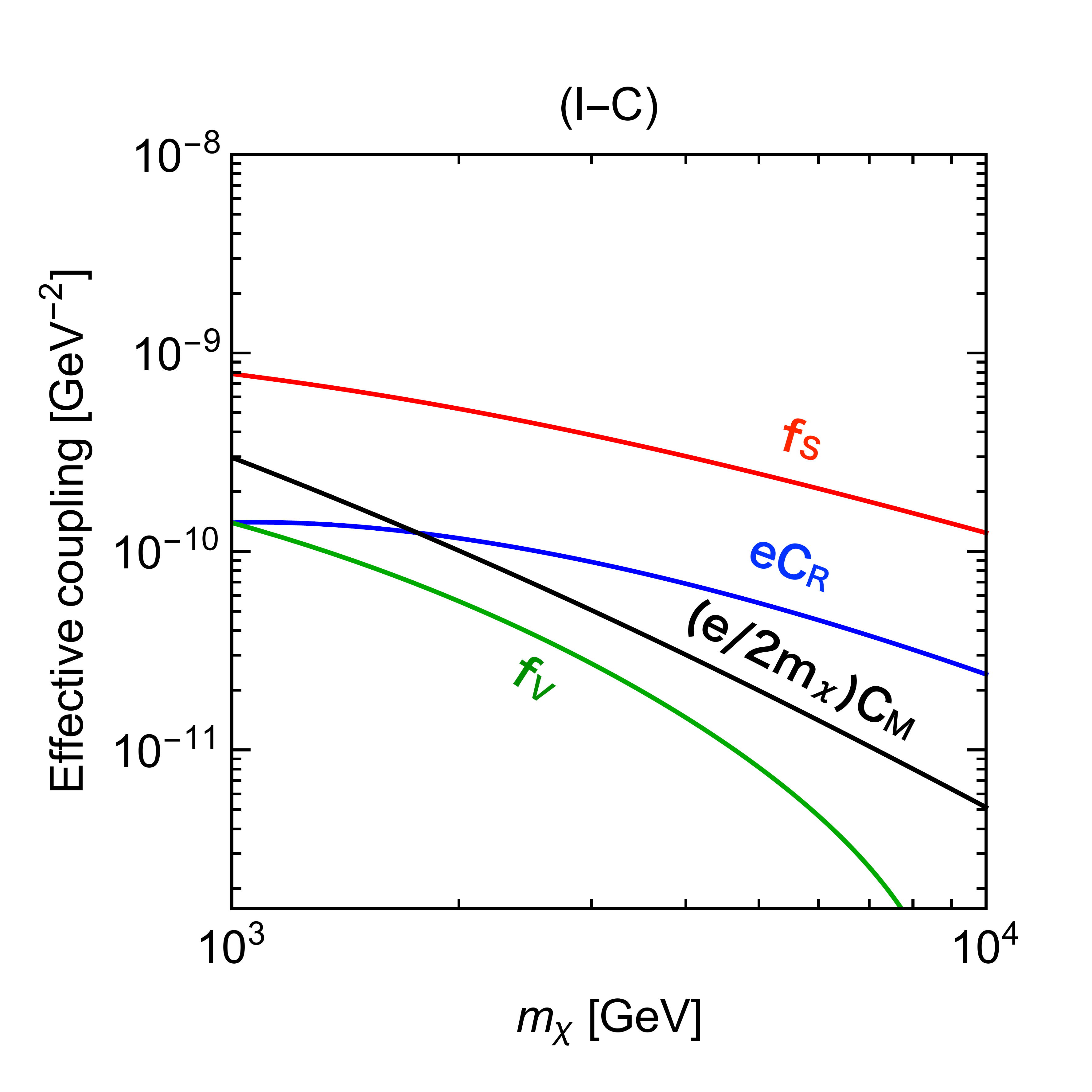}
  \\
   \includegraphics[width=50mm]{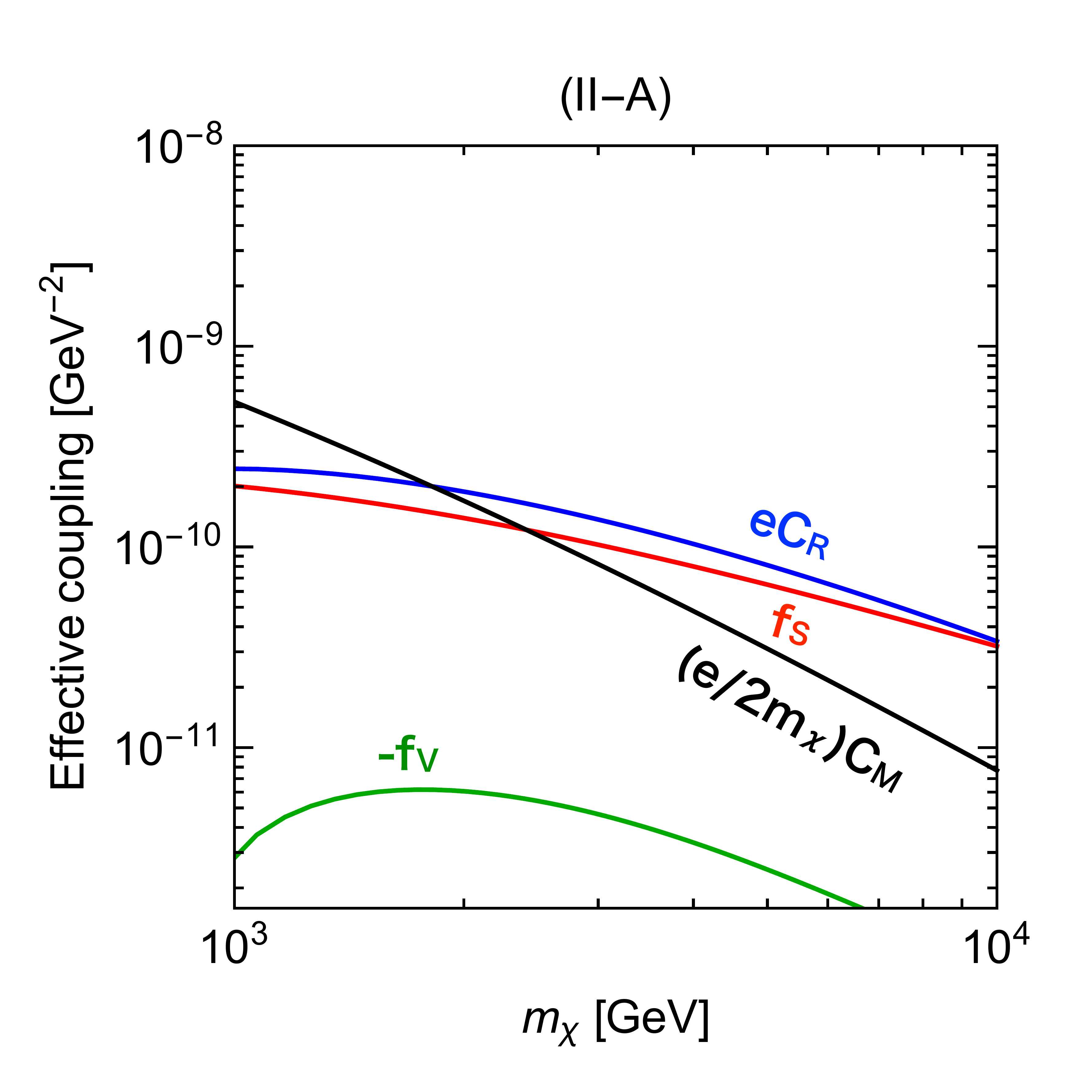}~~
  \includegraphics[width=50mm]{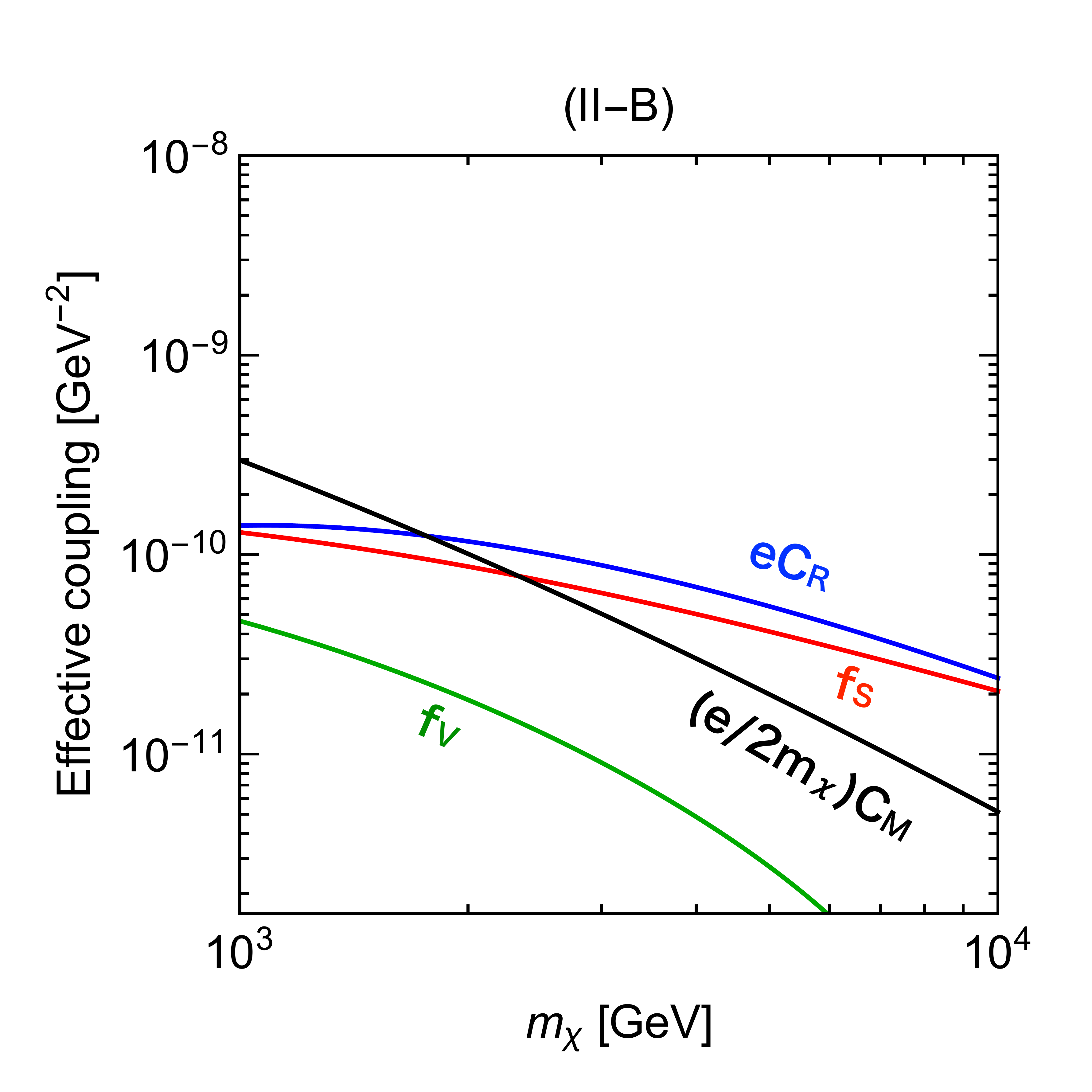}
 \end{center}
 \caption{Green, red, blue, and black lines show the absolute values of
   $f^{(p)}_V$, $f^{(p)}_S$, $eC^{\gamma}_R$, and
 $eC_{M}^\gamma/(2m_\chi)$ as functions of the DM mass, respectively,
 with the minus sign indicated if the corresponding quantity is negative. Here
   we take the same parameter sets as in Fig.~\ref{fig:gME}. }
 \label{fig:SVR}
\end{figure}

We further integrate out the $Z$ and Higgs bosons to obtain the
effective DM-quark/gluon operators. At low energies with three flavor
quarks, the relevant interactions are\footnote{We include the factor
$\alpha_s/\pi$ in the definition of the gluon scalar operator such that
it is invariant under the renormalization group (RG) flow at one-loop level
\cite{Hisano:2015bma, Hisano:2015rsa}. This operator runs and mixes with
the quark scalar operators at higher orders in $\alpha_s$, while the
vector operator is invariant under the RG flow. 

If we turn on the electroweak (EW) radiative corrections,
the vector operator mixes with other operators which we have not shown here \cite{Crivellin:2014qxa,DEramo:2014nmf,Brod:2018ust,Bishara:2018vix}.
For example, the $\bar{\chi}\gamma^\mu\chi \bar{q}\gamma_\mu\gamma_5\chi$ effective interaction induces the vector operator through the EW RG effect, which is order $\frac{y^2_t}{\pi^2}\ln(v/m_{\rm{med}})\sim\mathcal{O}(10\%)$.
It is however found that such contributions rarely affect the resultant 
scattering cross section since the contribution of the vector operator
is always subdominant, as we see in Fig.~\ref{fig:SVR}.
We thus ignore such effects in our analysis.
}

\begin{equation}
 {\cal L}_{\text{eff--}q/g} = 
C^q_V \overline{\chi} \gamma^\mu \chi \overline{q} \gamma_\mu q 
+C^q_S m_q \overline{\chi} \chi \overline{q}  q 
+ C_S^g \overline{\chi} \chi \cdot \frac{\alpha_s}{\pi} G^A_{\mu\nu}
G^{A\mu\nu} ~,
\label{eq:leffqg}
\end{equation}
where $m_q$ are the quark masses, $\alpha_s \equiv g_s^2/(4\pi)$ with
$g_s$ the strong gauge coupling constant, and $G^A_{\mu\nu}$ is the
gluon field strength tensor. 
The vector and scalar couplings $C_V^q$ and $C_S^q$ are generated via
the $Z$ and Higgs boson exchange processes, respectively, while the
gluonic interaction is induced by the Higgs exchange as well as the vector-like
quark loop diagrams shown in Fig.~\ref{fig:dmg}. 
The Wilson coefficients of these operators are given in
Appendix~\ref{app:DM-lowen}.

These quark/gluon operators then induce the spin-independent DM-nucleon
interactions, which are described by 
\begin{align}
\mathcal{L}_{\text{eff--}N}
=
f^{(N)}_V\overline{\chi}\gamma^\mu\chi \overline{N}\gamma_\mu N
+
f^{(N)}_S\overline{\chi}\chi \overline{N}N ~,
\end{align}
where $N=p,n$ represents the nucleon field. The DM-nucleon vector
coupling $f_V^{(N)}$ is readily obtained from the DM-quark couplings
$C_V^q$ as
\begin{equation}
 f_V^{(p)} = 2 C_V^u + C_V^d ~, \qquad
 f_V^{(n)} =  C_V^u + 2 C_V^d ~.
\end{equation}
On the other hand, to obtain the DM-nucleon scalar coupling $f_S^{(N)}$ 
from the quark/gluon scalar couplings $C_S^q$ and $C_S^g$, we need the
nucleon matrix elements of the quark scalar operators defined by
$f^{(N)}_{T_q} \equiv \langle N| m_q \bar{q}q |N\rangle /m_N$ with $m_N$
the nucleon mass. A recent compilation \cite{Ellis:2018dmb} gives 
\begin{align}
 f_{T_u}^{(p)} &= 0.018(5) ~, \qquad 
 f_{T_d}^{(p)}  = 0.027(7) ~, \qquad
 f_{T_s}^{(p)}  = 0.037(17) ~, \nonumber  \\
 f_{T_u}^{(n)} &= 0.013(3) ~, \qquad 
 f_{T_d}^{(n)}  = 0.040(10) ~, \qquad
 f_{T_s}^{(n)}  = 0.037(17) ~. 
\end{align}
At leading order in $\alpha_s$, we then have \cite{Shifman:1978zn} 
\begin{equation}
 \frac{f_S^{(N)}}{m_N} = \sum_{q=u,d,s} C^q_S f_{T_q}^{(N)}
- \frac{8}{9} C_S^g f_{T_G}^{(N)} ~,
\end{equation}
with $f_{T_G}^{(N)} \equiv 1- \sum_{q=u,d,s} f_{T_q}^{(N)}$.

%%%figure%%%
%%%figure%%%
%%%figure%%%
%%%%%%%%%

To see the significance of the contributions of the vector and scalar
interactions as well as that of the charge radius, in
Fig.~\ref{fig:SVR}, we plot the absolute values of $f^{(p)}_V$,
$f^{(p)}_S$, $eC^{\gamma}_R$, and $eC_{M}^\gamma/(2m_\chi)$ as
functions of $m_\chi$ in the green, red, blue, and black lines,
respectively, with the minus sign indicated if the corresponding
quantity is negative. 
As we see below (Eq.~\eqref{eq:dsigma}), these terms interfere with each
other to give a contribution to the spin-independent charge-charge
scattering of the DM with protons. We will see below that there is cancellation
among these contributions. 
We also note that although $f_V^{(N)}$ and $f_S^{(N)}$ are induced by the
dimension-less effective DM-$Z$ and DM-Higgs couplings, they are
suppressed as the DM mass gets large. This is because the generation of
these couplings requires the electroweak-symmetry-breaking effect
as mentioned in Appendix~\ref{app:DM-eff}, which results in an
additional suppression factor with the heavy mass scale. From the dimensional analysis,
$f_S^{(N)}$ is proportional to $1/m_\chi$ for $m_\chi\gg m_Z$ ($m_Z$:$Z$ boson mass), while $e C_R^\gamma$, $e C_M^\gamma/2m_\chi$, 
and $f_V^{(N)}$ are scaled as $1/m_\chi^2$. In Fig.~\ref{fig:SVR}, $e C_R^\gamma$ looks protortional 
to $1/m_\chi$ though we found it accidental.

%%%%%%%%%%%%%%%%%%%%%%%%%%%%%%%%%%%%%%%%%%%%%
\subsection{Scattering cross section}
%%%%%%%%%%%%%%%%%%%%%%%%%%%%%%%%%%%%%%%%%%%%%

By using the effective couplings obtained above, we evaluate the
differential scattering cross section of the DM with a target nucleus
with respect to the recoil energy $E_R$: 
\begin{align}
\frac{d\sigma_{\chi T}}{dE_R}
&= F^2_c(E_R)
\biggl[
\frac{Z^2e^2}{4\pi}\left(\frac{1}{E_R}
-\frac{1}{E^{\rm{max}}_R(v^2_{\rm{rel}})}\right)(C^\gamma_M)^2  
+
\frac{Z^2e^2}{4\pi v^2_{\rm{rel}}}\frac{1}{E_R}(C^\gamma_E)^2
\nonumber\\
&+
\frac{m_T}{2\pi v^2_{\rm{rel}}}\biggl|
Z\left(f^{(p)}_S+f^{(p)}_V-e C^\gamma_R-\frac{e}{2m_\chi}C^\gamma_M\right)
+
(A-Z)\left(f^{(n)}_S+f^{(n)}_V\right)
\biggr|^2 \biggr] ~,
\label{eq:dsigma}
\end{align}
where $A$, $Z$, and $m_T$ are the atomic number,  mass number, and mass of the target
nucleus, respectively. $\bm{v}_{\rm{rel}}$ is the relative velocity
of the DM and the target nucleus with $v_{\text{rel}} \equiv |\bm{v}_{\rm{rel}}|$, and
$E^{\rm{max}}_R(v^2_{\rm{rel}})={2m^2_\chi m_T
v^2_{\rm{rel}}}/{(m_\chi+m_T)^2}$ is the maximum recoil energy for a given
$v_{\rm{rel}}$. $F_c(E_R)$ is the nuclear form factor, for which we
exploit the Lewin-Smith parametrization \cite{Lewin:1995rx} of the Helm
form factor \cite{Helm:1956zz}:
\begin{equation}
 F_c^2 (E_R) = \biggl(\frac{3 j_1 (qR)}{qR}\biggr)^2 e^{-q^2 s^2} ~,
\end{equation}
with $j_1(x)$ a spherical Bessel function of the first kind, $q =
\sqrt{2 m_T E_R}$, $s=0.9$~fm, $R =\sqrt{c^2 + \frac{7}{3} \pi^2 a^2 -5s^2}$, $a=
0.52$~fm,  and $c= (1.23 A^{1/3} -0.60)$~fm.

%%%figure%%%
%%%figure%%%
\begin{figure}[t]
 \begin{center}
  \includegraphics[width=50mm]{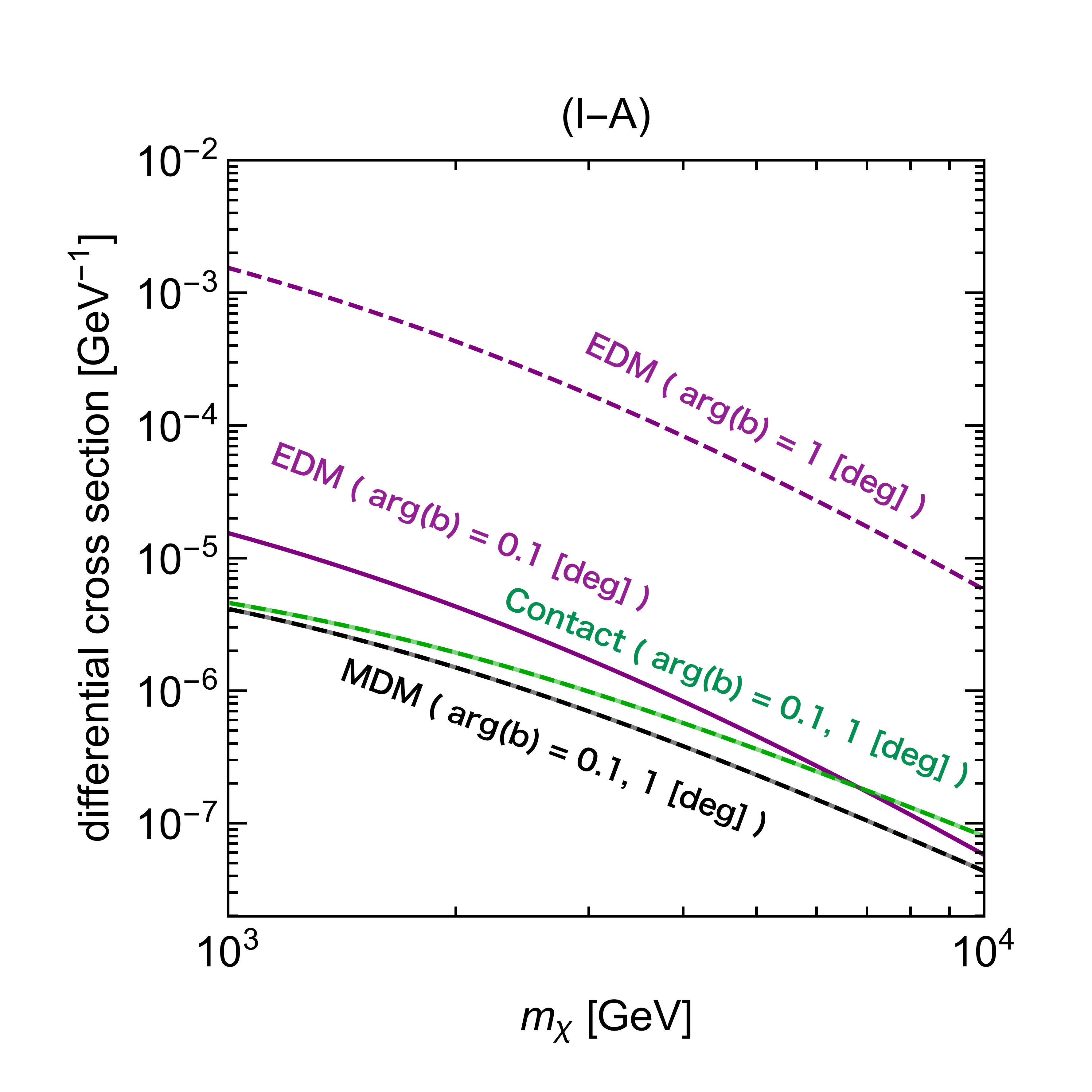}~~
  \includegraphics[width=50mm]{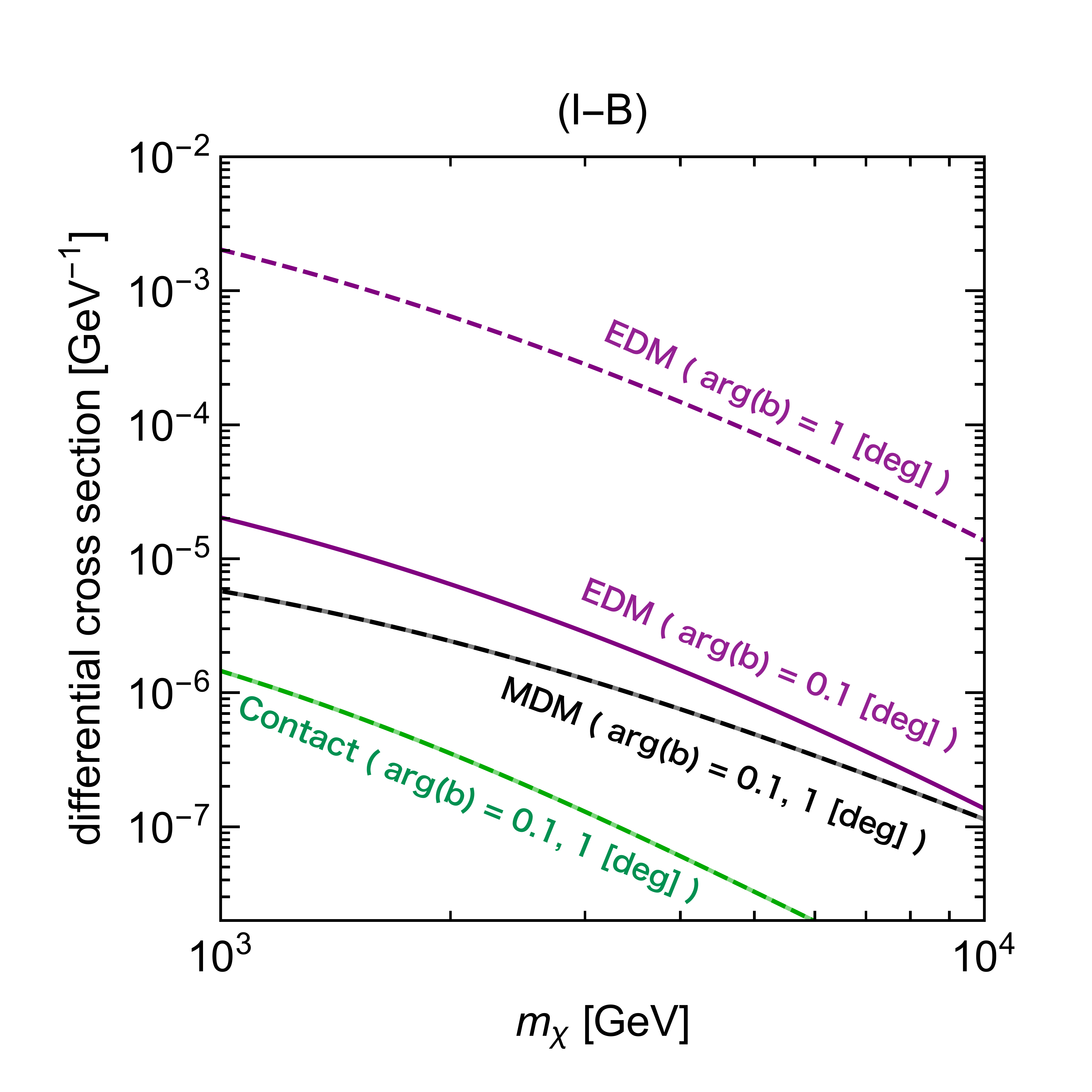}~~
  \includegraphics[width=50mm]{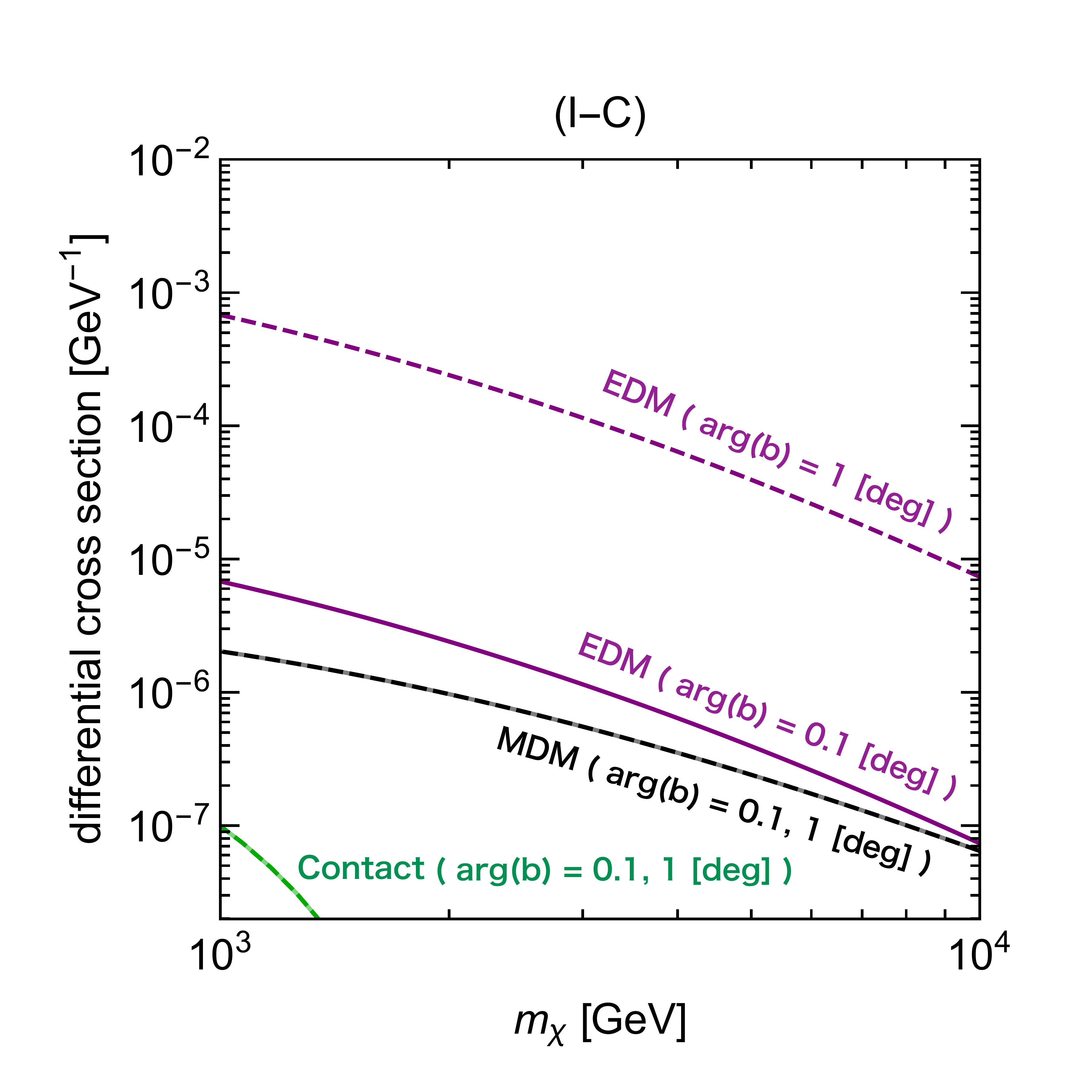}
  \\
   \includegraphics[width=50mm]{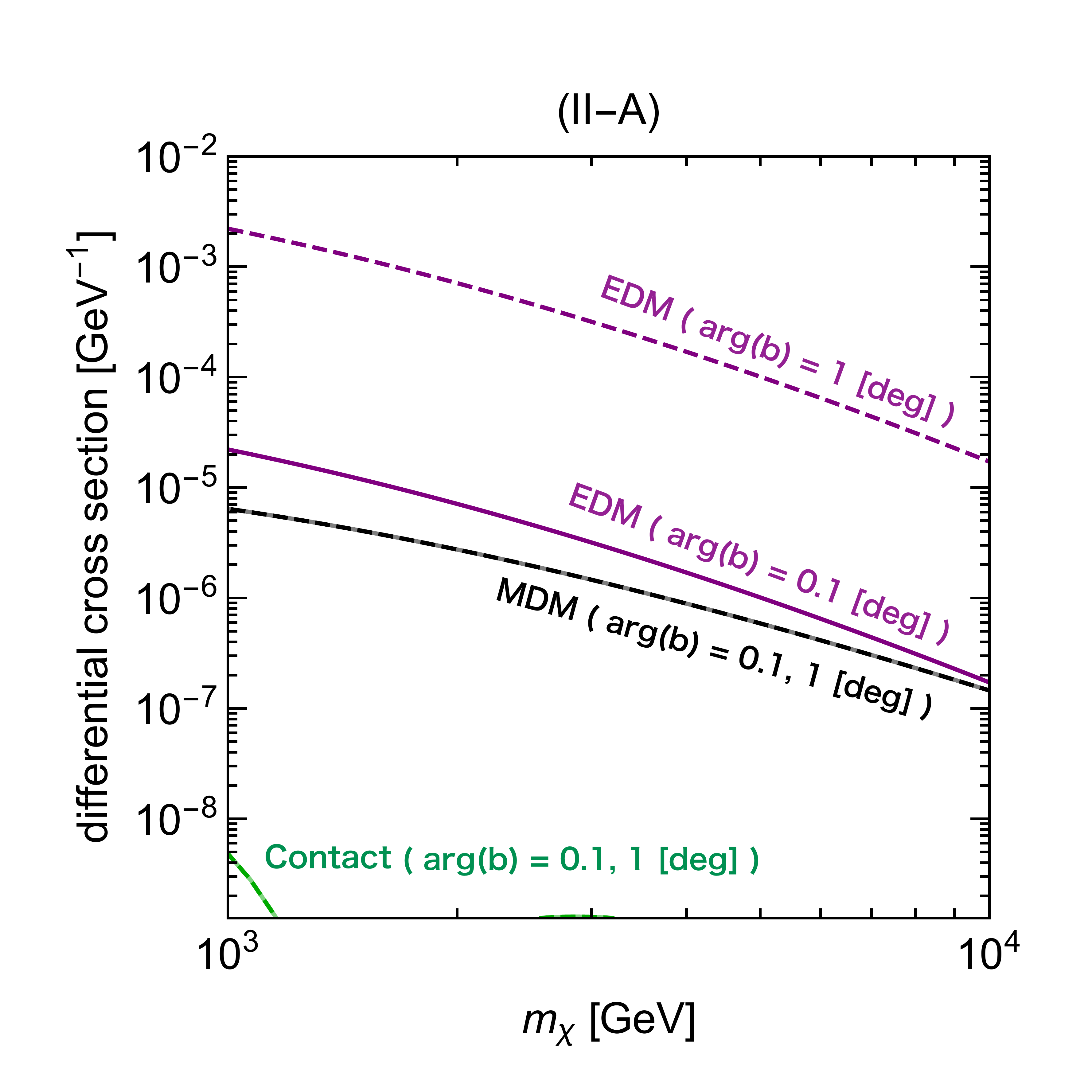}~~
  \includegraphics[width=50mm]{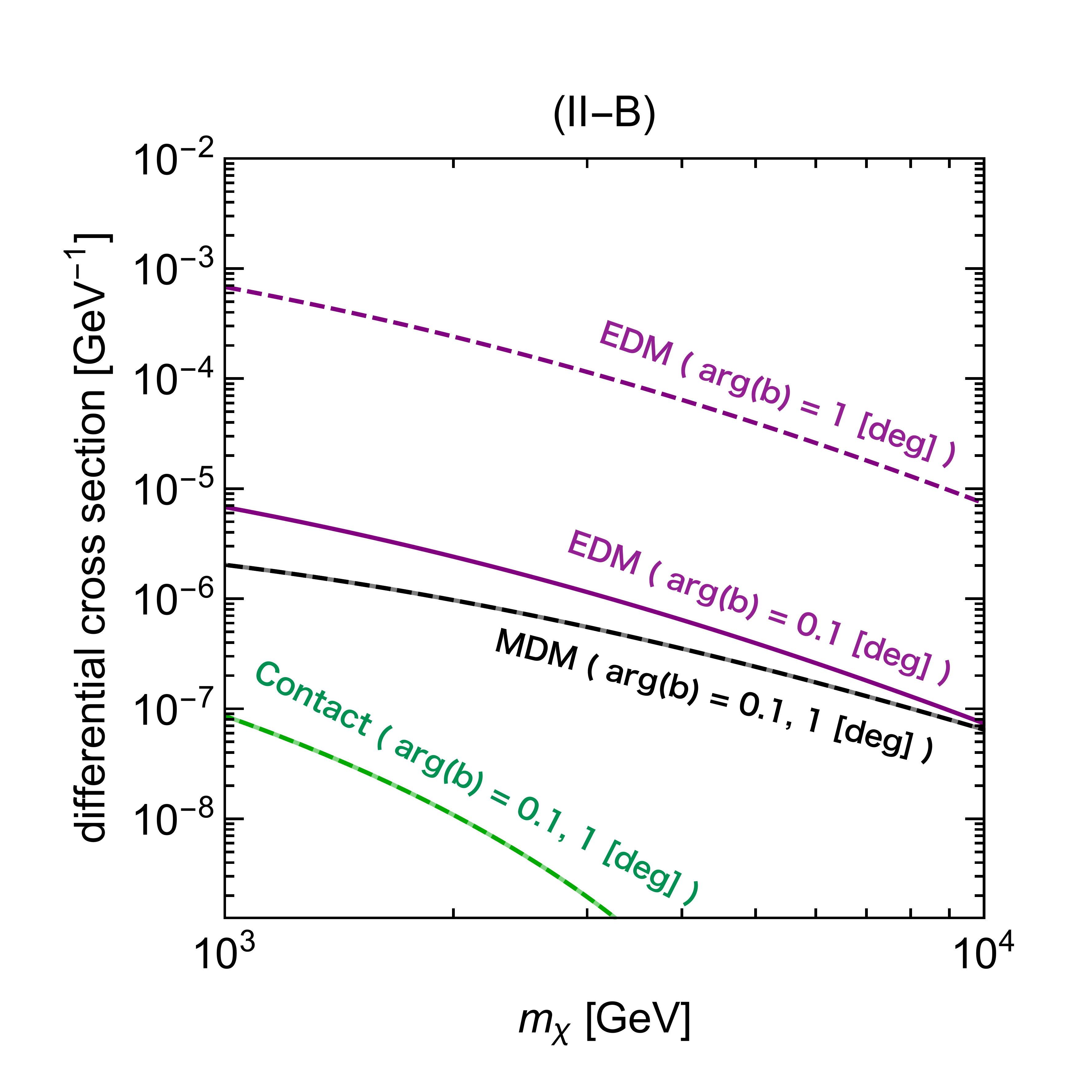}
 \end{center}
 \caption{The first, second, and third terms in the square bracket in
 Eq.~\eqref{eq:dsigma} as functions of DM mass in black, purple,
 and green lines, respectively. Here we take the same parameter sets as
 in Fig.~\ref{fig:gME}, with the solid (dashed) lines correspond to
 $\text{arg}(b) = 0.1$ (1) degrees, and fix the rest of the parameters in
Eq.~\eqref{eq:dsigma} to be $E_R = 30$~keV, $v_{\text{rel}} =
232~\text{km}/\text{s}$, $Z = 54$, $A=131$, and $m_T \simeq 122$~GeV (for $^{131}$Xe).  }
 \label{fig:diffxsec}
\end{figure}
%%%figure%%%
%%%%%%%%%

Figure~\ref{fig:diffxsec} shows the first, second, and third terms in
the square bracket in Eq.~\eqref{eq:dsigma} as functions of the DM mass
in the black, purple, and green lines, respectively;
we refer to them as the MDM, EDM, and Contact contributions,
respectively. Here we take the same parameter sets as in
Fig.~\ref{fig:gME}, with the solid (dashed) lines correspond to $\text{arg}(b)
= 0.1$ (1) degrees, and fix the rest of the parameters in
Eq.~\eqref{eq:dsigma} to be $E_R = 30$~keV, $v_{\text{rel}} =
232~\text{km}/\text{s}$, $Z = 54$, $A=131$, and $m_T \simeq 122$~GeV (for $^{131}$Xe). 
As we see, the Contact contributions in Model I-C, II-A, and II-B are much smaller than the MDM and EDM contributions 
due to the cancellation found in Fig.~\ref{fig:SVR}. Moreover, if there
is a sizable CP violation in the DM couplings, the EDM contribution
dominates the other contributions, since this contribution is strongly
enhanced at low recoil energy because of the long-range photon exchange,
as can be seen in Eq.~\eqref{eq:dsigma}.

%%%%%%%%%%%%%%%%%%%%%%%%%%
\subsection{Event rate}
\label{sec:eventrate}
%%%%%%%%%%%%%%%%%%%%%%%%%%

We are now ready to estimate the expected number of events in direct
detection experiments. The differential event rate of the DM-target
nuclei scattering process per
unit detector mass is 
\begin{align}
\frac{dR}{dE_R}
&=
\frac{\rho_{\rm{DM}}}{m_Tm_\chi}\int^\infty_{v_{\rm{min}}(E_R)}d^3
 \bm{v}_{\rm{rel}} \, 
 v_{\rm{rel}} \,f_{\oplus}(\bm{v}_{\rm{rel}})\frac{d\sigma_{\chi
 T}}{dE_R} ~, 
\label{eq:drder}
\end{align}
where $\rho_{\text{DM}}$ is the local DM density, 
$f_{\oplus}(\bm{v}_{\rm{rel}})$ is the DM velocity distribution in the lab frame,
and $v_{\rm{min}}(E_R)$ is the minimum speed required to yield 
recoil energy $E_R$:  
\begin{align}
v_{\rm{min}}(E_R)
=
\sqrt{\frac{(m_\chi+m_T)^2E_R}{2m^2_\chi m_T}} ~.
\end{align}
For the DM velocity distribution, we assume a Maxwell-Boltzmann
distribution in the galactic frame with a maximum speed that is set to
be equal to the galaxy escape velocity $v_{\text{esc}}$; namely,
\begin{equation}
 f_{\oplus}(\bm{v}) = f (\bm{v} + \bm{v}_\text{E}) ~,
\end{equation}
with $\bm{v}_{\text{E}}$ denotes the velocity of the Earth with respect
to the galactic frame and 
\begin{equation}
 f (\bm{v}) = 
\begin{cases}
 \frac{1}{N} e^{-v^2/v_0^2}  & (|\bm{v}| < v_{\text{esc}}) \\ 
 0& (|\bm{v}| > v_{\text{esc}}) 
\end{cases}
~,
\end{equation}
with 
\begin{equation}
 N = \pi^{3/2} v_0^3 \biggl[
\text{erf} \biggl(\frac{v_{\text{esc}}}{v_0}\biggr)
- \frac{2 v_{\text{esc}}}{\sqrt{\pi} v_0}  e^{-
\frac{v_{\text{esc}}^2}{v_0^2}} 
\biggr]~.
\end{equation}
From Eqs.~\eqref{eq:dsigma} and \eqref{eq:drder}, we see that to obtain
the differential event rate we need to perform the following integrals
with respect to the relative velocity: 
\begin{equation}
 \zeta (E_R) = \int_{v_\text{min}(E_R)}^\infty
\frac{d^3 \bm{v}_{\text{rel}}}{v_{\text{rel}}} 
f (\bm{v}_{\text{rel}} + \bm{v}_\text{E}) ~,
\qquad 
\xi (E_R) =  \int_{v_\text{min}(E_R)}^\infty
d^3 \bm{v}_{\text{rel}}\,
v_{\text{rel}}\,
f (\bm{v}_{\text{rel}} + \bm{v}_\text{E}) ~.
\end{equation}
We give analytical expressions for these integrals in
Appendix~\ref{app:integ}.

%%%%%%%%%%%%%%%%%%%%%%%%%%%%%%%%%%%%%%
\subsection{Direct detection bound}
%%%%%%%%%%%%%%%%%%%%%%%%%%%%%%%%%%%%%%

When we consider limits from direct detection experiments on a given
model, it is often the case that we compute the DM-nucleon scattering
cross section in the model and compare this with the bounds on this
quantity obtained by direct detection experiments. In the present case,
however, we are unable to adopt this strategy due to the following
reasons. First, as can be seen in Eq.~\eqref{eq:dsigma}, the
differential scattering cross section in our models may have several
components that have different dependence on recoil energy, while in
obtaining the direct detection limits the recoil-energy dependence is
assumed to be the same as that of contact interactions (the second line
in Eq.~\eqref{eq:dsigma}). Second, the DM-proton scattering cross
section may be quite different from that of the DM-neutron scattering in
our model, while in the ordinary cases those two quantities are almost
the same. Consequently, to study the current limits on our models from
direct DM searches and their future prospects, we need to
compute the number of events $N_{\text{event}}$ in each direct detection
experiment by integrating out $dR/dE_R$ over the recoil energy $E_R$ with
taking account of the detection efficiency, and compare this with the
results from the existing experiments or with the sensitivities
expected in the future experiments.

The prescriptions for this procedure exploited in this paper are given
for each direct detection experiment in Appendix~\ref{app:eventnum}. In
this work, we use the latest result from the XENON1T experiment
\cite{Aprile:2018dbl} to obtain the current bound. For the estimate of
future prospects, we consider the expected sensitivity of the XENONnT
\cite{Aprile:2015uzo}.\footnote{We have checked that the expected
sensitivity of the LUX-ZEPLIN (LZ) \cite{Akerib:2018lyp} is quite
similar to that of the XENONnT. }
To obtain the expected number of events, we use
$\rho_{\text{DM}} = 0.3~\text{GeV} \cdot \text{cm}^{-3}$, $v_0 =
220~\text{km}/\text{s}$, $v_{\text{esc}} = 544~\text{km}/\text{s}$, and
$v_\text{E} = 232~\text{km}/\text{s}$ as in Ref.~\cite{Aprile:2018dbl}.

%%%figure%%%%%%%%%%%%%%%%%%%%%%%%%%%%%%%%%%%%%%%%%%%%%%%%%%%%
\begin{figure}[t]
 \centering
  \subcaptionbox{DM-MDM}{
  \includegraphics[width=0.45\columnwidth]{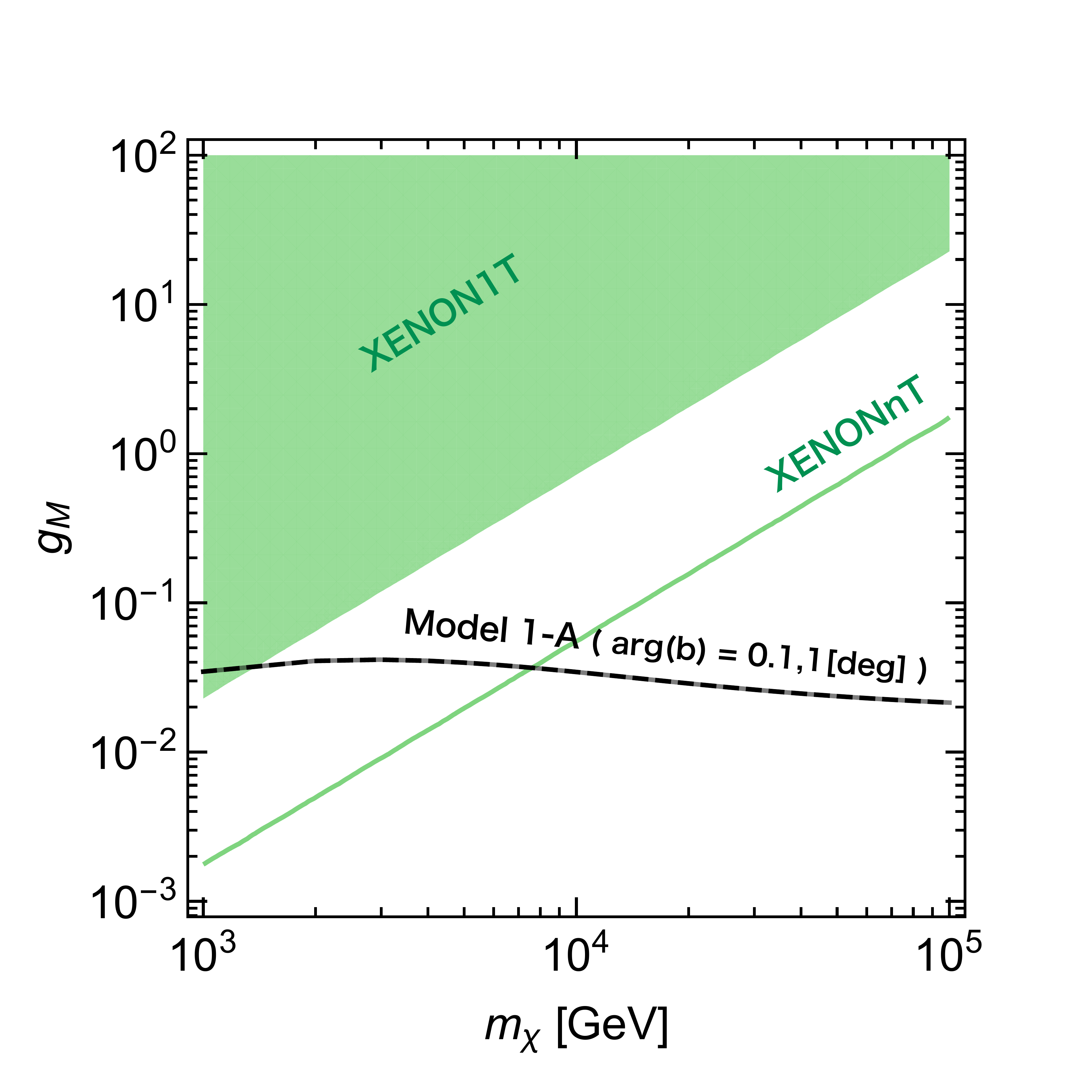}}
  \subcaptionbox{DM-EDM}{
  \includegraphics[width=0.45\columnwidth]{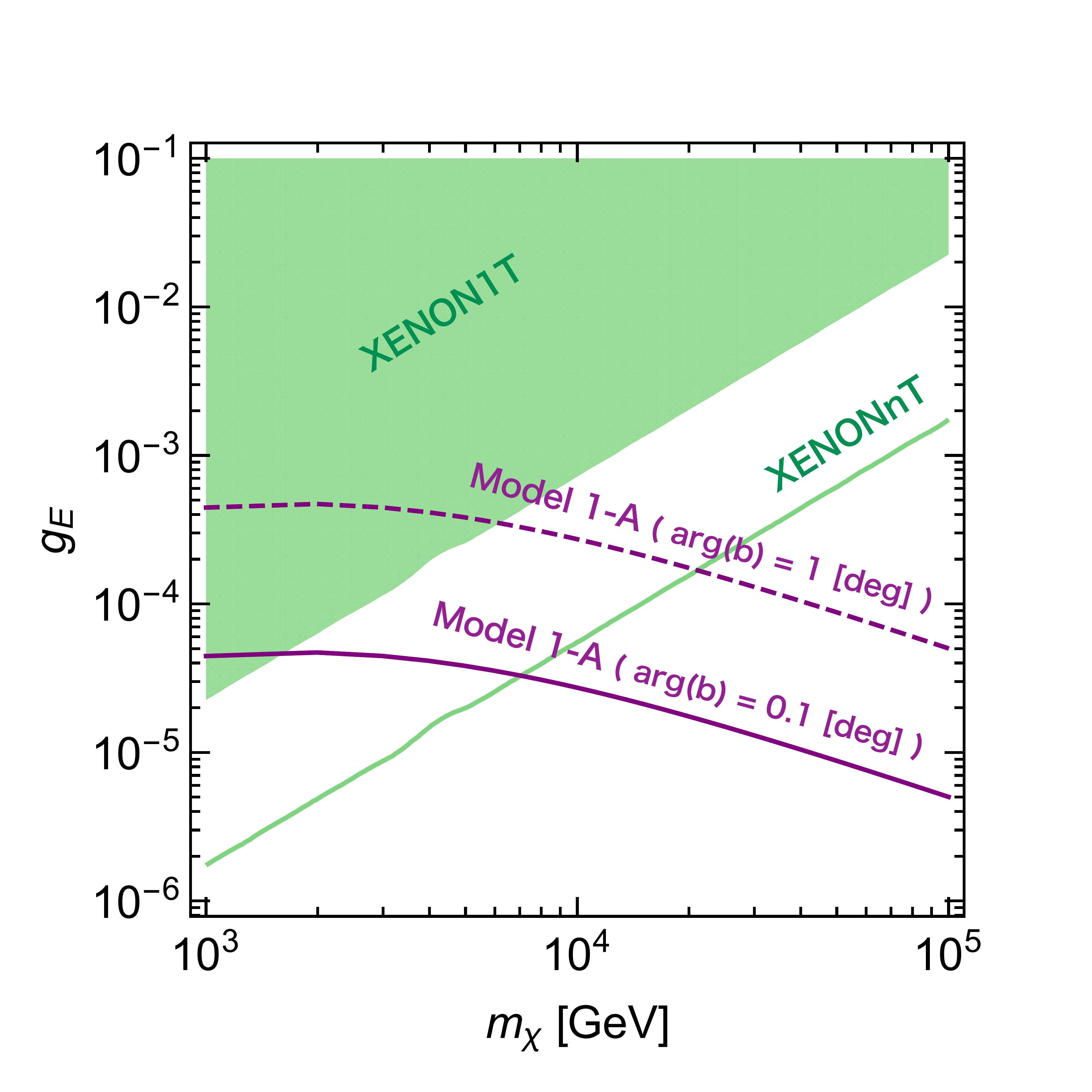}}
 \caption{Current constraints from XENON1T \cite{Aprile:2018dbl} on the
 DM-MDM and DM-EDM as functions of the DM mass shown in the green shaded
 regions, as well as the sensitivity of XENONnT \cite{Aprile:2015uzo} in
 the green solid line. We also show values of $g_M$ and $g_E$ predicted
 in Model I-A, which have already been shown in Fig.~\ref{fig:gME}. }
 \label{fig:const1}
\end{figure}
%%%%%%%%%%%%%%%%%%%%%%%%%%%%%%%%%%%%%%%%%%%%%%%%%%%%%%%%%%%%

In Fig.~\ref{fig:const1}, we show the current constraints on the DM-MDM
and DM-EDM as functions of the DM mass by the green shaded regions,
which correspond to the XENON1T $90\%$ C.L. exclusion limit
\cite{Aprile:2018dbl}. Here, we adopt the normalization in
Eq.~\eqref{eq:gmgedef}, and assume
$f^{(p)}_S=f^{(n)}_S=f^{(p)}_V=f^{(n)}_V=C^\gamma_R=0$. We also show the
sensitivity of XENONnT \cite{Aprile:2015uzo} in the green solid line. In
addition, for comparison, we show the values of $g_M$ and $g_E$
predicted in Model I-A, which have already been shown in
Fig.~\ref{fig:gME}. It is found that the DM direct detection experiments
are able to probe the singlet Dirac fermion DM at the TeV scale through
the MDM and EDM interactions; in particular, the EDM interaction may
allow us to probe the DM with a mass of as large as ${\cal O}(10)$~TeV
if there is a sizable CP violation in the DM couplings. More detailed
analysis for each model is given in Sec.~\ref{sec:results}.

%%%%%%%%%%%%%%%%%%%%%%%%%%%%%%%%%%%%%%%%
\section{Electric Dipole Moments}
\label{sec:edms}
%%%%%%%%%%%%%%%%%%%%%%%%%%%%%%%%%%%%%%%%

As we mentioned above, the DM-EDM is induced only in the presence of CP
phases in the couplings of new particles. Such CP phases may also induce
CP-odd quantities in the SM sector, which can be probed via the
measurements of the EDMs of electron and nucleons. In this section, we
evaluate these EDMs induced in our models.

%%%%%%%%%%%%%%%%%%%%%%%%%%%%%%%%%%%%%%%%%%%%%%%%%%%%
\subsection{CP violation in the DM interactions}
\label{sec:CPVDM}
%%%%%%%%%%%%%%%%%%%%%%%%%%%%%%%%%%%%%%%%%%%%%%%%%%%%

%%%figure%%%%%%%%%%%%%%%%%%%%%%%%%%
\begin{figure}[t]
 \begin{center}
   \includegraphics[width=45mm]{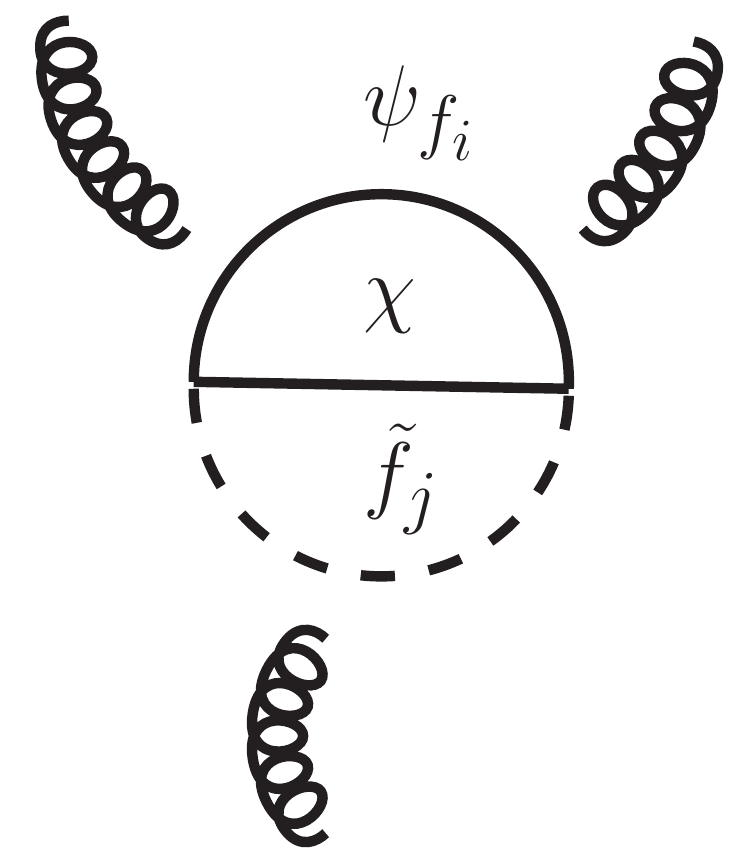}
     \end{center}
 \caption{Feynman diagram that induces the Weinberg operator. The gluon
 lines are attached to the vector-like quark and/or scalar lines. }
 \label{fig:GGGDM}
\end{figure}
%%%figure%%%%%%%%%%%%%%%%%%%%%%%%%%%%%%%%%%

Let us first consider the case with $\varphi_{\mu_f}\neq 0$. In this
case, in Model I, the non-zero CP phase $\varphi_{\mu_f}$ induces the
Weinberg operator \cite{Weinberg:1989dx} at the two-loop level through
the diagram depicted in
Fig.~\ref{fig:GGGDM}:\footnote{$\varphi_{A_{\bar{f}}}$ also induces
the Weinberg operator, but this contribution is smaller than that from
$\varphi_{\mu_f}$ as it is always accompanied by the insertion of Higgs
VEVs. }
\begin{align}
\mathcal{L}_{\cancel{\rm CP}}
=
-\frac{1}{6}w\, f^{ABC}\epsilon^{\mu\nu\rho\sigma}G^A_{\mu\lambda}G^{B\lambda}_\nu
 G^C_{\rho\sigma} ~ ,
\label{eq:weinbop}
\end{align} 
where $f^{ABC}$ denotes the structure constants of SU(3) and
$\epsilon^{\mu\nu\rho\sigma}$ is the totally antisymmetric tensor with
$\epsilon^{0123} = 1$. We give an analytical expression for the Wilson
coefficient $w$ in Appendix~\ref{sec:weinbop}, which we derive from
generic results obtained in Ref.~\cite{Abe:2017sam}.

This Weinberg operator induces non-zero EDMs of nucleons, which are
estimated by means of the QCD sum rules \cite{Demir:2002gg}:
\begin{align}
d_N(w)/e= \pm (10-30)~\mbox{MeV}\times w(1~\mbox{GeV})~,\label{eq:dNmatchingw} 
\end{align}
for $N= n, p$, where $w(1~\mbox{GeV})$ denotes the Wilson coefficient of
the Weinberg operator at the scale of 1~GeV.  In our numerical analysis,
we use $d_N(w)/e=+20~\mbox{MeV}\times w(1~\mbox{GeV})$ as a reference
value. We obtain $w(1~\mbox{GeV})$ by
evolving the Wilson coefficient from the matching scale down to 1~GeV
according to the renormalization group equation (RGE). The RGE for $w$
at the leading order is given by 
\cite{Degrassi:2005zd}  
\begin{align}
\frac{d}{d \ln\mu_R}w(\mu_R)
=
\frac{\alpha_s(\mu_R)}{4\pi}(N_c+2N_f)\,w(\mu_R)~,
\end{align}
where $N_c=3$ and $N_f$ are the numbers of colors and quark flavors,
respectively, and $\mu_R$ denotes the renormalization scale. 

%%%%%%%%%
%%%figure%%%
\begin{figure}[t]
 \centering
 \includegraphics[width=80mm]{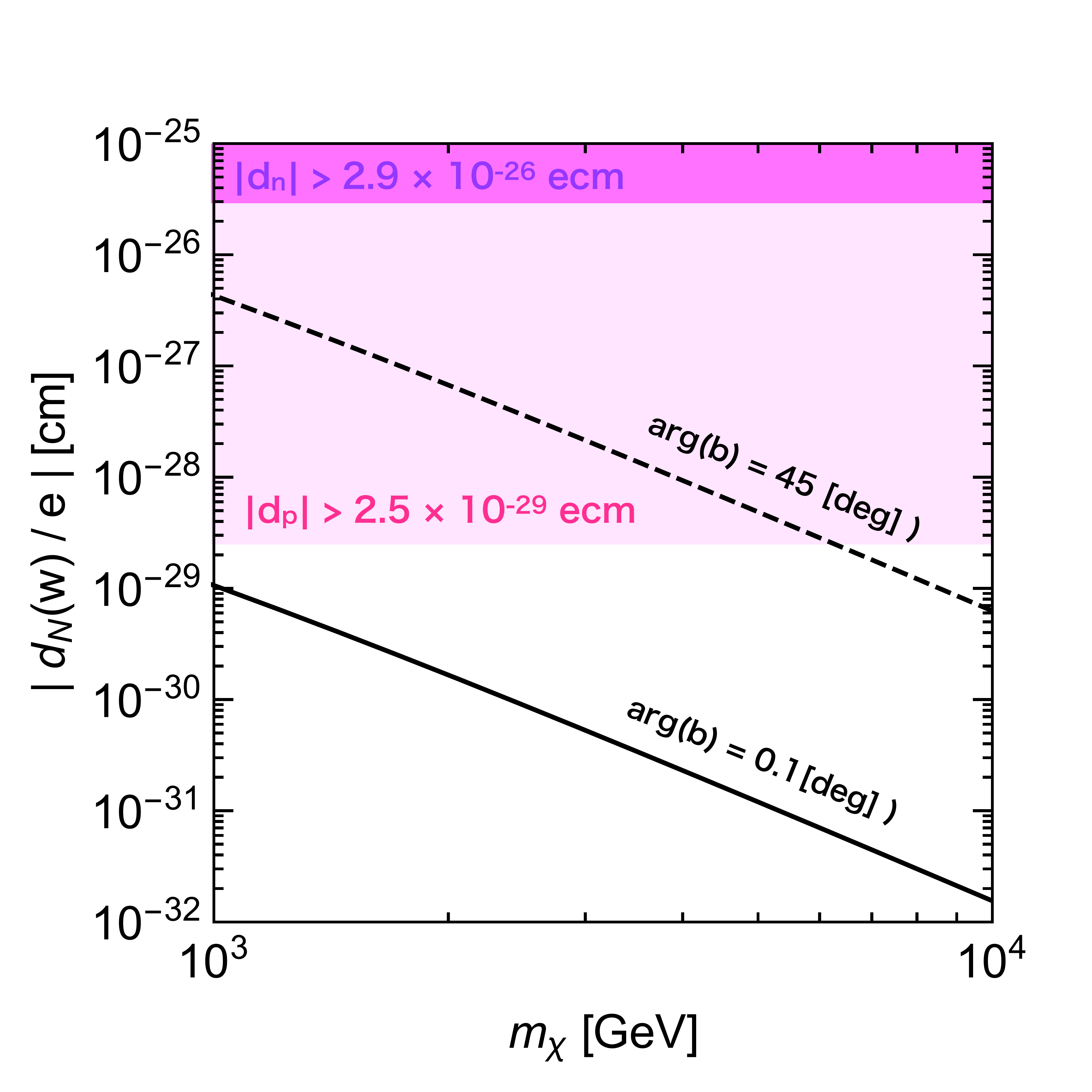}
 \caption{Nucleon EDM induced by the CP-violating interactions between
 the DM and the mediator fields. We take the same parameter sets as in
 Model I-A in Fig.~\ref{fig:gME}, with $\text{arg}(b) = 0.1$ and 45
 degrees in black solid and dashed lines, respectively. Dark and light
 magenta regions correspond to the current bound on the neutron EDM
 ($|d_n|\geq 2.9\times 10^{-26}~e \cdot \text{cm}$ \cite{Baker:2006ts}) and the
 future prospects for the observation of the proton EDM ($|d_p|\geq
 2.5\times 10^{-29}~e\cdot \text{cm}$ \cite{Lehrach:2012eg}), respectively.}
 \label{fig:nEDMGGGDM}
\end{figure}
%%%figure%%%
%%%%%%%%%

In Fig.~\ref{fig:nEDMGGGDM}, we show the nucleon EDM induced by the
CP-violating couplings between the DM and the mediator fields. Here, we
take the same parameter sets as in Model I-A in Fig.~\ref{fig:gME}, with
$\text{arg}(b) = 0.1$ and 45 degrees in the black solid and dashed lines,
respectively. The dark and light magenta regions correspond to the
current bound on the neutron EDM ($|d_n|\geq 2.9\times 10^{-26}~e \cdot
\text{cm}$ \cite{Baker:2006ts}) and the future prospects for the
observation of the proton EDM ($|d_p|\geq 2.5\times 10^{-29}~e\cdot
\text{cm}$ \cite{Lehrach:2012eg}), respectively. As we see, although the
present limit is not sensitive to the CP phases in the DM couplings yet,
they may be probed in future measurements of the nucleon EDMs. Notice that
the Weinberg operator depends on the CP phases in the DM couplings in
the same manner as the DM EDM, as we see from
Eqs.~\eqref{eq:cgame} and \eqref{eq:wanalytic}. We thus expect a strong
correlation between the DM-nucleus scattering cross section and the
nucleon EDM with respect to the CP phases in the DM
couplings, which we actually see in the subsequent section.

%%%%%%%%%%%%%%%%%%%%%%%%%%%%%%%%%%%%%%%%
\subsection{CP violation in the vector-like fermion-Higgs couplings}
%%%%%%%%
%%%figure%%%
\begin{figure}[t]
 \begin{center}
   \includegraphics[width=50mm]{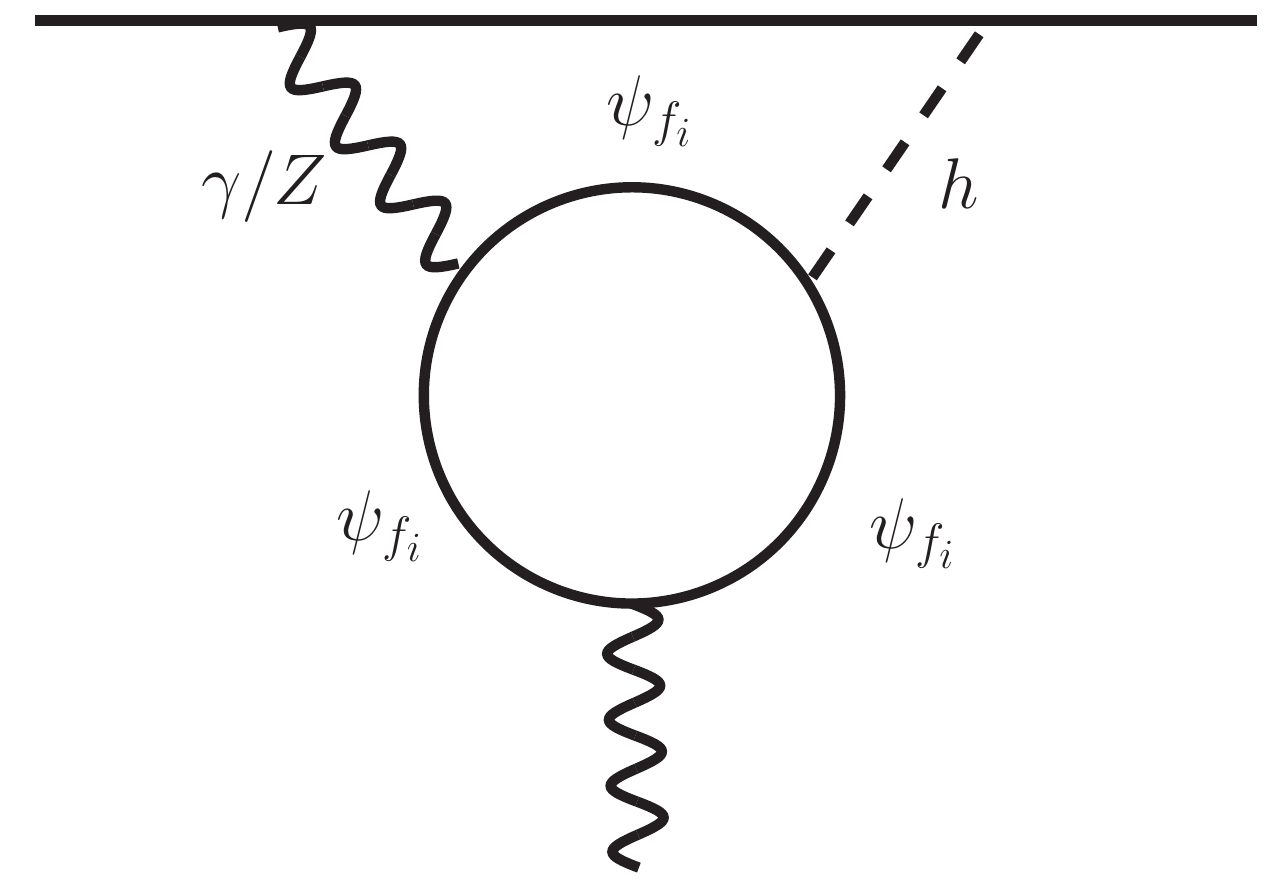}
   ~~~
   \includegraphics[width=50mm]{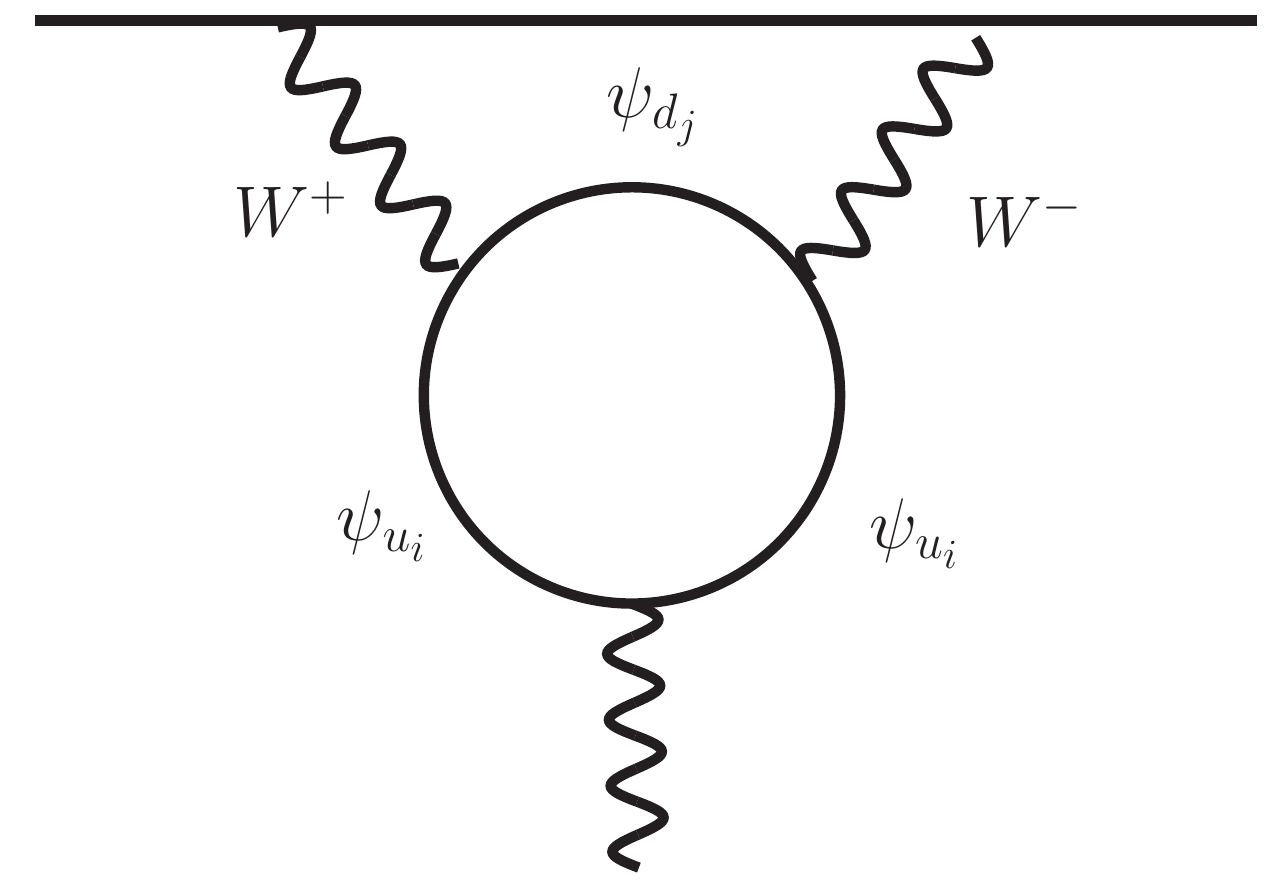}
   \\[5pt]
   \includegraphics[width=50mm]{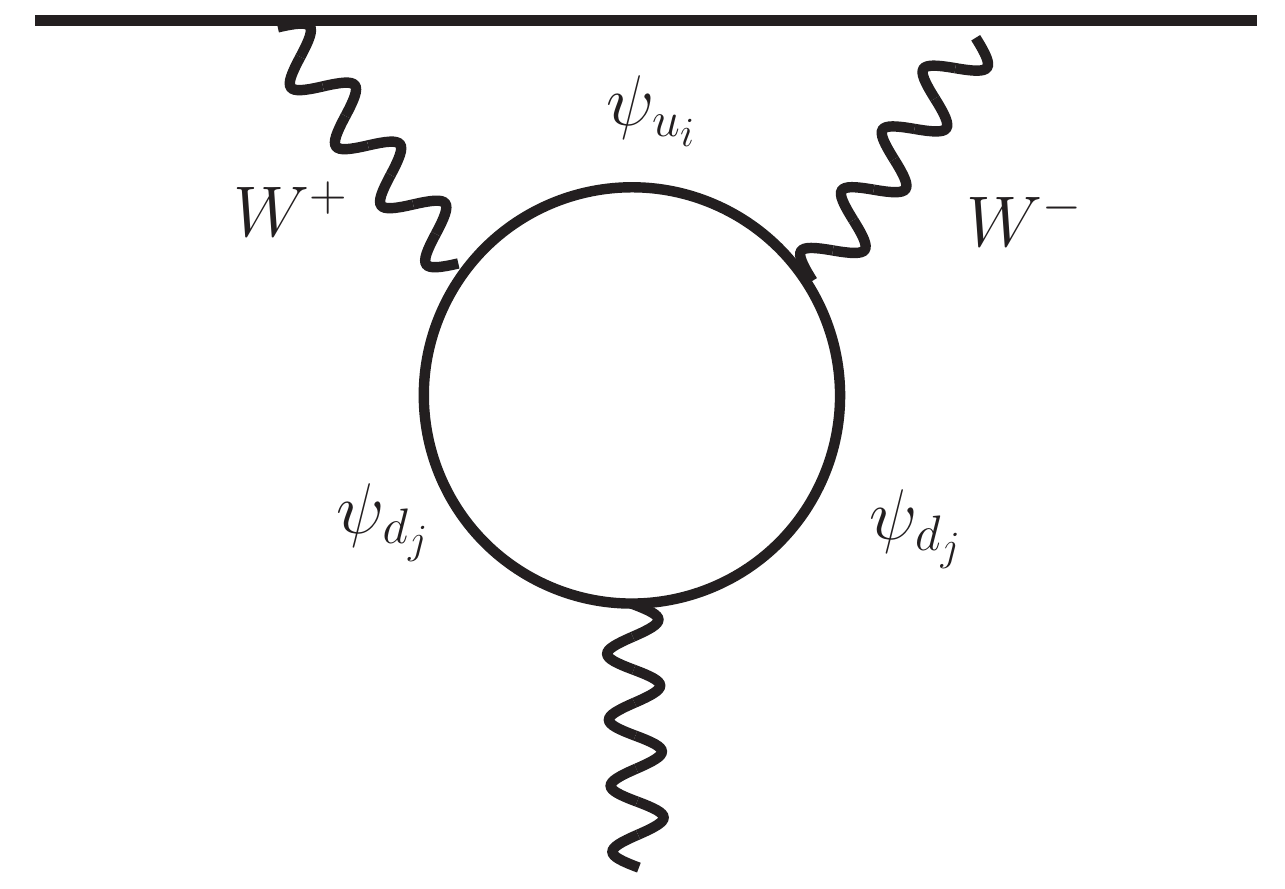}
   ~~~
   \includegraphics[width=50mm]{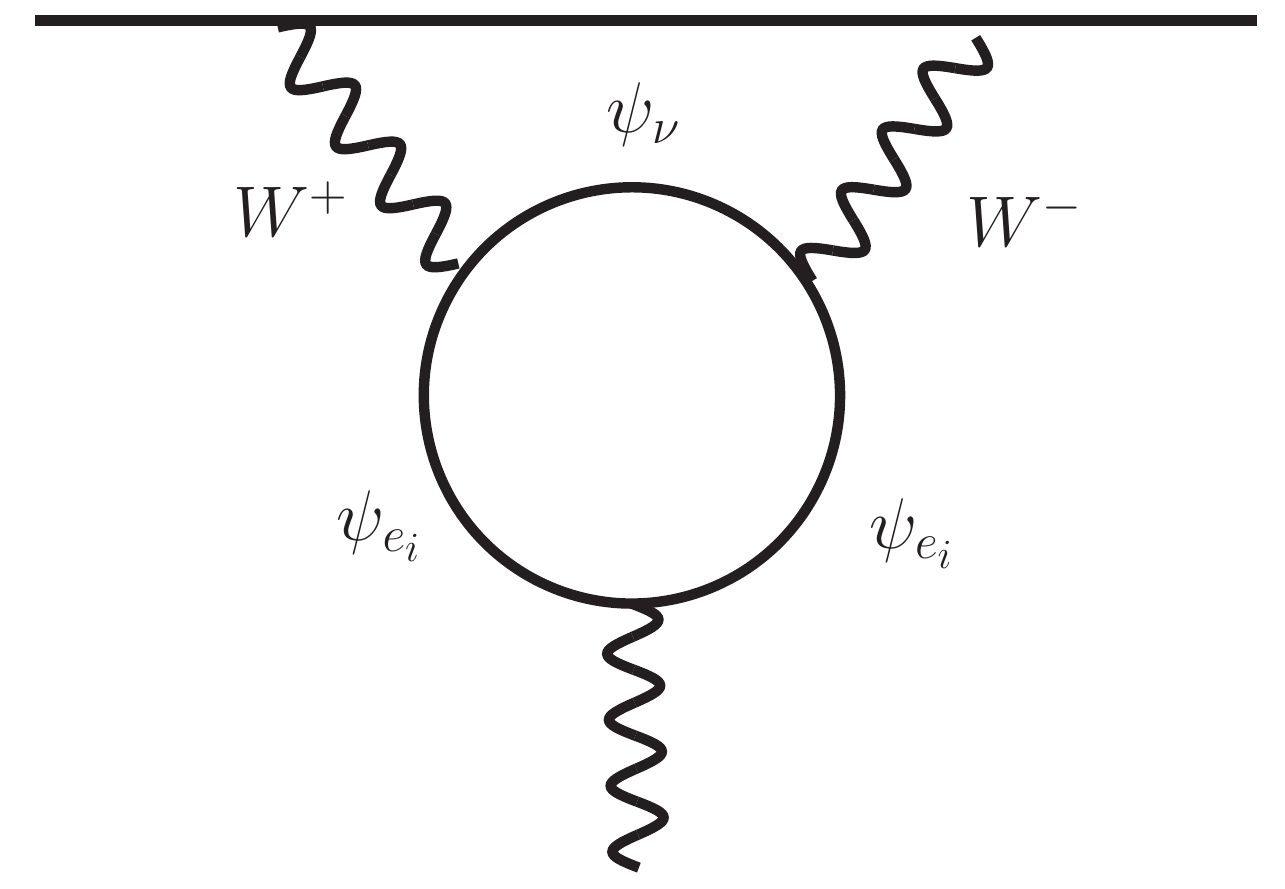}
     \end{center}
 \caption{Barr-Zee diagrams for electron EDM.}
 \label{fig:BZ}
\end{figure}
%%%figure%%%
%%%%%%%%%

%%%%%%%%%%%%%% FIGURE %%%%%%%%%%%%%%%%%%%%%%%%%%%%%%%%%%%%
\begin{figure}[t]
  \centering
  \subcaptionbox{Model I}{
  \includegraphics[width=0.45\columnwidth]{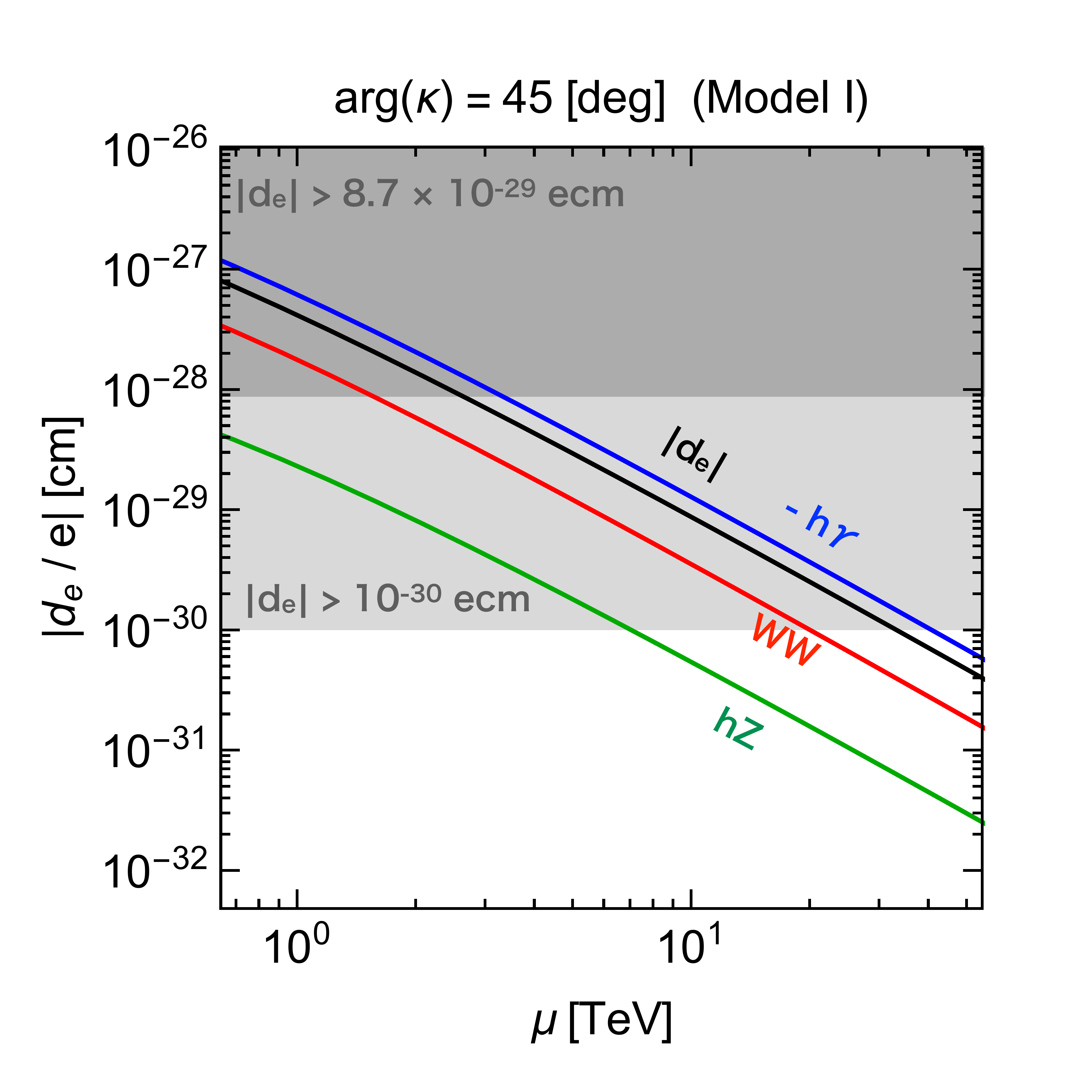}}
  \subcaptionbox{Model II}{
  \includegraphics[width=0.45\columnwidth]{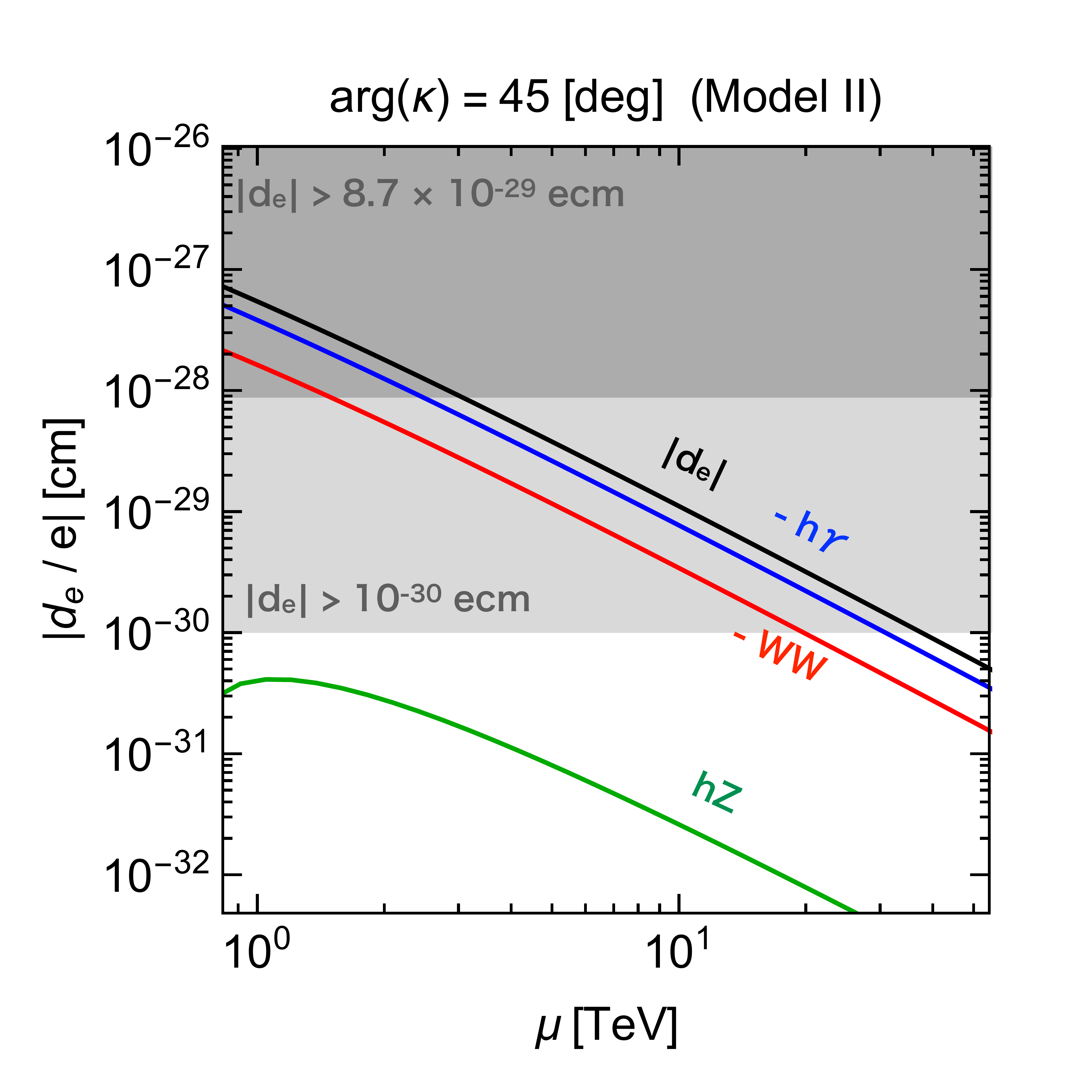}}
\caption{Each contribution to the electron EDM as a function of the
 fermion mass parameter $\mu$. Black, red, blue, and green lines show
 the total, $WW$, $h\gamma$, and $hZ$ contributions, respectively, with
 the sign of each contribution indicated explicitly. 
 We set $\mu_Q = \mu_L = \mu + 100$~GeV, $\mu_{\bar{u}} = \mu + 50$~GeV,
 $\mu_{\bar{d}} = \mu_{\bar{e}} = \mu$, $\kappa = \kappa_{\bar{f}} =
 \kappa_{\bar{f}}^\prime$, $|\kappa| = 0.5$, and $\text{arg}(\kappa) =
 45$~deg. 
 Dark and light gray areas correspond to the present bound on the
 electron EDM ($|d_e|\geq 8.7\times 10^{-29} ~e \cdot \text{cm}$
 \cite{Baron:2013eja}) and the future prospects for the observation of
 the electron EDM ($|d_e|\geq1.0\times  10^{-30} ~e \cdot \text{cm}$
 \cite{2012NJPh14j3051K,Kawall:2011zz}), respectively.  }
  \label{fig:eedmcomp}
\end{figure}
%%%%%%%%%%%%%%%%%%%%%%%%%%%%%%%%%%%%%%%%%%%%%%%%%%%%%%%%%%

%%%%%%%%%%%%%% FIGURE %%%%%%%%%%%%%%%%%%%%%%%%%%%%%%%%%%%%
\begin{figure}[t]
  \centering
  \subcaptionbox{Model I}{
  \includegraphics[width=0.45\columnwidth]{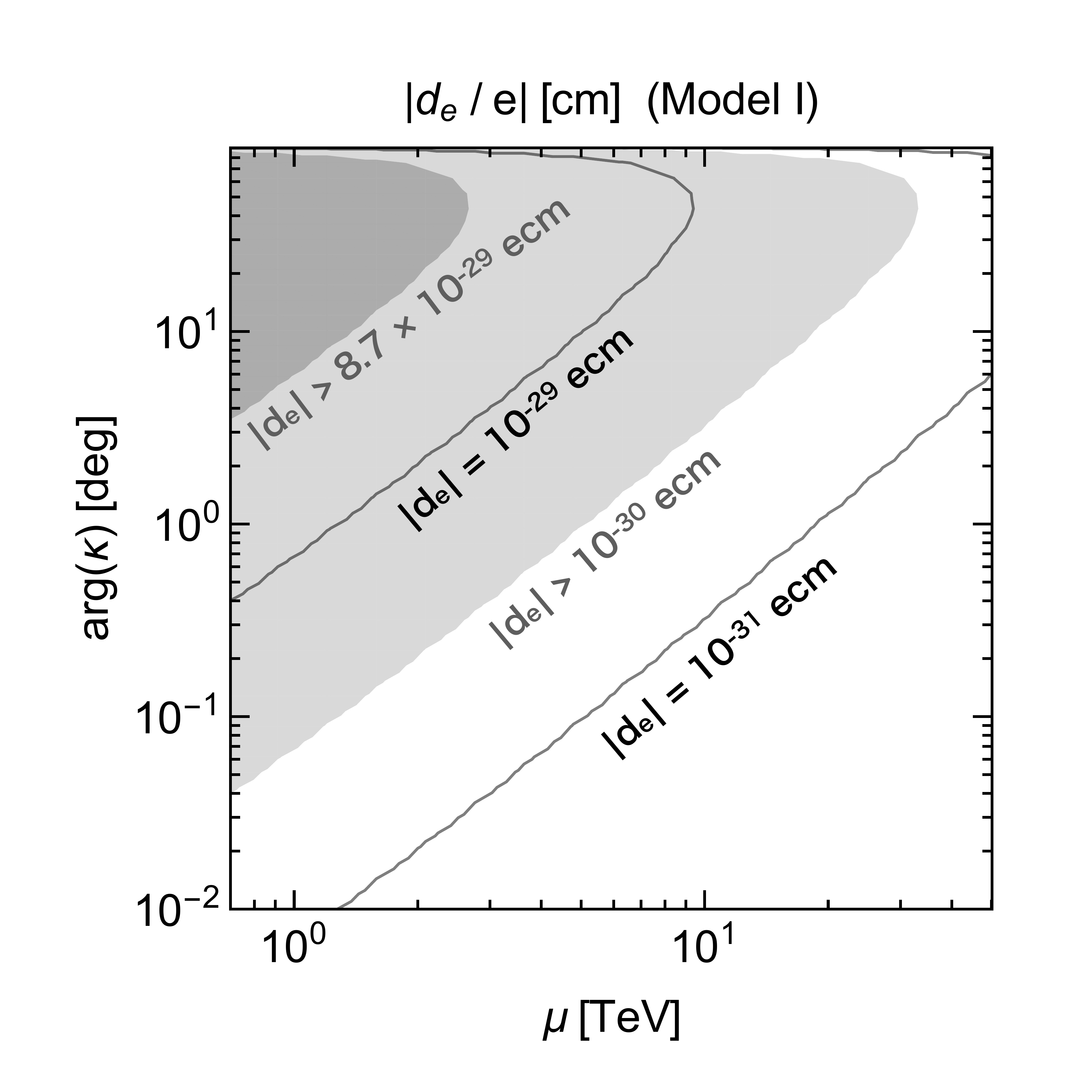}}
  \subcaptionbox{Model II}{
  \includegraphics[width=0.45\columnwidth]{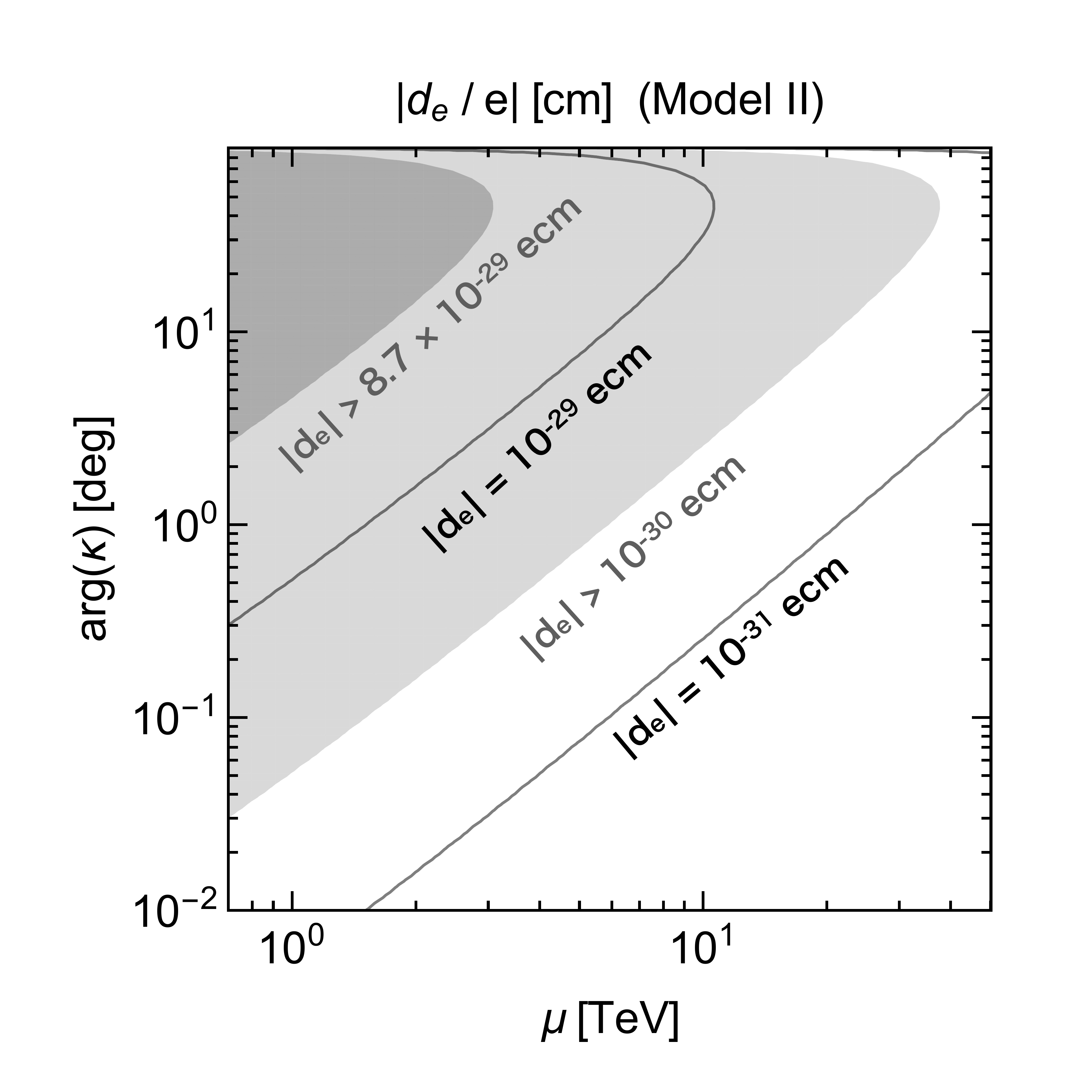}}
\caption{Contour plots for the electron EDM. We take the same parameters
 as in Fig.~\ref{fig:eedmcomp} except for $\text{arg}(\kappa)$.
 Dark and light gray areas correspond to the present bound on the
 electron EDM ($|d_e|\geq 8.7\times 10^{-29} ~e \cdot \text{cm}$
 \cite{Baron:2013eja}) and the future prospects for the observation of
 the electron EDM ($|d_e|\geq1.0\times  10^{-30} ~e \cdot \text{cm}$
 \cite{2012NJPh14j3051K,Kawall:2011zz}), respectively.} 
  \label{fig:eedmcont}
\end{figure}
%%%%%%%%%%%%%%%%%%%%%%%%%%%%%%%%%%%%%%%%%%%%%%%%%%%%%%%%%%

The CP phases in the vector-like fermion-Higgs couplings, $\varphi_{\kappa_{\bar{f}}}$,
can also be probed in EDM experiments, as they induce the electron EDM
at two-loop level through the Barr-Zee diagrams \cite{Barr:1990vd} shown
in Fig.~\ref{fig:BZ}: 
\begin{align}
\mathcal{L}_e 
=
-\frac{i}{2}d_e\,\bar{e}\sigma^{\mu\nu}\gamma_5 e F^{\mu\nu} ~.
\end{align}
The resultant expressions for these two-loop contributions are
summarized in Appendix~\ref{app:BZ}. The predicted values of the
electron EDM in Model I and II are then given in
Figs.~\ref{fig:eedmcomp} and \ref{fig:eedmcont}.  In
Fig.~\ref{fig:eedmcomp}, we show each contribution to the electron EDM
as a function of the fermion mass parameter $\mu$. The black, red, blue,
and green lines represent the total, $WW$, $h\gamma$, and $hZ$
contributions, respectively, with the sign of each contribution
indicated explicitly. Here, the $h\gamma$ and $hZ$ contributions come
from the upper left diagram while the $WW$ contribution from the other
diagrams in Fig.~\ref{fig:BZ}. We set $\mu_Q = \mu_L = \mu + 100$~GeV,
$\mu_{\bar{u}} = \mu + 50$~GeV, $\mu_{\bar{d}} = \mu_{\bar{e}} = \mu$,
$\kappa = \kappa_{\bar{f}} = \kappa_{\bar{f}}^\prime$, $|\kappa| = 0.5$,
and $\text{arg}(\kappa) = 45$~deg. The dark and light gray areas
correspond to the present bound on the electron EDM ($|d_e|\geq
8.7\times 10^{-29} ~e \cdot \text{cm}$ \cite{Baron:2013eja}) and the
future prospects for the observation of the electron EDM
($|d_e|\geq1.0\times  10^{-30} ~e \cdot \text{cm}$
\cite{2012NJPh14j3051K,Kawall:2011zz}), respectively. As we see, the
signs of the $WW$ contributions in Model I and II are different from
each other. We then show the contour plots for the electron EDM in the
$\mu$-arg($\kappa$) plane in Fig.~\ref{fig:eedmcont}, with the same
parameter sets except for arg($\kappa$). These plots show that even the
present limit gives a strong constraint on the vector-like fermions at
the TeV scale if there is an ${\cal O}(1)$ CP phase in their
couplings, which may be pushed down to as strong as $\mathcal{O}(0.01)$
in the future. 

In Model I, a non-zero value of $\varphi_{\kappa_{\bar{f}}}$ also
induces the EDM and chromo-EDM of quarks through the Barr-Zee
diagrams,\footnote{For the contribution to the quark chromo-EDM, the
relevant diagram is obtained from the upper left diagram in
Fig.~\ref{fig:BZ} by replacing the $\gamma/Z$ lines with gluon lines. } 
which then give rise to the nucleon EDMs. We see however that the limit
from the neutron EDM is weaker than that from the electron EDM; we hence do
not give further analyses on this here, though we provide an analytical
expression of the quark chromo-EDM in Appendix~\ref{app:BZ} just for
completeness. Nevertheless, we note in passing that the quark EDM and 
chromo-EDM can be important when we discuss the effect of the
Weinberg operator considered in the previous subsection, as all of these
quantities can give comparable contributions to nucleon EDM.

%%%%%%%%%%%%%%%%%%%%%%%%%%%%%%%%%%%%%%%%
\section{Results}
\label{sec:results}
%%%%%%%%%%%%%%%%%%%%%%%%%%%%%%%%%%%%%%%%

Now we study the experimental constraints discussed above on the
parameter space of our models and assess their testability. We again
consider the simplified scenarios described in Sec.~\ref{sec:dd} with
parameters taken to be
$a=|b|=1$, 
$\lambda^\prime_Q=\lambda^\prime_L=0$,
$\lambda_f=\kappa_{\bar{f}}=\kappa^\prime_{\bar{f}}=0.5$, 
$\mu_Q= \mu_L = 800$~GeV, $\mu_{\bar{u}} = 750$~GeV, $\mu_{\bar{d}} =
\mu_{\bar{e}} = 700$~GeV, 
$\widetilde{m}_Q = \widetilde{m}_L = 1.2 M$,
$\widetilde{m}_{\bar{u}} = 1.1 M$,
$\widetilde{m}_{\bar{d}} = \widetilde{m}_{\bar{e}}=M$, 
$A_{\bar{f}} = 2 M$, and
$M= 1.1 m_\chi$.

%%%figure%%%
\begin{figure}[t]
 \begin{center}
  \includegraphics[width=50mm]{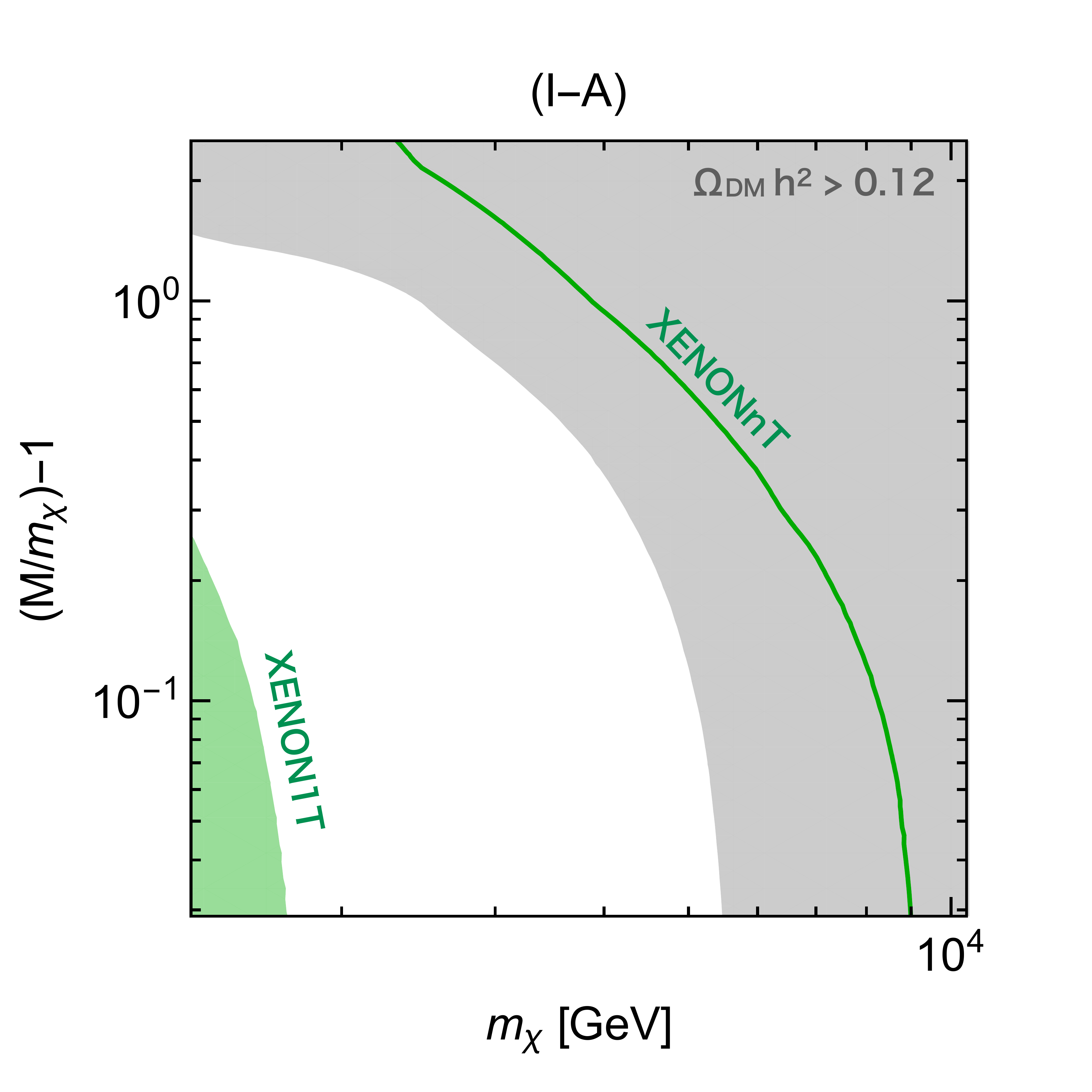}
  ~~
  \includegraphics[width=50mm]{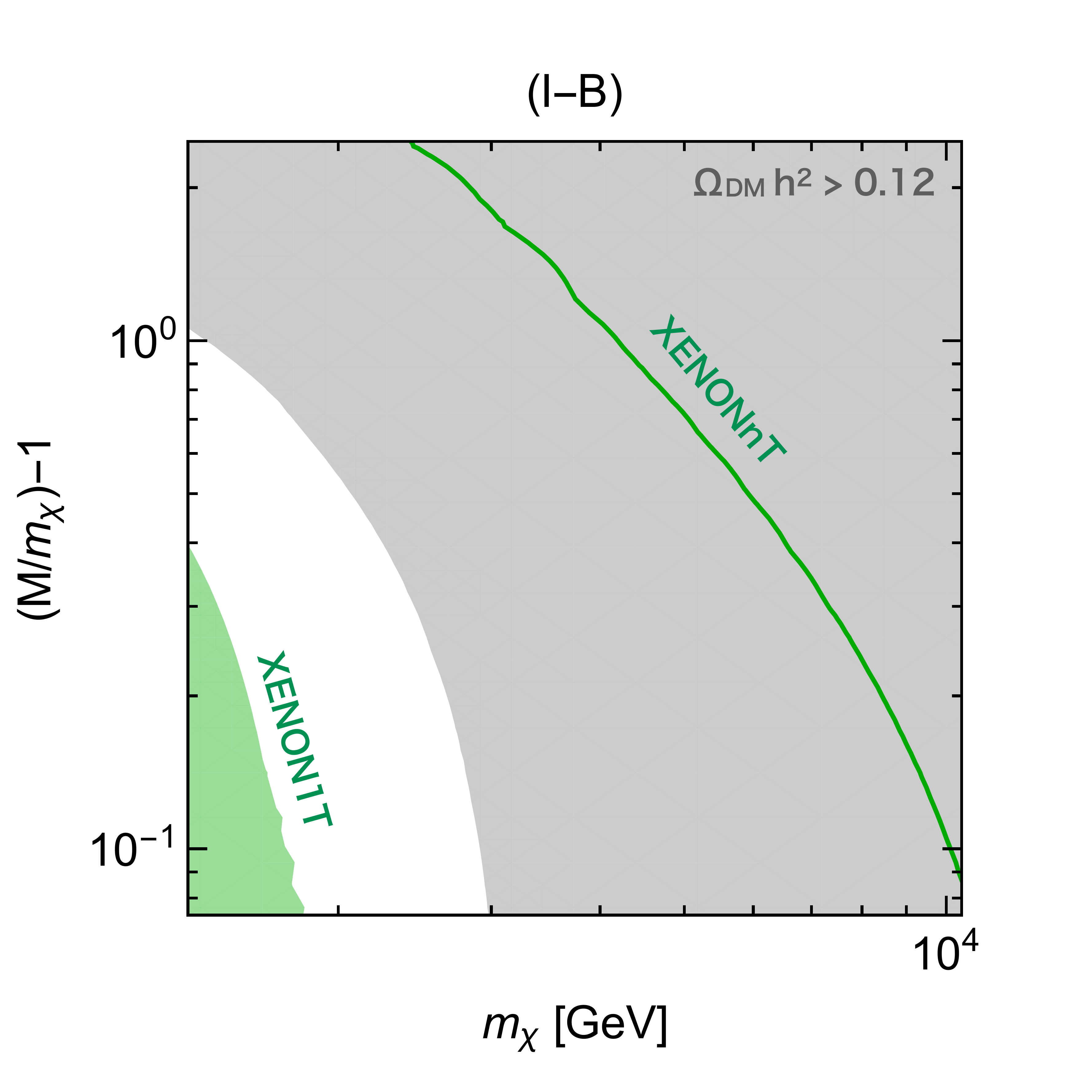}
  ~~
  \includegraphics[width=50mm]{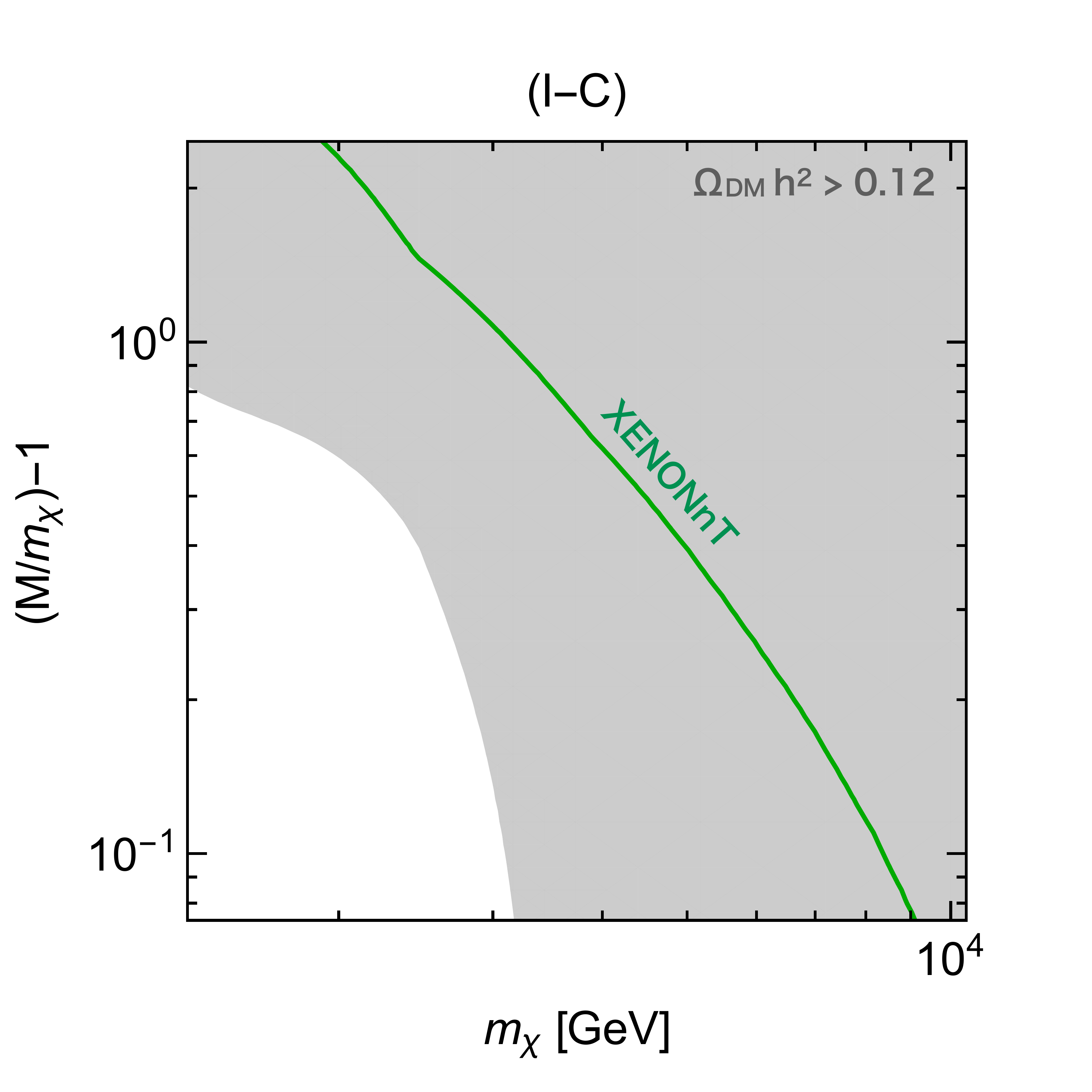}
  \\
  \includegraphics[width=50mm]{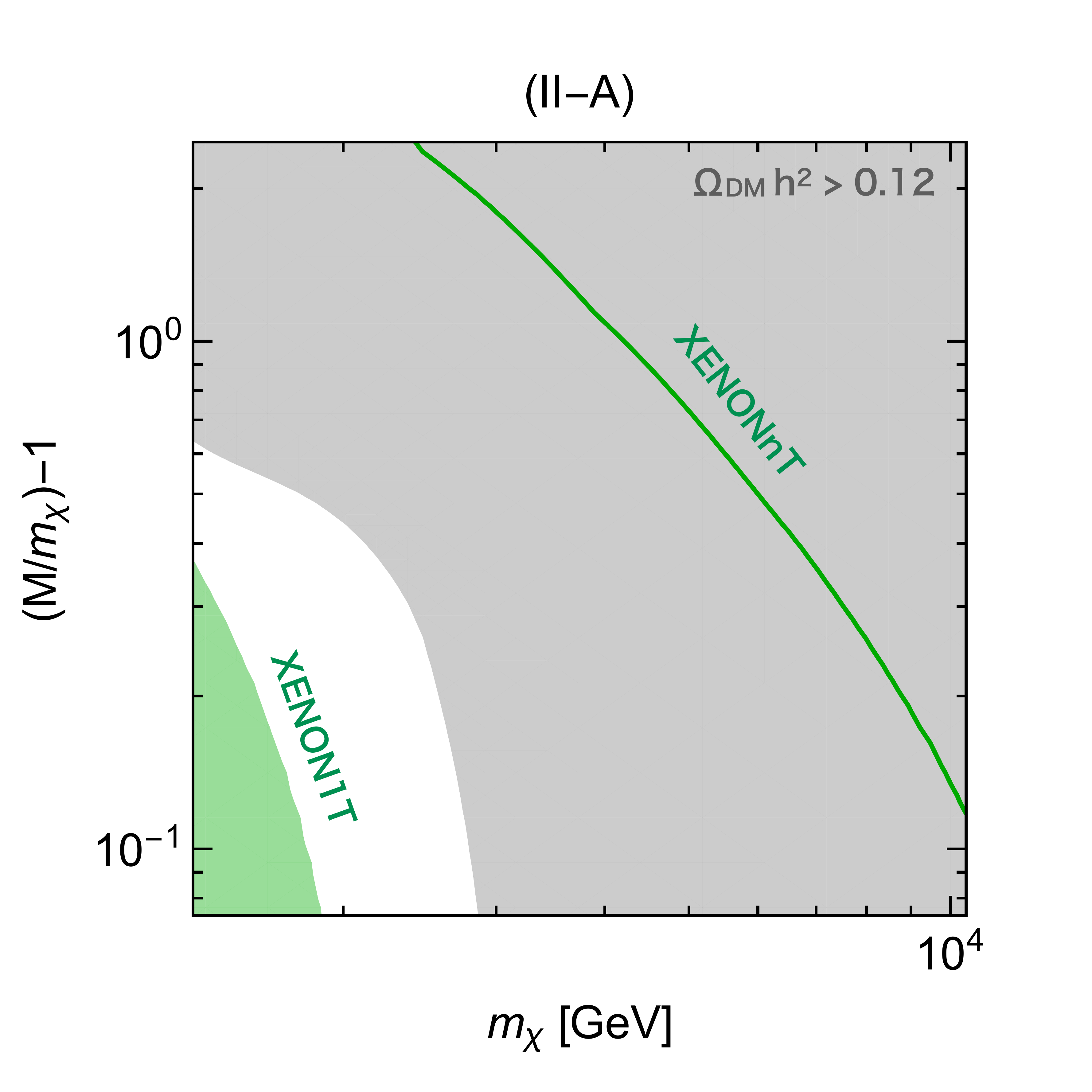}
  ~~
  \includegraphics[width=50mm]{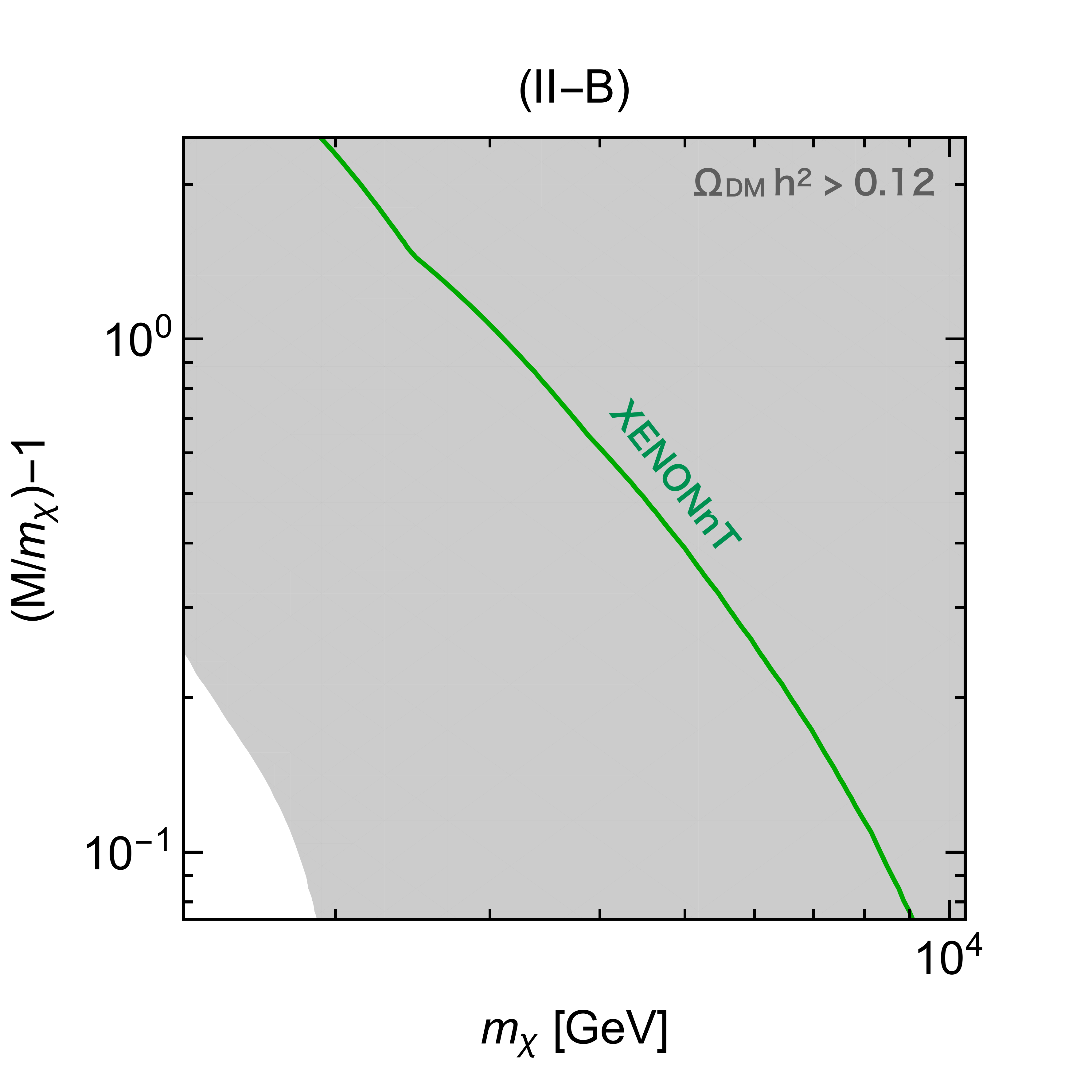}
   \end{center}
 \caption{Current limits provided by XENON1T \cite{Aprile:2018dbl}
 (green shaded area) and expected sensitivity of XENONnT
 \cite{Aprile:2015uzo} (green solid line) in the parameter space of each
 model. The gray shaded area represents the region where the DM density
 exceeds the observed value: $\Omega_{\text{DM}} h^2\geq 0.12$. Here, we
 take $\text{arg}(b) = 0$.}
 \label{fig:res-noEDM}
\end{figure}
%%%figure%%%
%%%%%%%%%
%%%%%%%%
%%%%%%%%

Let us begin with the CP-conserving cases. Figure~\ref{fig:res-noEDM}
shows the current limits provided by XENON1T \cite{Aprile:2018dbl}
(green shaded area) and the expected sensitivity of XENONnT
\cite{Aprile:2015uzo} (green solid line) in the parameter space of each
model. The gray shaded area represents the region where the DM density
exceeds the observed value: $\Omega_{\text{DM}} h^2\geq 0.12$. Here, we
take $\text{arg}(b) = 0$. We find that at present the allowed region
with $\Omega_{\rm{DM}} h^2\leq 0.12$ exists in all of the models, where
the masses of both the DM and new scalar particles are predicted to be
at the TeV scale. All of these allowed regions can be fully
probed in the future direct detection experiments.

%%%%%%%%
%%%figure%%%
\begin{figure}[t]
 \begin{center}
  \includegraphics[width=50mm]{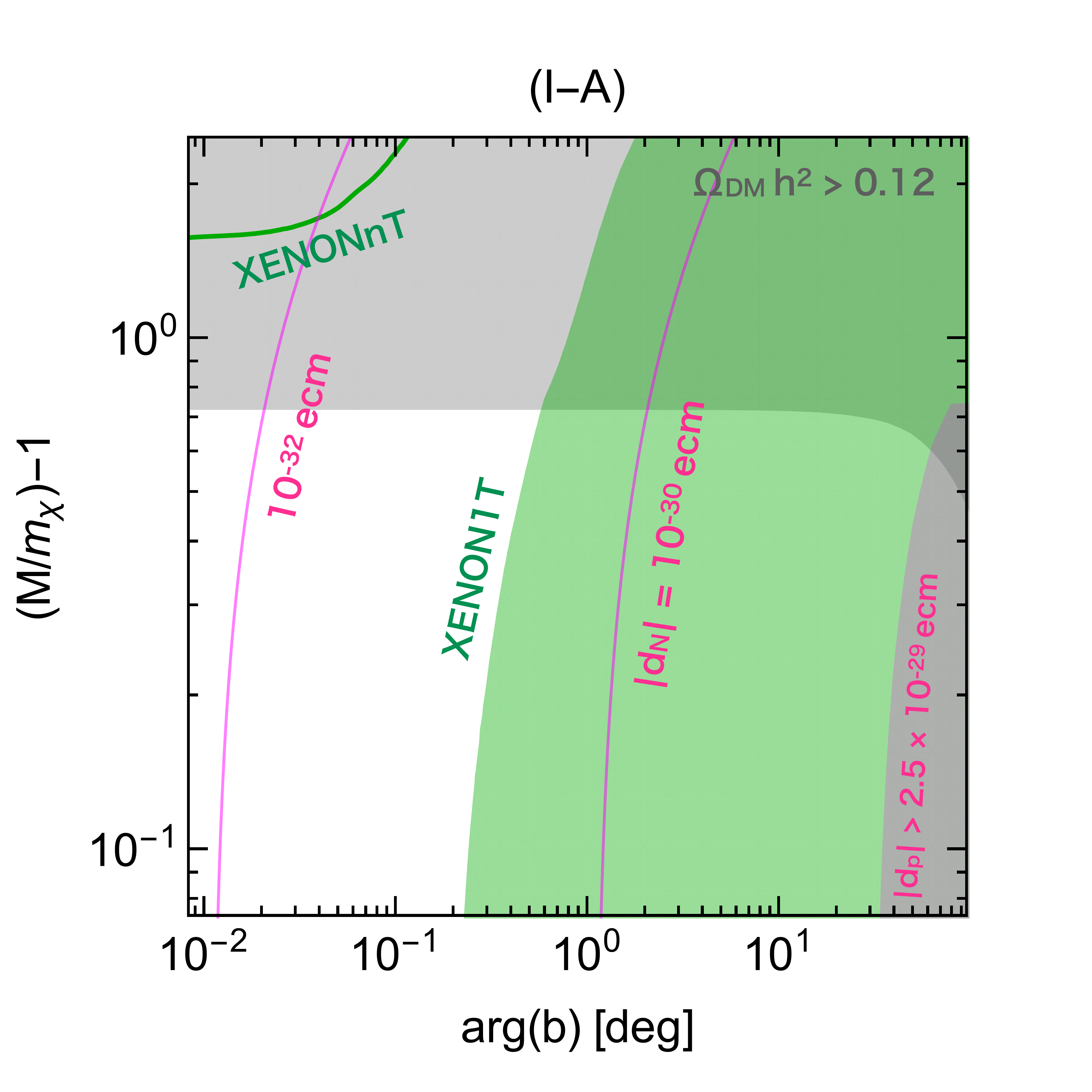}
  ~~
  \includegraphics[width=50mm]{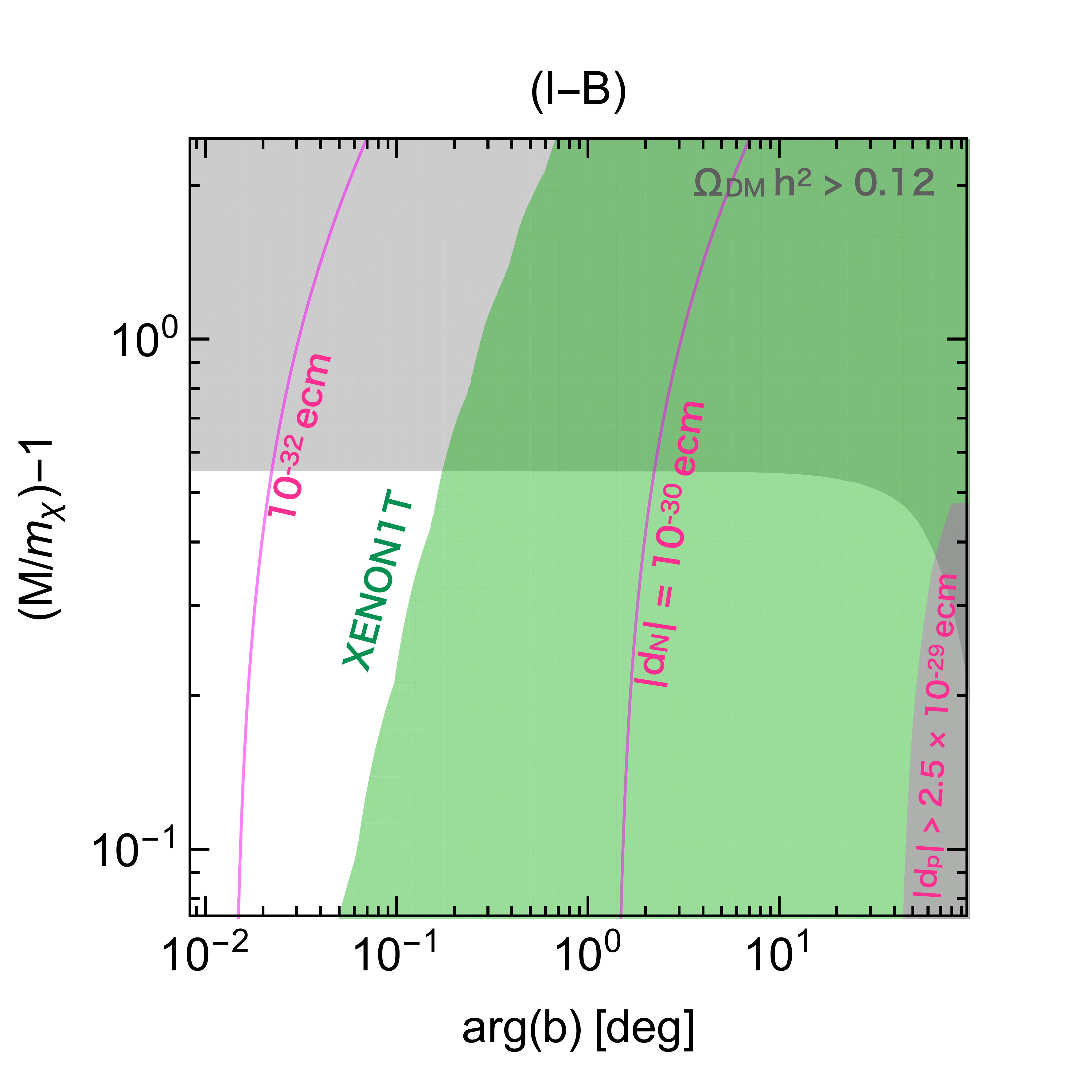}
  ~~
  \includegraphics[width=50mm]{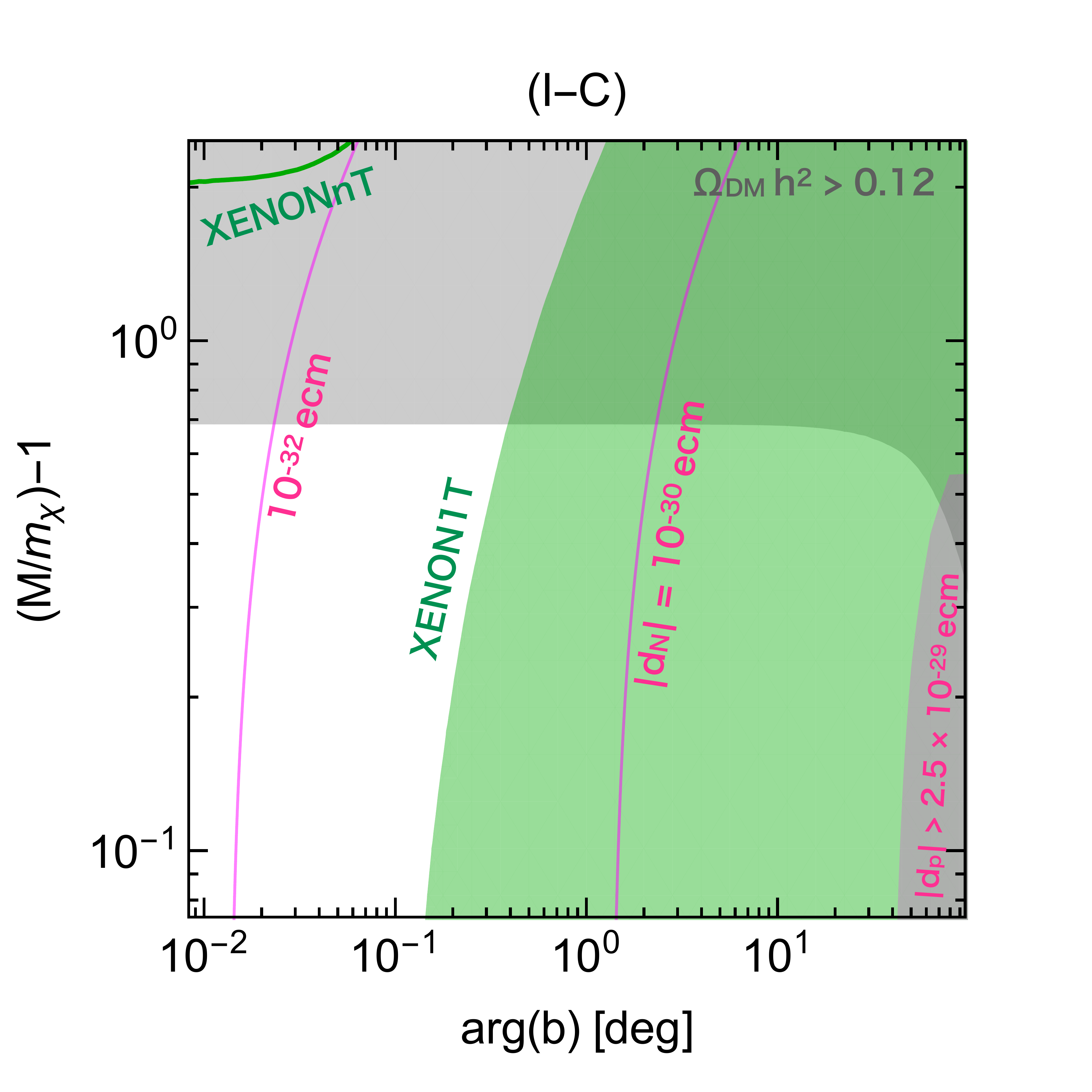}
  \\
  \includegraphics[width=50mm]{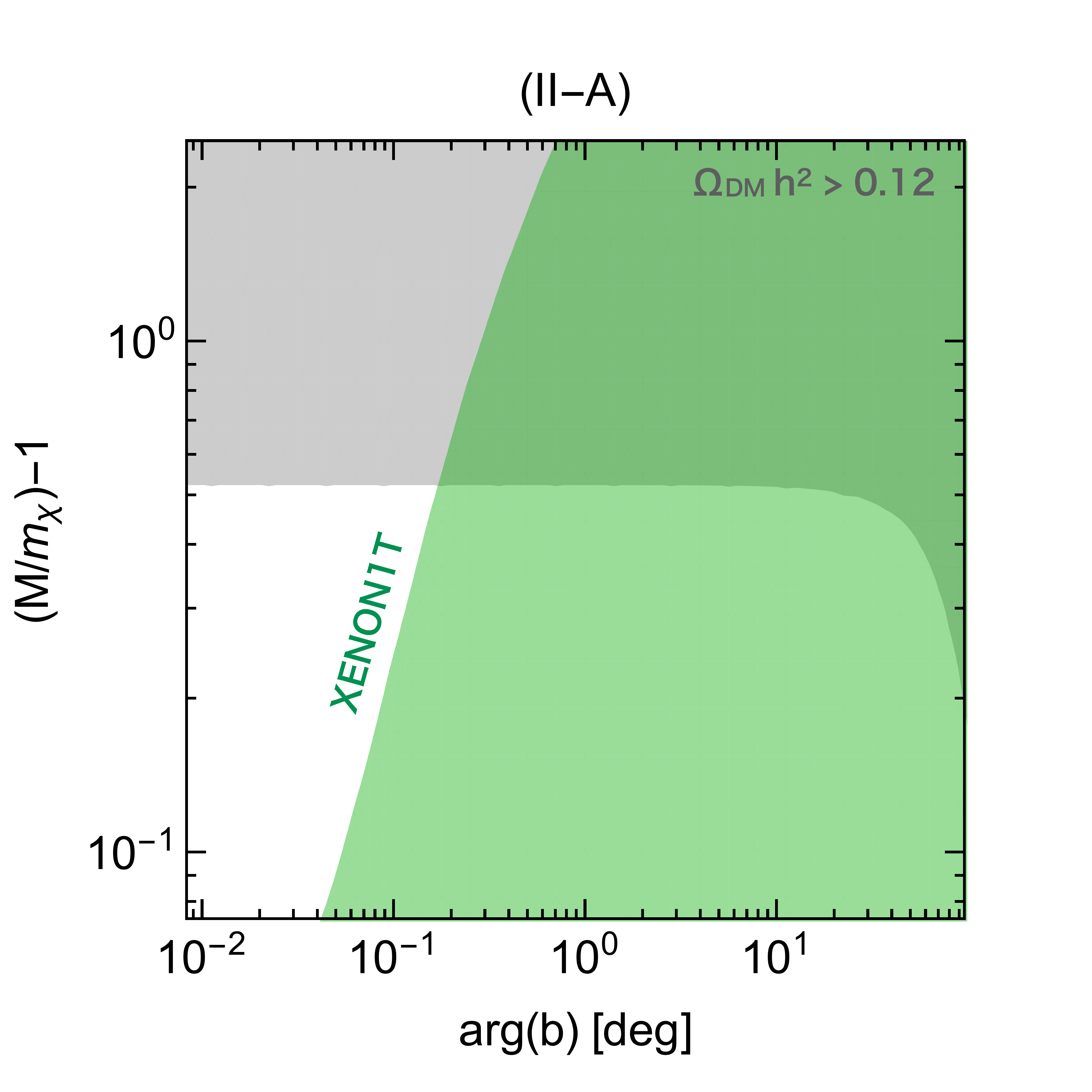}
  ~~
  \includegraphics[width=50mm]{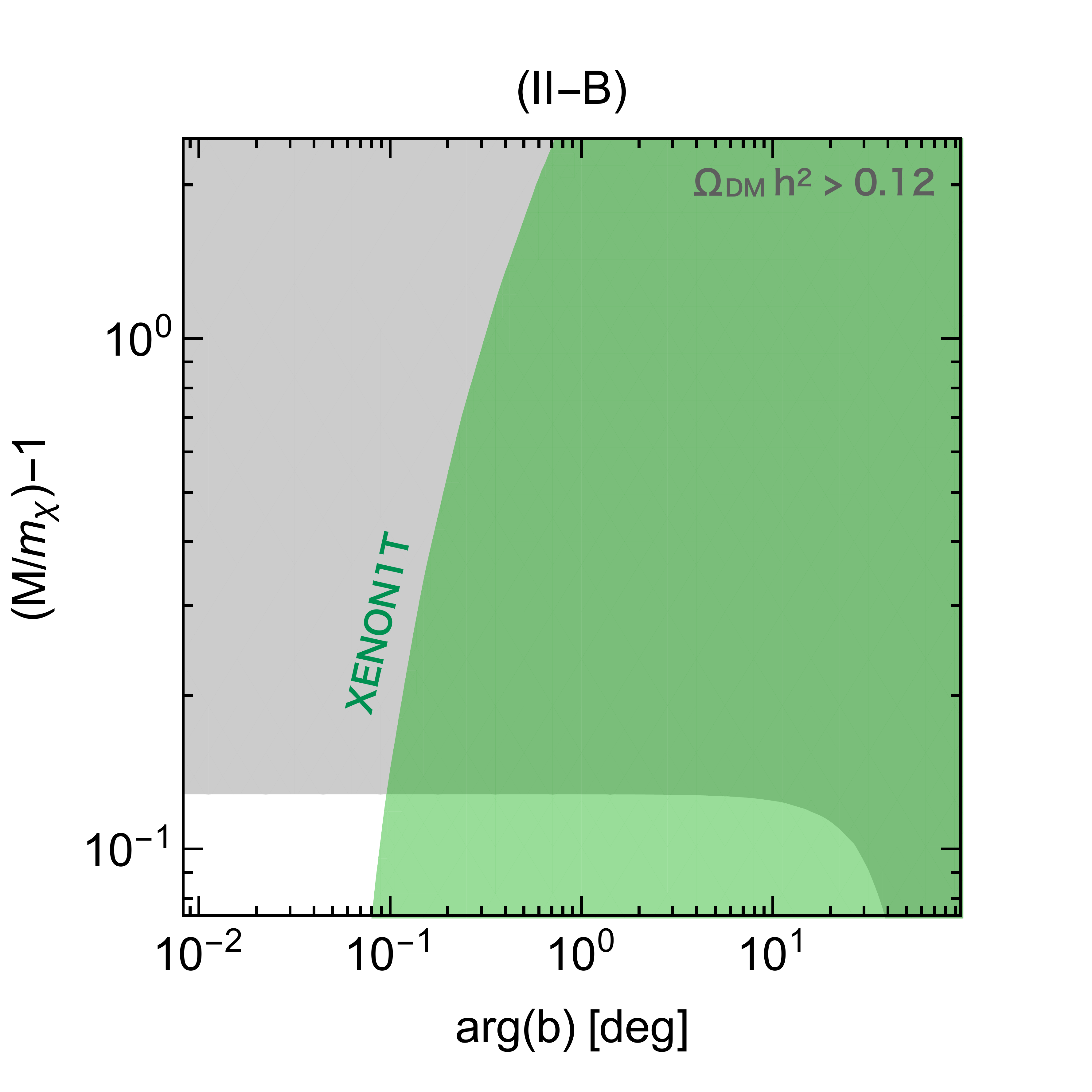}
   \end{center}
 \caption{Current limits provided by XENON1T \cite{Aprile:2018dbl}
 (green shaded area) and expected sensitivity of XENONnT
 \cite{Aprile:2015uzo} (green solid line) in the parameter space of each
 model. We also show the predicted values of the nucleon EDM in the
 magenta solid lines, as well as the expected reach provided by
 the future nucleon EDM search, $|d_p|\geq 2.5\times 10^{-29}~e \cdot
 \text{cm}$ \cite{Lehrach:2012eg}, in the magenta shaded region. 
The gray shaded area represents the region
 where the DM density exceeds the observed value: $\Omega_{\text{DM}}
 h^2\geq 0.12$. Here we take $m_\chi=3$~TeV, 2~TeV, 2~TeV, 2~TeV, and
 1.5~TeV in the cases of Model I-A, I-B, I-C, II-A, and II-B,
 respectively.  }
 \label{fig:res-EDM}
\end{figure}
%%%figure%%%
%%%%%%%%%
%%%%%%%%
%%%%%%%%

Next we discuss the cases where there is a non-zero CP phase in the
DM-mediator interactions. In Fig.~\ref{fig:res-EDM}, we again show current
limits provided by XENON1T \cite{Aprile:2018dbl} (green shaded area) and
the expected sensitivity of XENONnT \cite{Aprile:2015uzo} (green solid
line) in the parameter space of each model, where we now vary
$\text{arg}(b)$. We also show the predicted values of the nucleon EDM in the
magenta solid lines, as well as the expected reach provided by
the future nucleon EDM search, $|d_p|\geq 2.5\times 10^{-29}~e \cdot
\text{cm}$ \cite{Lehrach:2012eg}, in the magenta shaded region. The gray shaded
area represents the region where the DM density exceeds the observed
value: $\Omega_{\text{DM}} h^2\geq 0.12$. Here we take $m_\chi=3$~TeV,
2~TeV, 2~TeV, 2~TeV, and 1.5~TeV in the cases of Model I-A, I-B, I-C,
II-A, and II-B, respectively. We see that the limits from the direct
detection experiments get quite strong in the presence of the CP
phases in the DM-mediator interactions due to the large contribution of
the DM EDM. Indeed, if the CP phases are ${\cal O}(1)$, the current
limits are so severe that most of the parameter space with
$\Omega_{\rm{DM}} h^2\leq 0.12$ has already been disfavored. In
addition, we see a correlation between the direct detection bound and
the predicted value of the nucleon EDM in the case of Model I, as
expected, though it turns out that the regions which can be covered by the future
nucleon EDM searches have already been excluded by the XENON1T limit. We however note that EDM experiments may still be able to
probe both Model I and II since CP phases in the
vector-like-fermion-Higgs couplings induce the electron and nucleon EDMs
without generating the DM EDM, as we see in the previous section. All in
all, even though the relevant physical observables (such as the DM
detection rate and electron/nucleon EDMs) are always induced at
loop level in the singlet Dirac fermion DM scenarios discussed in this
paper, they are still detectable in the forthcoming experiments, and thus
these scenarios could be put on the table to be thoroughly investigated
in the future.

%%%%%%%%%%%%%%%%%%%%%%%%%%%%%
\section{Conclusion}
\label{sec:concl}
%%%%%%%%%%%%%%%%%%%%%%%%%%%%%

In this paper, 
we have studied simplified models for a singlet Dirac fermion DM which interacts 
with the SM particles only through radiative corrections. 
We have considered the cases where certain sets of vector-like fermions
and complex scalars are introduced as mediator fields, which are
supposed to interact with the SM gauge and Higgs bosons at the
renormalizable level. The Dirac fermion DM has phenomenologically
distinct features as it is able to have vector and tensor couplings, which in
general result in a large DM-nucleus scattering rate. In addition, the
Dirac fermion DM particles can pair-annihilate in the $s$-wave
processes without suffering from a chirality suppression, which
allows the DM to avoid overproduction even if the DM mass is $> 1
$~TeV. Indeed, we have found that the right amount of DM abundance is
obtained for a DM mass of ${\cal O}(1)$~TeV with ${\cal O}(1)$
DM-mediator couplings, if vector-like fermions are lighter than the
Dirac fermion DM.

These simplified models generically introduce new CP phases in the
couplings. If there is a CP phase in the DM-mediator couplings, the DM
acquires an EDM, which strongly enhances the DM-nucleus scattering cross
section. Such a CP phase also induces nucleon EDMs via the Weinberg
operator if the mediator fields have color charges (Model I). Even if the mediators
are non-colored particles, CP phases in the couplings between the
vector-like fermions and the SM Higgs field generate the electron and
nucleon EDMs through the Barr-Zee diagrams.

To see the experimental implications of the rich phenomenology of the
Dirac fermion DM, in this paper, we consider the DM direct detection
experiments and the EDM measurements, with particular emphasis on the
effect of CP phases on the observables in these experiments. For direct
detection experiments, it turns out that the current limit set by the
XENON1T experiment has already started to be excluded the parameter
space. In particular, if there is a sizable CP phase in the DM-mediator
couplings, the DM direct detection rate is significantly enhanced due to
the DM-EDM contribution via the photon exchange so that even the current
limit excludes a wide range of the parameter space. We also find that
the future DM experiments can basically probe all of the parameter
region favored by the thermal relic abundance of the DM. As mentioned
above, CP phases in the DM-mediator couplings, which generate the
DM-EDM, also induce the nucleon EDMs in Model I, and in fact we have
found a correlation between these observables. It is found that the
future measurements of nucleon EDMs are less promising compared with the
DM direct detection experiments if these phases are the only source
of the extra CP violation. If, on the other hand, there are CP phases in
the couplings between the vector-like fermions and the SM Higgs field,
these phases induce the electron EDM, which may be probed in the future
experiments if the vector-like fermions are at ${\cal
O}(1)$~TeV. Consequently, even though the relevant observables are
generated only at the loop level, both the DM direct searches and the
EDM experiments are quite promising for testing the singlet Dirac DM
scenario discussed in this paper, which makes this scenario a suitable
benchmark for the forthcoming experiments.

%%%%%%%%%%%%%%%%%%%%%%%%%%%%%%%%%%%%
\section*{Acknowledgments}
%%%%%%%%%%%%%%%%%%%%%%%%%%%%%%%%%%%%

This work is supported in part by the Grant-in-Aid for Innovative Areas
(16H06492 [JH], 16H06490 [RN], 18H05542 [NN]) and Young
Scientists B (17K14270 [NN]). The work of JH is also supported by
World Premier International Research Center Initiative (WPI Initiative),
MEXT, Japan.  NN would like to thank E-ken at Nagoya University for
hospitality while this work was initiated.

%%%%%%%%%%%%%%%%%%%%%%%%%%%%%%%%%%%%%%%%%%%%%%
\newpage
\section*{Appendix}
\appendix
%%%%%%%%%%%%%%%%%%%%%%%%%%%%%%%%%%%%%%%%%%%%%

%%%%%%%%%%%%%%%%%%%%%%%%%%%%%%%%%%%%%%%%%%%%%%%%%%%%%%
\section{Interactions in Mass Eigenbasis}
\label{sec:interactions}
%%%%%%%%%%%%%%%%%%%%%%%%%%%%%%%%%%%%%%%%%%%%%%%%%%%%%%

In this section, we write down the interactions of the new particles in
the mass eigenbasis. We here use the four-component notation, with
vector-like fermion mass eigenstates defined by 
\begin{equation}
 \psi_{f_i} \equiv
\begin{pmatrix}
 \psi_{f_iL} \\ \psi_{f_iR}
\end{pmatrix}
~.
\end{equation}

%%%%%%%%%%%%%%%%%%%%%%%%%%%%%%%
\subsection{Gauge interactions}
%%%%%%%%%%%%%%%%%%%%%%%%%%%%%%%

We list the gauge interaction terms that are relevant for the
discussion in this paper.  
\begin{align}
  {\cal L}_{\rm gauge} &= -e A_\mu \sum_{f,i}
 Q_f \overline{\psi}_{f_i} \gamma^\mu \psi_{f_i} 
 -g_Z Z_\mu \sum_{f,i,j} 
 \overline{\psi}_{f_i} \gamma^\mu 
\left(
C^{ij}_{fZL} P_L + C^{ij}_{fZR} P_R
\right)
\psi_{f_j}
\nonumber \\[2pt]
- \frac{g}{\sqrt{2}}&
\biggl[
\overline{\psi}_{u_i} \gamma^\mu W_\mu^+ 
\left(C^{ij}_{QWL} P_L + C^{ij}_{QWR} P_R\right) \psi_{d_j}
+
\overline{\psi}_{\nu} \gamma^\mu W_\mu^+ 
\left(C^{i}_{LW L} P_L + C^{i}_{LWR} P_R\right) \psi_{e_i}
+{\rm h.c.}
\biggr]
\nonumber \\[2pt]
&-ie A_\mu \sum_{f,i} Q_f\widetilde{f}^*_i
 \overleftrightsmallarrow{\partial}^\mu \widetilde{f}_i^{}
-ig_Z Z_\mu \sum_{f,i,j} \widetilde{C}^{ij}_{fZ} \, \widetilde{f}^*_i
 \overleftrightsmallarrow{\partial}^\mu \widetilde{f}_j^{} +\dots~,
\end{align}
with
\begin{align}
 C_{fZL/R}^{ij} &= \left(V_{fL/R}\right)^*_{1i}
 \left(V_{fL/R}\right)^{}_{1j} T_{3f} -  Q_f \sin^2
 \theta_W \delta_{ij}~, \nonumber\\[2pt]
 C_{QWL/R}^i&= \left(V_{uL/R}\right)^*_{1i}
 \left(V_{dL/R}\right)^{}_{1j}~,\qquad
 C_{LWL/R}^i = e^{\pm\frac{i}{2}\theta_L}
 \left(V_{eL/R}\right)^{}_{1i}~,
\nonumber\\[2pt]
\widetilde{C}^{ij}_{fZ}  &=
\bigl(\widetilde{V}_f\bigr)^*_{1i}
\bigl(\widetilde{V}_f\bigr)_{1j} T_{3f}-  Q_f \sin^2
 \theta_W \delta_{ij}~,
\end{align}
where $e>0$ is the electric charge of positron, $g_Z\equiv \sqrt{g^{\prime
2} + g^2}$, $g^\prime$ and $g$ are the gauge coupling constants of
U(1)$_Y$ and SU(2)$_L$, respectively, $P_{L/R} \equiv (1\mp
\gamma_5)/2$, $Q_f$ and $T_{3f}$ are the electric charge and the weak
isospin of the fermion $f$, respectively, $\theta_W$ is the weak mixing
angle,
and $A\overleftrightsmallarrow{\partial}_\mu B \equiv A (\partial_\mu B)
- (\partial_\mu A) B$.

%%%%%%%%%%%%%%%%%%%%%%%%%%%%%%%
\subsection{Higgs couplings}
%%%%%%%%%%%%%%%%%%%%%%%%%%%%%%%

The relevant part of the Higgs interactions is 
\begin{align}
 {\cal L}_{\rm Higgs} &= -\frac{1}{\sqrt{2}}\sum_{f,i,j}
h\overline{\psi}_{f_i} \left(C^{ij}_{fhL} P_L +C^{ij}_{fhR} P_R \right)
 \psi_{f_j} 
-\frac{1}{\sqrt{2}}\sum_{f,i,j}\widetilde{C}^{ij}_{fh} \,h \widetilde{f}^*_i \widetilde{f}_j^{} ~,
\end{align}
with 
\begin{align}
C^{ij}_{fhL} &= 
\left[
\kappa_{\bar{f}}^{} \bigl(V_{fR}\bigr)_{2i}^* \bigl(V_{fL}\bigr)_{1j}^{} 
+
\kappa_{\bar{f}}^\prime \bigl(V_{fR}\bigr)_{1i}^* \bigl(V_{fL}\bigr)_{2j}^{} 
\right] ~, \nonumber \\[2pt]
C^{ij}_{fhR} &= (C^{ji}_{fhL})^* ~, \nonumber \\[2pt]
\widetilde{C}^{ij}_{fh}&= 
\left[
A_{\bar{f}}^{} \bigl(\widetilde{V}_{f}\bigr)^*_{2i} 
\bigl(\widetilde{V}_{f}\bigr)^{}_{1j}
+A_{\bar{f}}^* \bigl(\widetilde{V}_{f}\bigr)^*_{1i} 
\bigl(\widetilde{V}_{f}\bigr)^{}_{2j}
\right]\nonumber \\
&+v\left[
(\lambda_f-2T_{3f} \lambda_{f}^\prime) 
\bigl(\widetilde{V}_{f}\bigr)^*_{1i} 
\bigl(\widetilde{V}_{f}\bigr)^{}_{1j}
+\lambda_{\bar{f}}\bigl(\widetilde{V}_{f}\bigr)^*_{2i} 
\bigl(\widetilde{V}_{f}\bigr)^{}_{2j}
\right]~,
\end{align}
where $\lambda_u^{(\prime)} = \lambda_d^{(\prime)} \equiv
\lambda_Q^{(\prime)}$, $\lambda_\nu^{(\prime)} = \lambda_e^{(\prime)} \equiv
\lambda_L^{(\prime)}$, and $\lambda_{\bar{\nu}} = 0$.

%%%%%%%%%%%%%%%%%%%%%%%%%%%%%%%
\subsection{Dark matter couplings}
\label{app:dmcoups}
%%%%%%%%%%%%%%%%%%%%%%%%%%%%%%%

In the mass eigenbasis, the terms in Eq.~\eqref{eq:lagdmint} are given
as follows:
\begin{align}
 {\cal L}_{\chi f \tilde{f}} = \overline{\chi} 
\left(C^{ij}_{f\chi L} P_L +C^{ij}_{f\chi R} P_R\right)
\psi_{f_i} \widetilde{f}^*_j +{\rm h.c.} ~,
\end{align}
with 
\begin{align}
 C^{ij}_{f\chi L} &= e^{-\frac{i}{2} \theta_\chi}
\left[a_f \bigl(V_{fL}\bigr)_{1i}^{} \bigl(\widetilde{V}_f\bigr)^*_{1j}
+a_{\bar{f}} \bigl(V_{fL}\bigr)_{2i}^{}
 \bigl(\widetilde{V}_f\bigr)^*_{2j} 
\right] ~, \nonumber \\[2pt]
C^{ij}_{f\chi R} &= e^{\frac{i}{2} \theta_\chi}
\left[b_f^* \bigl(V_{fR}\bigr)_{1i}^{} \bigl(\widetilde{V}_f\bigr)^*_{1j}
+b^*_{\bar{f}} \bigl(V_{fR}\bigr)_{2i}^{}
 \bigl(\widetilde{V}_f\bigr)^*_{2j} 
\right] ~,
\end{align}
where $a_u = a_d \equiv a_Q$, $b_u = b_d \equiv b_Q$, $a_\nu = a_e  \equiv a_L$, $b_\nu =
b_e \equiv b_L$, and $a_{\bar{\nu}} = b_{\bar{\nu}} = 0$.

%%%%%%%%%%%%%%%%%%%%%%%%%%%%%%%%%%%%%%%%%%%%%%%%%%%%%%%%%%%%%
\section{Effective Interactions of Singlet Dirac Fermion DM}
\label{app:DM-eff}
%%%%%%%%%%%%%%%%%%%%%%%%%%%%%%%%%%%%%%%%%%%%%%%%%%%%%%%%%%%%%

In this appendix, we summarize the one-loop formulae for the Wilson
coefficients of the effective operators considered in Sec.~\ref{sec:dd}. 

%%%%%%%%%%%%%%%%%%%%%%%%%%%%%%%%%%%%%%%%%%%%%%%%%%%%%%
\subsection{DM-photon effective interactions}
\label{app:DM-photon}
%%%%%%%%%%%%%%%%%%%%%%%%%%%%%%%%%%%%%%%%%%%%%%%%%%%%%%

The Wilson coefficients of the DM-photon effective interactions given in
Eq.~\eqref{eq:Effint_photon}, which are induced by the diagrams shown in
Fig.~\ref{fig:DM-gamma}, are computed as 
\begin{align}
C^{\gamma}_M
&=
\frac{e}{(4\pi)^2 m_\chi}\sum_{f,i,j}N_{c}Q_f
 \scalebox{0.9}{$\displaystyle
\biggl\{C^{ij}_{f\chi SS} \,
 g_{M1}(m_\chi,\widetilde{m}_{f_j} 
 ,m_{f_i})+
\frac{m_{f_i}}{m_\chi}
\text{Re}\left[C^{ij}_{f\chi L} C^{ij*}_{f\chi
 R}\right] \,g_{M2}(m_\chi,\widetilde{m}_{f_j} ,m_{f_i})\biggr\}
$}
 ~,\\
C^{\gamma}_E
&=
\frac{e}{(4\pi)^2 m_\chi^2}\sum_{f,i,j}N_{c}Q_f
m_{f_i}
\mbox{Im}\left[C^{ij}_{f\chi L}
 C^{ij*}_{f\chi R}\right]g_{E1}(m_\chi,\widetilde{m}_{f_j} ,m_{f_i}) ~,
\label{eq:cgame}
\\
C^{\gamma}_R
&=\frac{e}{(4\pi)^2 m_\chi^2}\sum_{f,i,j}N_{c}Q_f
 \scalebox{0.9}{$\displaystyle
\biggl\{C^{ij}_{f\chi SS} \,
 g_{R1}(m_\chi,\widetilde{m}_{f_j} 
 ,m_{f_i})+
\frac{m_{f_i}}{m_\chi}
\mbox{Re}\left[C^{ij}_{f\chi L} C^{ij*}_{f\chi
 R}\right] \, {g}_{R2}(m_\chi,\widetilde{m}_{f_j} ,m_{f_i})\biggr\}
$}
 ~,
\end{align}
where
\begin{equation}
 C^{ij}_{f\chi SS} \equiv \frac{1}{2} 
\left\{|C^{ij}_{f\chi L}|^2+|C^{ij}_{f\chi R}|^2\right\}
~, 
\label{eq:cijfchiss}
\end{equation}
and $N_c$ is the color factor: $N_c=3$ (1) for Model I (II). The mass
functions in the above expressions are given by
\begin{align}
 g_{M1} (m_\chi, M, m) &= 
1 - \frac{M^2 -m^2}{2m_\chi^2} \ln \biggl(\frac{M^2}{m^2}\biggr)
+ 
\frac{\Delta +m_\chi^2 (M^2-m_\chi^2 + m^2)}{2 m_\chi^2}
L 
~, \\[3pt]
 g_{M2} (m_\chi, M, m) &= 
\frac{1}{2} \biggl[
\ln \biggl(\frac{M^2}{m^2}\biggr)
-(M^2 + m_\chi^2 -m^2) L
\biggr] ~, \\[3pt]
 g_{E1} (m_\chi, M, m) &=  g_{M2} (m_\chi, M, m) ~, \\[3pt]
 g_{R1} (m_\chi, M, m) &=
\frac{1}{12} 
\biggl[
\frac{8(M^2 -m^2) + m_\chi^2}{m_\chi^2} \ln \biggl(
\frac{M^2}{m^2}
\biggr)
- \frac{4}{\Delta} \left\{
4\Delta + m_\chi^2 (M^2 + 3m^2)-m_\chi^4
\right\}
\nonumber \\[3pt]
 -&\frac{1}{m_\chi^2 \Delta}
\left\{
8\Delta^2 + (9M^2 -5m_\chi^2 + 7m^2) m_\chi^2 \Delta 
- 4m^2 m_\chi^4 (3M^2 -m_\chi^2 + m^2) 
\right\}
L
\biggr] ~, \\[3pt]
 g_{R2} (m_\chi, M, m) &=\frac{1}{3}
\biggl[- \ln \biggl(\frac{M^2}{m^2}\biggr) + 
\frac{2 m_\chi^2 (M^2 - m^2)}{\Delta}
\nonumber \\[3pt]
&\qquad +\frac{M^2 -m^2}{\Delta} \left\{
\Delta + m_\chi^4 -m_\chi^2 (M^2 +m^2)
\right\}L
\biggr]
~,
\end{align}
where\footnote{$\Delta$ can be factorized as 
 $\Delta = (M-m_\chi -m)(M-m_\chi + m) (M + m_\chi -m) (M+m_\chi +m)$.
} 
\begin{align}
 \Delta (m_\chi^2, M^2, m^2) &\equiv 
m_\chi^4 - 2 m_\chi^2 (M^2 + m^2) +(M^2 -m^2)^2 ~,
\end{align}
and 
\begin{equation}
 L(m_\chi^2, M^2, m^2) \equiv  
\begin{cases}
 \frac{1}{\sqrt{\Delta}} \ln \biggl(
\frac{M^2 + m^2 -m_\chi^2 + \sqrt{\Delta}}{M^2 + m^2 -m_\chi^2 -
 \sqrt{\Delta}} 
\biggr) & (\Delta > 0)\\
 \frac{2}{\sqrt{|\Delta|}} \mathrm{arctan}
\biggl(\frac{\sqrt{|\Delta|}}{M^2 + m^2 -m_\chi^2 }\biggr)
& (\Delta < 0)
\end{cases}
~.
\end{equation}
We have checked that $\mu_\chi$ and $b_\chi$ in
Ref.~\cite{Ibarra:2015fqa} are reproduced from the above expressions for
$C^\gamma_M$ and $C^\gamma_R$, respectively, by taking $C_{f\chi L}^{ij}
= 0$, and $C^\gamma_M$, $C^\gamma_E$, and $C^\gamma_R$ are consistent with those given in Ref.~\cite{Herrero-Garcia:2018koq}.
From Eq.~\eqref{eq:cgame}, we see that the DM-EDM can be generated only
in the presence of the left-right mixing as well as a non-zero imaginary
component of the product $C^{ij}_{f\chi L} C^{ij*}_{f\chi R}$, which are
not required for the generation of DM-MDM and DM charge radius. 

%%%%%%%%%%%%%%%%%%%%%%%%%%%%%%%%%%%%%%%%%%%%%%%%%%%%%%
\subsection{DM-$Z$ effective interactions}
\label{app:DM-z}
%%%%%%%%%%%%%%%%%%%%%%%%%%%%%%%%%%%%%%%%%%%%%%%%%%%%%%

The diagrams in Fig.~\ref{fig:DM-gamma} also generate the effective
DM-$Z$ vector coupling,\footnote{The same diagrams also induce the
DM-$Z$ axial-vector couplings. These couplings result in the DM-quark
interactions that are velocity-suppressed or spin-dependent, both
of which are not considered in our analysis. }
which gives rise to the DM-quark vector
interactions. In the broken phase, this coupling is represented by
\begin{equation}
 {\cal L}_{\chi Z} = -g_Z C_{\chi Z} \, \overline{\chi} \gamma^\mu \chi Z_\mu ~,
\end{equation}
where the coupling $C_{\chi Z}$ is given by
\begin{align}
 C_{\chi Z} &=\frac{1}{(4\pi)^2} \sum_{f,i,j,k} N_c 
\bigl[
C^{kij}_{fZ1} \,  g_{Z1} (m_\chi, \widetilde{m}_{f_k}, m_{f_i}, m_{f_j})
\nonumber \\
&+C^{kij}_{fZ2} \,  g_{Z2} (m_\chi, \widetilde{m}_{f_k}, m_{f_i},
 m_{f_j})
+C^{kij}_{fZ3} \,  g_{Z3} (m_\chi, \widetilde{m}_{f_k}, m_{f_i},
 m_{f_j})
\nonumber \\[3pt]
&+
\widetilde{C}^{jki}_{fZ1} \,  \widetilde{g}_{Z1} (m_\chi,
 \widetilde{m}_{f_j}, \widetilde{m}_{f_k}, m_{f_i})
+
\widetilde{C}^{jki}_{fZ2} \,  \widetilde{g}_{Z2} (m_\chi,
 \widetilde{m}_{f_j}, \widetilde{m}_{f_k}, m_{f_i})
\bigr] ~,
\end{align}
with 
\begin{align}
 C^{kij}_{fZ1} &= 
\frac{1}{2} \bigl[
C^{ij}_{fZR} C^{ik}_{f\chi R} C^{jk*}_{f\chi R}
+
C^{ij}_{fZL} C^{ik}_{f\chi L} C^{jk*}_{f\chi L}
\bigr] ~, \\[3pt]
 C^{kij}_{fZ2} &= 
\frac{m_{f_i} m_{f_j}}{2 m_\chi^2}  \bigl[
C^{ij}_{fZR} C^{ik}_{f\chi L} C^{jk*}_{f\chi L}
+
C^{ij}_{fZL} C^{ik}_{f\chi R} C^{jk*}_{f\chi R}
\bigr] ~, \\[3pt]
 C^{kij}_{fZ3} &= 
\frac{1}{2 m_\chi }
\bigl[
C_{fZR}^{ij}
\bigl\{
m_{f_i} C_{f\chi L}^{ik} C_{f\chi R}^{jk*}
+ 
m_{f_j} C_{f\chi R}^{ik} C_{f\chi L}^{jk*}
\bigr\}
\nonumber \\
&\qquad + 
C_{fZL}^{ij}
\bigl\{
m_{f_i} C_{f\chi R}^{ik} C_{f\chi L}^{jk*}
+ 
m_{f_j} C_{f\chi L}^{ik} C_{f\chi R}^{jk*}
\bigr\}
\bigr]~,\\[3pt]
 \widetilde{C}^{jki}_{fZ1} &= \frac{1}{2}
\widetilde{C}^{jk}_{fZ} \bigl[
C^{ik}_{f\chi L} C^{ij *}_{f\chi L} +
C^{ik}_{f\chi R} C^{ij *}_{f\chi R}
\bigr] ~,\\[3pt]
 \widetilde{C}^{jki}_{fZ2} &= \frac{m_{f_i}}{2m_\chi }
\widetilde{C}^{jk}_{fZ} \bigl[
C^{ik}_{f\chi L} C^{ij *}_{f\chi R} +
C^{ik}_{f\chi R} C^{ij *}_{f\chi L}
\bigr] ~,
\end{align}
and 
\begin{align}
& g_{Z1} (m_\chi, M, m_i, m_j) 
= -\frac{1}{2} \Delta_{\epsilon, \mu} 
+\frac{1}{2} \ln M + \frac{m_i^2 \ln m_i - m_j^2 \ln m_j}{2(m_i^2
 -m_j^2)}
+ \frac{m_i^2 + m_j^2 -2 M^2 -m_\chi^2}{2 m_\chi^2}
\nonumber \\[2pt]
&+ \frac{m_i^4 + m_j^4 - m_i^2 m_j^2 -3M^2(m_i^2 + m_j^2-M^2) 
-m_\chi^2 (m_i^2 + m_j^2 + 2M^2)}{4 m_\chi^4} \ln \biggl(\frac{M^2}{m_i
 m_j}\biggr)
\nonumber \\[2pt]
& - \frac{
1
}{4m_\chi^4 (m_i^2 -m_j^2)} 
\bigl[
2 m_\chi^6 -6 m_\chi^4 M^2 + m_\chi^2 \bigl\{6 M^4 -2 M^2 (m_i^2 + m_j^2) -
 m_i^4 - m_j^4\bigr\} \nonumber \\
&-2M^6 + 3 M^4 (m_i^2 + m_j^2) 
- 3 M^2 (m_i^4 + m_j^4) + m_i^6 + m_j^6
\bigr]
\ln \biggl(\frac{m_i}{m_j}\biggr) 
\nonumber \\[2pt]
&
+ \frac{\{(M^2 - m_i^2)^2 + (M^2-m_\chi^2)^2 - M^4\} \Delta
 (m_\chi^2, M^2, m_i^2) }{4 m_\chi^4
 (m_i^2 - m_j^2)}
 L(m_\chi^2, M^2, m_i^2) \nonumber \\
&
- \frac{\{(M^2 - m_j^2)^2 + (M^2-m_\chi^2)^2 - M^4\} \Delta
 (m_\chi^2, M^2, m_j^2) }{4 m_\chi^4
 (m_i^2 - m_j^2)}
 L(m_\chi^2, M^2, m_j^2) ~,
\end{align}
\begin{align}
 & g_{Z2} (m_\chi, M, m_i, m_j)
= \frac{M^2 -m_\chi^2}{m_i^2 -m_j^2} \ln
 \biggl(\frac{m_i}{m_j}\biggr)
+ \frac{1}{2} 
\biggl[\ln \biggl(\frac{M^2}{m_i m_j}\biggr)
- \frac{m_i^2 + m_j^2}{m_i^2 -m_j^2} \ln \biggl(\frac{m_i}{m_j}\biggr)
\biggr] 
\nonumber \\[2pt]
& + \frac{1}{2  (m_i^2 -m_j^2)} \bigl[
\Delta (m_\chi^2, M^2, m_i^2) L (m_\chi^2, M^2, m_i^2) 
-\Delta (m_\chi^2, M^2, m_j^2) L (m_\chi^2, M^2, m_j^2)
\bigr] ~,
\end{align}
\begin{align}
  & g_{Z3} (m_\chi, M, m_i, m_j)
= \frac{1}{2} + \frac{m_i^2 + m_j^2 -2 M^2}{4 m_\chi^2}  \ln 
\biggl(\frac{M^2}{m_i m_j}\biggr)
\nonumber \\[2pt]
&- \frac{m_i^4 + m_j^4 - 2M^2 (m_i^2 + m_j^2) + 2(M^2 -m_\chi^2)^2 }{4
 m_\chi^2 (m_i^2 -m_j^2)} \ln \biggl(\frac{m_i}{m_j}\biggr)
\nonumber \\[2pt]
&+ \frac{\scalebox{0.8}{$\displaystyle
(m_i^2-M^2 + m_\chi^2) \Delta (m_\chi^2, M^2, m_i^2) L (m_\chi^2, M^2,
 m_i^2)
-(m_j^2-M^2 + m_\chi^2) \Delta (m_\chi^2, M^2, m_j^2) L (m_\chi^2, M^2,
 m_j^2)
$}
}{4 m_\chi^2 (m_i^2 -m_j^2)} ~,
\end{align}
\begin{align}
& \widetilde{g}_{Z1} (m_\chi, M_i, M_j, m)
= \frac{1}{2} \Delta_{\epsilon, \mu}
- \frac{1}{2} \ln m  - \frac{M_i^2 \ln M_i - M_j^2 \ln M_j}{2(M_i^2 -
 M_j^2)}
+ \frac{M_i^2 + M_j^2 -2m^2 + m_\chi^2}{2 m_\chi^2} \nonumber \\[2pt]
&
+ \frac{3 m^2 (m^2 - M_i^2 -M_j^2) -m_\chi^2 (M_i^2 + M_j^2)
+ M_i^4 + M_j^4 + M_i^2 M_j^2
}{4 m_\chi^4} \ln \biggl(\frac{m^2}{M_i M_j}\biggr)
\nonumber \\[2pt] 
& \scalebox{0.95}{$\displaystyle - \frac{M_i^6 + M_j^6-(3m^2 + m_\chi^2) (M_i^4 + M_j^4)
+ 3 m^4 (M_i^2 + M_j^2) -2(m^2 -m_\chi^2)^2 (m^2 + m_\chi^2)
}{4 m_\chi^4
 (M_i^2 -M_j^2)} \ln \biggl(\frac{M_i}{M_j}\biggr) $}
\nonumber \\[2pt]
&+ \frac{
\scalebox{0.8}{$\displaystyle
\{(M_i^2 -m^2)^2 -m_\chi^4\}\Delta (m_\chi^2, M_i^2, m^2)
 L(m_\chi^2, M_i^2, m^2)
-\{(M_j^2 -m^2)^2 -m_\chi^4\}\Delta (m_\chi^2, M_j^2, m^2)
 L(m_\chi^2, M_j^2, m^2)
$} 
}{4 m_\chi^4 (M_i^2-M_j^2)}  ~,
\end{align}
\begin{align}
 & \widetilde{g}_{Z2} (m_\chi, M_i, M_j, m)
= -1 - \frac{M_i^2 + M_j^2 -2m^2}{2 m_\chi^2} \ln \biggl(\frac{m^2}{M_i
 M_j}\biggr) 
\nonumber \\[2pt]
&+ \frac{M_i^4 + M_j^4 -2 m^2 (M_i^2 + M_j^2)
+ 2 (m^2 -m_\chi^2)^2
}{2m_\chi^2 (M_i^2 - M_j^2)}
\ln \biggl(\frac{M_i}{M_j}\biggr) 
\nonumber \\[2pt]
&+
\frac{
\scalebox{0.85}{$\displaystyle
(m^2-m_\chi^2 -M_i^2) \Delta (m_\chi^2, M_i^2, m^2) L(m_\chi^2, M_i^2,
 m^2)
-(m^2-m_\chi^2 -M_j^2) \Delta (m_\chi^2, M_j^2, m^2) L(m_\chi^2, M_j^2,
 m^2)
$} 
}{2m_\chi^2(M_i^2 -M_j^2)}
~,
\end{align}
where $\Delta_{\epsilon, \mu}$ denotes a divergent constant term. This
divergent constant does not appear in $C_{\chi Z}$ since 
\begin{equation}
 \sum_{i, j, k} C_{fZ1}^{kij} =  \sum_{i, j, k}
  \widetilde{C}^{jki}_{fZ1} 
\end{equation}
follows from the unitarity of $V_{fL/R}$ and $\widetilde{V}_f$, and thus
the divergent terms in $g_{Z1}$ and $\widetilde{g}_{Z1}$ cancel with
each other. We further note that the effective DM-$Z$ coupling $C_{\chi
Z}$ can be generated only after
the electroweak symmetry is spontaneously broken; since the DM is
singlet under the electroweak gauge symmetry, it can couple to the $Z$
boson only via the electroweak symmetry breaking effects. 
Diagrammatically, such
effects are represented by the insertion of the Higgs VEVs into the
diagrams in Fig.~\ref{fig:DM-gamma}, and they are represented by effective interactions including the Higgs field with more than dimention six. We have checked that the above
expressions are consistent with the result given in
Ref.~\cite{Ibarra:2015fqa}.

%%%%%%%%%%%%%%%%%%%%%%%%%%%%%%%%%%%%%%%%%%%%%%%%%%%%%
\subsection{DM-Higgs effective interactions}
\label{app:DM-h}
%%%%%%%%%%%%%%%%%%%%%%%%%%%%%%%%%%%%%%%%%%%%%%%%%%%%%%

The DM-Higgs effective scalar coupling is induced by the one-loop
diagrams shown in Fig.~\ref{fig:DM-Higgs}. We parametrize the coupling
in the broken phase as 
\begin{equation}
 {\cal L}_{\chi h} = \frac{1}{\sqrt{2}} C_{\chi h} \overline{\chi} \chi h ~.
\end{equation}
Again, the generation of this coupling requires the electroweak symmetry
breaking. It is represented by effective operators including the Higgs field with more than dimension-five. 
We neglect the pseudo-scalar coupling as it is always suppressed by the
DM velocity. The coupling $C_{\chi h}$ is computed as follows:
\begin{align}
 C_{\chi h} &=\frac{1}{(4\pi)^2} \sum_{f,i,j,k} N_c 
\bigl[
C^{kij}_{fh1} \,  g_{h1} (m_\chi, \widetilde{m}_{f_k}, m_{f_i}, m_{f_j})
\nonumber \\
&+C^{kij}_{fh2} \,  g_{h2} (m_\chi, \widetilde{m}_{f_k}, m_{f_i},
 m_{f_j})
+C^{kij}_{fh3} \,  g_{h3} (m_\chi, \widetilde{m}_{f_k}, m_{f_i},
 m_{f_j})
\nonumber \\[3pt]
&+
\widetilde{C}^{jki}_{fh1} \,  \widetilde{g}_{h1} (m_\chi,
 \widetilde{m}_{f_j}, \widetilde{m}_{f_k}, m_{f_i})
+
\widetilde{C}^{jki}_{fh2} \,  \widetilde{g}_{h2} (m_\chi,
 \widetilde{m}_{f_j}, \widetilde{m}_{f_k}, m_{f_i})
\bigr] ~,
\end{align}
with 
\begin{align}
 C^{kij}_{fh1} &= 
\frac{1}{2} \bigl[
C^{ij}_{fhR} C^{ik}_{f\chi L} C^{jk*}_{f\chi R}
+
C^{ij}_{fhL} C^{ik}_{f\chi R} C^{jk*}_{f\chi L}
\bigr] ~, \\[3pt]
 C^{kij}_{fh2} &= 
\frac{m_{f_i} m_{f_j}}{2 m_\chi^2}  \bigl[
C^{ij}_{fhR} C^{ik}_{f\chi R} C^{jk*}_{f\chi L}
+
C^{ij}_{fhL} C^{ik}_{f\chi L} C^{jk*}_{f\chi R}
\bigr] ~, \\[3pt]
 C^{kij}_{fh3} &= 
\frac{1}{2 m_\chi }
\bigl[
C_{fhR}^{ij}
\bigl\{
m_{f_i} C_{f\chi L}^{ik} C_{f\chi L}^{jk*}
+ 
m_{f_j} C_{f\chi R}^{ik} C_{f\chi R}^{jk*}
\bigr\}
\nonumber \\
&\qquad + 
C_{fhL}^{ij}
\bigl\{
m_{f_i} C_{f\chi R}^{ik} C_{f\chi R}^{jk*}
+ 
m_{f_j} C_{f\chi L}^{ik} C_{f\chi L}^{jk*}
\bigr\}
\bigr]~,\\[3pt]
 \widetilde{C}^{jki}_{fh1} &= \frac{1}{2m_\chi }
\widetilde{C}^{jk}_{fh} \bigl[
C^{ik}_{f\chi L} C^{ij *}_{f\chi L} +
C^{ik}_{f\chi R} C^{ij *}_{f\chi R}
\bigr] ~,\\[3pt]
 \widetilde{C}^{jki}_{fh2} &= \frac{m_{f_i}}{2m_\chi^2 }
\widetilde{C}^{jk}_{fh} \bigl[
C^{ik}_{f\chi L} C^{ij *}_{f\chi R} +
C^{ik}_{f\chi R} C^{ij *}_{f\chi L}
\bigr] ~,
\end{align}
and the mass functions are given by
\begin{align}
& g_{h1} (m_\chi, M, m_i, m_j)
= 2 + \Delta_{\epsilon, \mu} -\ln M - \frac{m_i^2 \ln m_i - m_j^2 \ln
 m_j}{m_i^2 -m_j^2} 
\nonumber \\[2pt]
&+\frac{m_i^2 + m_j^2 -M^2}{2 m_\chi^2} \ln \biggl(\frac{M^2}{m_i
 m_j}\biggr)
-\frac{m_i^4 + m_j^4 -M^2 (m_i^2 + m_j^2)}{2 m_\chi^2 (m_i^2 -m_j^2)}
 \ln \biggl(\frac{m_i}{m_j}\biggr)
\nonumber \\[2pt]
&+ \frac{
m_i^2 \Delta (m_\chi^2, M^2, m_i^2) L (m_\chi^2, M^2, m_i^2)
-m_j^2 \Delta (m_\chi^2, M^2, m_j^2) L (m_\chi^2, M^2, m_j^2)
}{2 m_\chi^2 (m_i^2 - m_j^2)} ~,
\end{align}
\begin{align}
 g_{h2} (m_\chi, M, m_i, m_j) &= g_{Z2} (m_\chi, M, m_i, m_j) ~, 
\\[3pt]
 g_{h3} (m_\chi, M, m_i, m_j) &= g_{Z3} (m_\chi, M, m_i, m_j) ~, 
\end{align}
\begin{align}
& \widetilde{g}_{h1} (m_\chi, M_i, M_j, m)
= \frac{1}{2} + 
\frac{M_i^2 + M_j^2 -2 (m^2 + m_\chi^2)}{4 m_\chi^2}
\ln \biggl(\frac{m^2}{M_i M_j}\biggr)
\nonumber \\[2pt]
&- \frac{M_i^4 + M_j^4 -2 (M_i^2 + M_j^2) (m^2 + m_\chi^2) - 2(m_\chi^4
 - m^4)}{4 m_\chi^2 (M_i^2 - M_j^2)} \ln \biggl(\frac{M_i}{M_j}\biggr)
\nonumber \\[2pt]
& + \frac{
\scalebox{0.85}{$\displaystyle
(M_i^2 -m^2 - m_\chi^2) \Delta (m_\chi^2, M^2_i, m^2) L (m_\chi^2,
 M^2_i, m^2)
-(M_j^2 -m^2 - m_\chi^2) \Delta (m_\chi^2, M^2_j, m^2) L (m_\chi^2,
 M^2_j, m^2)
$}
}{4 m_\chi^2 (M_i^2 - M_j^2)}
~,
\end{align}
\begin{align}
  \widetilde{g}_{h2} (m_\chi, M_i, &M_j, m)
= -\frac{1}{2} \ln \biggl(\frac{m^2}{M_i M_j}\biggr)
+ \frac{M_i^2 + M_j^2 + 2 (m_\chi^2 - m^2)}{2(M_i^2 - M_j^2)}
\ln \biggl(\frac{M_i}{M_j}\biggr) \nonumber \\[2pt]
&- \frac{ \Delta (m_\chi^2, M^2_i, m^2) L (m_\chi^2,
 M^2_i, m^2)
-\Delta (m_\chi^2, M^2_j, m^2) L (m_\chi^2,
 M^2_j, m^2)
}{2 (M_i^2 -M_j^2)} ~.
\end{align}
Again, the divergent constant $\Delta_{\epsilon, \mu}$ in $g_{h1}$ does
not contribute to $C_{\chi h}$ since 
\begin{equation}
 \sum_{i, j, k} C^{kij}_{fh1} = 0 ~.
\end{equation}
We have checked that the above expressions are consistent with the
results given in Refs.~\cite{Ibarra:2015fqa, Herrero-Garcia:2018koq}.

%%%%%%%%%%%%%%%%%%%%%%%%%%%%%%%%%%%%%%%%%%%%%%%%%%%%%%
\subsection{DM-gluon effective interactions}
\label{app:DM-g}
%%%%%%%%%%%%%%%%%%%%%%%%%%%%%%%%%%%%%%%%%%%%%%%%%%%%%%

The diagrams in Fig.~\ref{fig:dmg} give rise to the effective DM-gluon
scalar coupling.\footnote{It is found to be convenient to compute these
diagrams in the Fock-Schwinger gauge \cite{Fock:1937dy,
*Schwinger:1973rv, *Cronstrom:1980hj, *Novikov:1983gd}, where the
diagrams (a) and (c) vanish \cite{Hisano:2010ct}, the diagram (b) gives
$g_{g1}$ and $g_{g3}$, and the diagram (d) yields $g_{g2}$ and
$g_{g4}$. }
At the vector-like fermion mass threshold, the Wilson
coefficient of the gluon operator in Eq.~\eqref{eq:leffqg}, $C^g_S (m_{\text{vec}})$, is
given by \cite{Hisano:2010ct}
\begin{align}
 C^g_S (m_{\text{vec}}) &= \frac{1}{16}
\biggl[
C^{ij}_{f\chi SS} \,
m_\chi 
\bigl\{
g_{g1} (m_\chi,\widetilde{m}_{f_j},m_{f_i}) 
+g_{g2} (m_\chi,\widetilde{m}_{f_j},m_{f_i}) 
\bigr\}
\nonumber \\[2pt]
& +\text{Re} \bigl[C^{ij}_{f\chi L} C^{ij*}_{f\chi R}\bigr]
m_{f_i} 
\bigl\{
g_{g3} (m_\chi,\widetilde{m}_{f_j},m_{f_i}) 
+g_{g4} (m_\chi,\widetilde{m}_{f_j},m_{f_i}) 
\bigr\}
\biggr] ~,
\end{align}
where $C^{ij}_{f\chi SS} $ is defined in Eq.~\eqref{eq:cijfchiss}
and
\begin{align}
 g_{g1} (m_\chi, M, m) 
&= 
-\frac{(\Delta - 6 M^2 m^2) (M^2 + m^2 -m_\chi^2)}{6 \Delta^2 M^2}
- \frac{2M^2 m^4}{\Delta^2} L ~, \\[3pt]
g_{g2} (m_\chi, M, m) &= -\frac{\Delta +12 M^2 m^2}{6 \Delta^2}
+ \frac{m^2 M^2 (M^2 + m^2 -m_\chi^2)}{\Delta^2} L ~, \\[3pt]
g_{g3} (m_\chi, M, m) &=
- \frac{3 \Delta M^2 - (\Delta -6m^2 M^2)(M^2 -m^2 + m_\chi^2)}{6
 \Delta^2 M^2} \nonumber \\[2pt]
&+ \frac{m^2 M^2 (M^2 -m^2 -m_\chi^2)}{\Delta^2} L ~,\\[3pt]
g_{g4} (m_\chi, M, m) &=
\frac{3 \Delta m^2 +2(\Delta + 3 m^2 M^2) (M^2 -m^2
 -m_\chi^2)}{6\Delta^2 m^2}  \nonumber \\[2pt]
& - \frac{M^2 \{\Delta + m^2 (M^2 -m^2 + m_\chi^2)\}}{\Delta^2} L ~.
\end{align}

%%%%%%%%%%%%%%%%%%%%%%%%%%%%%%%%%%%%%%%%%%%%%%%%%%%%%%
\subsection{Low-energy DM-quark/gluon couplings}
\label{app:DM-lowen}
%%%%%%%%%%%%%%%%%%%%%%%%%%%%%%%%%%%%%%%%%%%%%%%%%%%%%%

The DM-quark low-energy effective interactions in Eq.~\eqref{eq:leffqg}
are obtained from the DM-$Z$ and DM-Higgs effective couplings, $C_{\chi
Z}$ and $C_{\chi h}$, by integrating out the $Z$ and Higgs boson fields,
respectively: 
\begin{align}
 C^q_V &= - \frac{2}{v^2} (T_{3q} - 2Q_q \sin^2 \theta_W) C_{\chi Z} ~,
 \\ 
 C^q_S &= - \frac{1}{\sqrt{2} \,m_h^2 v} C_{\chi h} ~,
\end{align}
where $T_{3q} =+1/2$ $(-1/2)$ and $Q_q =2/3 $ ($-1/3$) for up-type
(down-type) quarks, $v\simeq 246$~GeV is the Higgs VEV,
$\theta_W$ is the weak-mixing angle, and $m_h$ is the Higgs boson mass. 

For the DM-gluon low-energy coupling $C^g_S$, not only the contribution
from the diagrams in Fig.~\ref{fig:dmg}, $ C^g_S (m_{\text{vec}})$, but
also those from the DM-heavy quark interactions induced by the Higgs coupling
$C_{\chi h}$ should be included. We thus have
\begin{equation}
 C^g_S = C^g_S (m_{\text{vec}}) + \frac{1}{4\sqrt{2} m_h^2 v} C_{\chi h} ~,
\end{equation}
where the second term in the right-hand side represents the contributions
from charm, bottom, and top quarks. These long-distance contributions
receive relatively large QCD corrections---for the inclusion of such
corrections, see Refs.~\cite{Vecchi:2013iza, Ellis:2018dmb}.

%%%%%%%%%%%%%%%%%%%%%%%%%%%%%%%%%%%%
\section{Velocity Integrals}
\label{app:integ}
%%%%%%%%%%%%%%%%%%%%%%%%%%%%%%%%%%%%

As we see in Sec.~\ref{sec:eventrate}, to obtain the differential event rate
we need to perform the following velocity integrals: 
\begin{equation}
 \zeta (E_R) = \int_{v_\text{min}}^\infty
\frac{d^3 \bm{v}}{v} 
f (\bm{v} + \bm{v}_\text{E}) ~,
\qquad 
\xi (E_R) =  \int_{v_\text{min}}^\infty
d^3 \bm{v}\,
v\,
f (\bm{v} + \bm{v}_\text{E}) ~,
\label{eq:intdefs}
\end{equation}
with
\begin{equation}
 f (\bm{v}) = 
\begin{cases}
 \frac{1}{N} e^{-v^2/v_0^2}  & (|\bm{v}| < v_{\text{esc}}) \\ 
 0& (|\bm{v}| > v_{\text{esc}}) 
\end{cases}
~,
\end{equation}
where
\begin{equation}
 N = \pi^{3/2} v_0^3 \biggl[
\text{erf} \biggl(\frac{v_{\text{esc}}}{v_0}\biggr)
- \frac{2 v_{\text{esc}}}{\sqrt{\pi} v_0}  e^{-
\frac{v_{\text{esc}}^2}{v_0^2}} 
\biggr]~.
\end{equation}
With this constant $N$, the distribution function $f(\bm{v})$ is
normalized such that 
\begin{equation}
  \int
d^3\bm{v} \, 
f (\bm{v} ) = 1 ~.
\end{equation}
In what follows, we summarize analytical expressions of the integrals in
Eq.~\eqref{eq:intdefs}.

The analytical expression for the integral $\zeta (E_R)$ is given in
Refs.~\cite{Savage:2006qr, McCabe:2010zh}: 
\begin{itemize}
 \item For $v_{\text{E}} + v_{\text{min}}  < v_{\text{esc}}$, 
\begin{equation}
  \zeta (E_R) =
 \frac{\pi^{{3}/{2}} v_0^3}{2N v_{\text{E}}}
\left[\text{erf}\left(\frac{v_{\text{min}}+v_{\text{E}}}{v_0} \right)
-\text{erf}\left(\frac{v_{\text{min}}-v_{\text{E}}}{v_0}  \right)
-\frac{4 v_{\text{E}}}{\sqrt{\pi} v_0} \exp\left(-
 \frac{v_{\text{esc}}^2}{v_0^2}\right) 
\right]~. 
\end{equation}
 \item For $v_{\text{min}} > |v_{\text{esc}} -v_{\text{E}} | $
and $v_{\text{E}} + v_{\text{esc}}  > v_{\text{min}}$, 
\begin{equation}
  \zeta (E_R) =
 \frac{\pi^{{3}/{2}} v_0^3}{2N v_{\text{E}}}
\left[\text{erf}\left(\frac{v_{\text{esc}}}{v_0} \right)
+\text{erf}\left(\frac{v_{\text{E}}-v_{\text{min}}}{v_0} \right)
-\frac{2}{\sqrt{\pi} } 
\left\{
\frac{v_{\text{esc}}+v_{\text{E}} -v_{\text{min}}}{v_0} 
\right\}
e^{-
 \frac{v_{\text{esc}}^2}{v_0^2}}
\right]~. 
\end{equation}
\item For $v_{\text{E}} > v_{\text{min}}  + v_{\text{esc}}$, 
\begin{equation}
 \zeta (E_R) = \frac{1}{v_{\text{E}}} ~.
\end{equation}
\item For $v_{\text{E}} + v_{\text{esc}}  < v_{\text{min}}$,
\begin{equation}
  \zeta (E_R) = 0~.
\end{equation} 
\end{itemize}

%%%%%%%%%%%%%%%%%%%%%%%%%%%%%%%%%%%%%%%%%%%
%\subsection{$\xi (E_R)$}
%%%%%%%%%%%%%%%%%%%%%%%%%%%%%%%%%%%%

An analytical expression for $\xi (E_R)$ is given as follows:
\begin{itemize}
 \item For $v_{\text{E}} + v_{\text{min}}  < v_{\text{esc}}$, 
\begin{align}
 \xi (E_R) &= \frac{\pi^{3/2} v_0^5 }{ 4 N v_{\text{E}}}
\biggl[\frac{v_0^2 + 2 v_{\text{E}}^2} {v_0^2}
\biggl\{
\text{erf}\biggl(\frac{v_{\text{min}} + v_{\text{E}}}{v_0}\biggr)
-\text{erf}\biggl(\frac{v_{\text{min}} - v_{\text{E}}}{v_0}\biggr)
\biggr\}
-\frac{8v_{\text{E}}}{\sqrt{\pi} v_0}
e^{- \frac{v_{\text{esc}}^2}{v_0^2}} 
\nonumber \\
&+\frac{2}{\sqrt{\pi}} \biggl\{
\frac{v_{\text{min}} + v_{\text{E}}}{v_0} \,
e^{-\frac{(v_{\text{min}}-v_{\text{E}})^2}{v_0^2}}
- 
\frac{v_{\text{min}} - v_{\text{E}}}{v_0} \,
e^{-\frac{(v_{\text{min}} + v_{\text{E}})^2}{v_0^2}}
\biggr\}
\nonumber \\
& + \frac{4}{3 \sqrt{\pi}} 
\biggl\{\frac{(v_{\text{esc}}- v_{\text{E}})^3 -(v_{\text{esc}}+
 v_{\text{E}})^3}{v_0^3}\biggr\} e^{- \frac{v_{\text{esc}}^2}{v_0^2}}
\biggr] ~.
\end{align}

 \item For $v_{\text{min}} > |v_{\text{esc}} -v_{\text{E}} | $
and $v_{\text{E}} + v_{\text{esc}}  > v_{\text{min}}$,

\begin{align}
  \xi (E_R) &= \frac{\pi^{3/2} v_0^5 }{ 4 N v_{\text{E}}}
\biggl[\frac{v_0^2 + 2 v_{\text{E}}^2} {v_0^2}
\biggl\{
\text{erf}\biggl(\frac{v_{\text{esc}} }{v_0}\biggr)
-\text{erf}\biggl(\frac{v_{\text{min}} - v_{\text{E}}}{v_0}\biggr)
\biggr\}
\nonumber \\
&+\frac{2}{\sqrt{\pi}} \biggl\{
\frac{v_{\text{min}} + v_{\text{E}}}{v_0} \,
e^{-\frac{(v_{\text{min}}-v_{\text{E}})^2}{v_0^2}}
- 
\frac{v_{\text{esc}} +2 v_{\text{E}}}{v_0} \,
e^{-\frac{v_{\text{esc}}^2}{v_0^2}}
\biggr\}
\nonumber \\
&+
\frac{4}{3 \sqrt{\pi}} 
\biggl\{\frac{v_{\text{min}}^3 -(v_{\text{esc}}+
 v_{\text{E}})^3}{v_0^3}\biggr\} e^{- \frac{v_{\text{esc}}^2}{v_0^2}}
\biggr] ~.
\end{align}

 \item For $v_{\text{E}} > v_{\text{min}}  + v_{\text{esc}}$,

\begin{equation}
 \xi (E_R) = 
v_{\text{E}}+
\frac{v_0^2}{2v_{\text{E}}} 
- \frac{2\pi v_0^2 v_{\text{esc}}^3}{3 N v_{\text{E}}} 
e^{- \frac{v_{\text{esc}}^2}{v_0^2}}
 ~.
\end{equation}

 \item For $v_{\text{E}} + v_{\text{esc}}  < v_{\text{min}}$,
\begin{equation}
  \xi (E_R) = 0~.
\end{equation}

\end{itemize}
We have checked that these results are consistent with those given in
Ref.~\cite{Barger:2010gv}.

%%%%%%%%%%%%%%%%%%%%%%%%%%%%%%%%%%%%%%%%%%%%%%%%%%%%%%%%%%%%%%%
\section{Direct Detection Limits/Prospects}
\label{app:eventnum}
%%%%%%%%%%%%%%%%%%%%%%%%%%%%%%%%%%%%%%%%%%%%%%%%%%%%%%%%%%%%%%%%

In this section, we show our prescription for the estimate of the
current limit and future prospects of the direct detection
experiments. For the current bound, we use the latest result from the
XENON1T experiment \cite{Aprile:2018dbl}, while for a future
experiment, we consider XENONnT \cite{Aprile:2015uzo}. We have
checked that the sensitivity of the LZ experiment \cite{Akerib:2018lyp}
is as good as that of XENONnT.

%%%%%%%%%%%%%%%%%%%%%%
%\subsection{XENON1T}
%%%%%%%%%%%%%%%%%%%%%%

Recently, the XENON1T collaboration reported their latest result of DM
direct search based on data with an exposure $w_{\text{exp}} = 278.8~ \text{days}
\times 1.30(1)$~ton. The expected number of DM-nuclei
scattering events in this experiment is estimated as follows
\cite{Aprile:2011hx}:
\begin{align}
 N_{\text{event}} = w_{\text{exp}}
 \int_{S^{\text{min}}_1}^{S^{\text{max}}_1} dS_1 
\sum_{n=1}^{\infty} \text{Gauss} (S_1|n, \sqrt{n} \sigma_{\text{PMT}})
\int_{0}^{\infty} dE_R \, \epsilon (E_R) \,
\text{Poiss} (n|\nu (E_R)) \frac{dR}{dE_R}
~,
\end{align}
where $S^{\text{min}}_1 = 3$ photoelectrons (PE), $S^{\text{max}}_1 =70$~PE,
$\sigma_{\text{PMT}}$ is the average single-PE
resolution of the photomultipliers, $\epsilon (E_R)$ is the detection
efficiency, and $\nu (E_R)$ is the expected number of PEs for a given
recoil energy $E_R$.
We conservatively take $\sigma_{\text{PMT}} = 0.4$ \cite{Aprile:2015lha,
Barrow:2016doe}, and read $\epsilon (E_R)$ from the black solid line in
Fig.~1 in Ref.~\cite{Aprile:2018dbl}. We obtain $\nu (E_R)$ from the S1
yield given in the lower left panel in Fig.~13 in
Ref.~\cite{Aprile:2015uzo}, which corresponds to $\nu
(E_R)/E_R$.\footnote{We can also estimate $\nu (E_R)$ using
\begin{equation}
 \nu (E_R) = E_R \cdot {\cal L}_{\text{eff}} \cdot {\cal L}_y \cdot
  S_{\text{NR}} ~,
\end{equation}
where  $S_{\text{NR}} = 0.95$ is the light yield suppression factor for
nuclear recoils due to the electric field, ${\cal L}_y = 7.7$~PE/keV
\cite{Aprile:2015uzo} is the average light yield, and ${\cal
L}_{\text{eff}}$ is the relative scintillation efficiency given in
Ref.~\cite{Aprile:2011hi}. We have checked that $\nu(E_R)$ obtained in
this manner is in a good agreement with that estimated from the S1
yield. 
}

To derive a bound from XENON1T, we consider the following Test Statistic
as in Ref.~\cite{DelNobile:2013sia}: 
\begin{equation}
 \text{TS} (m_\chi) = -2 \ln \biggl[\frac{{\cal L}(N_{\text{event}})}{{\cal
  L}_{\text{BG}}}\biggr]~,
\end{equation}
with
\begin{equation}
 {\cal L} (N_{\text{event}}) = \frac{1}{N_{\text{obs}}!} 
(N_{\text{event}} +N_{\text{BG}})^{N_{\text{obs}}} \exp \bigl\{
- (N_{\text{event}} +N_{\text{BG}})
\bigr\} ~,
\end{equation}
where $N_{\text{obs}}$ and $N_{\text{BG}}$ are the
numbers of the observed and background events,
respectively, and ${\cal L}_{\text{BG}} \equiv {\cal L} (0)$. We obtain
90\% CL limits from the condition $\text{TS}(m_\chi) > 2.71$, with
$N_{\text{obs}} = 14$ and $N_{\text{BG}} =7.36 (61)$
\cite{Aprile:2018dbl}, which corresponds to $N_{\text{event}} \lesssim
19.5$. We have checked that the bound obtained in this way is more
conservative than the XENON1T limit \cite{Aprile:2018dbl} by a factor of
$\sim 2$.

%%%%%%%%%%%%%%%%%%%%%%
%\subsection{XENONnT}
%%%%%%%%%%%%%%%%%%%%%%

To assess the future prospects, we consider XENONnT with an exposure
$w_{\text{exp}} = 20 ~\text{t} \cdot \text{yrs}$. We then apply the
maximum gap method \cite{PhysRevD.66.032005} on the assumption of zero
observed events, following Ref.~\cite{Witte:2017qsy}; namely, we require
$1- \exp(-N_{\text{event}}) \geq 0.9$, which corresponds to
$N_{\text{event}} \lesssim 2.3$.

%%%%%%%%%%%%%%%%%%%%%%%%%%%%%%%%%%%%
\section{Electric Dipole Moments}
%%%%%%%%%%%%%%%%%%%%%%%%%%%%%%%%%%%

In this section, we summarize the analytical expressions used in the
calculation of the nucleon and electron EDMs.

%%%%%%%%%%%%%%%%%%%%%%%%%%%%%%%%%%%%%%%%%%%%%%
\subsection{Weinberg operator}
\label{sec:weinbop}
%%%%%%%%%%%%%%%%%%%%%%%%%%%%%%%%%%%

The Wilson coefficient of the Weinberg operator in
Eq.~\eqref{eq:weinbop} is obtained by computing the diagram in
Fig.~\ref{fig:GGGDM}: 
\begin{equation}
 w = - \frac{3 g_s^3}{2(4\pi)^4} 
\sum_{f=u,d} \sum_{i,j} m_\chi m_{f_i} \text{Im} \left[C^{ij}_{f\chi L}
						  C^{ij*}_{f\chi
						  R}\right]
f_1 (m_\chi^2, m_{f_i}^2, \widetilde{m}_{f_j}^2) ~,
\label{eq:wanalytic}
\end{equation}
where the mass function $f_1 (m_\chi^2, m_{f_i}^2,
\widetilde{m}_{f_j}^2)$ is given in Eq.~(3.28) in
Ref.~\cite{Abe:2017sam}. The analytic expression of $f_1 (m_\chi^2, m_{f_i}^2,
\widetilde{m}_{f_j}^2)$ is obtained as
\begin{equation}
f_1(m^2_\chi,m^2_{f_i},\widetilde{m}^2_{f_j})
=
-\left\{
\frac{1}{3}\left(\frac{\partial}{\partial m^2_\chi}\right)^3
+
\frac{m^2_\chi}{6}\left(\frac{\partial}{\partial m^2_\chi}\right)^4
\right\}I(m^2_\chi,\widetilde{m}^2_{f_j},m^2_{f_i}),
\end{equation}
where $I(m^2_\chi,\widetilde{m}^2_{f_j},m^2_{f_i})$ is given in Eq.~(2.19) in Ref.~\cite{Martin:2001vx}.
Note that the above expression contains the
same factor as that in Eq.~\eqref{eq:cgame}: $\text{Im}[C^{ij}_{f\chi L}
C^{ij*}_{f\chi R}]$.

%%%%%%%%%%%%%%%%%%%%%%%%%%%%%%%%%%%%%%%%%%%
\subsection{Barr-Zee-type contribution}
\label{app:BZ}
%%%%%%%%%%%%%%%%%%%%%%%%%%%%%%%%%%%%%%%%%%%%%%%%%%%%%%

The electron EDM induced by the Barr-Zee diagrams in Fig.~\ref{fig:BZ} is given by
\begin{align}
d_e
=
d^{h\gamma}_e+d^{hZ}_e+d^{WW}_e ~,
\end{align}
with
\begin{align}
{d^{h\gamma}_e}
&=
-\frac{8N_{c}y_e e^3}{(4\pi)^4 m^2_h}
\sum_{f,i}
Q^2_f \, {m_{f_i}}\text{Im}[C^{ii}_{fhR}]\, f_{\rm BZ}
\left(0,\frac{m^2_{f_i}}{m^2_h},\frac{m^2_{f_i}}{m^2_h}\right)~
 ,\label{eq:frh}\\[3pt]
{d^{hZ}_e}
&=
-\frac{4N_{c}y_e g^2_Ze}{(4\pi)^4 m^2_h} \left(-\frac{1}{2}+2\sin^2
 \theta_W \right) 
\nonumber\\
&
\times
\sum_{f,i,j}
Q_f{m_{f_i}} \,
\text{Im}[(C^{ji}_{fZR}C^{ij}_{fhR}-C^{ji}_{fZL}C^{ij}_{fhL})]
f_{\rm BZ}\biggl(\frac{m^2_Z}{m^2_h},\frac{m^2_{f_i}}{m^2_h},\frac{m^2_{f_j}}{m^2_h}\biggr) 
~, \\[3pt]
%%
%{d^{hZ}_e}
%&=
%-\frac{4N_{c}y_e g^2_Ze}{(4\pi)^4 m^2_h} \left(-\frac{1}{2}+2\sin^2
% \theta_W \right) 
%\nonumber\\
%&
%\times
%\sum_{f,i,j}
%Q_f \,
%\biggl[
%\text{Im}[m^{(+)ij}_{fhZ}] \,
%f_{hZ}\biggl(\frac{m^2_Z}{m^2_h},\frac{m^2_{f_i}}{m^2_h},\frac{m^2_{f_j}}%{m^2_h}\biggr) 
%+
%\text{Im}[m^{(-)ij}_{fhZ}]\, 
%\tilde{f}_{hZ}\biggl(\frac{m^2_Z}{m^2_h},\frac{m^2_{f_i}}{m^2_h},
%\frac{m^2_{f_j}}{m^2_h}\biggr) 
%\biggr] ~, \\[3pt]
{d^{WW}_e}
&=
-\frac{3g^2 e}{(4\pi)^4}\frac{m_e}{m^4_W}
\nonumber
\sum_{i,j}{m_{u_i}m_{d_j}}
\biggl[
Q_u \,\text{Im}[C^{ij*}_{QWL}C^{ij}_{QWR}]\,f_{\rm BZ}\biggl(0,\frac{m^2_{d_j}}{m^2_W},
\frac{m^2_{u_i}}{m^2_W}\biggr) 
\nonumber\\
&~~~~~~~~~~~~~~~~~~~
+ Q_d\, \text{Im}[C^{ji *}_{QWL}C^{ji}_{QWR}]\,f_{\rm BZ}\biggl(0,\frac{m^2_{u_i}}{m^2_W},\frac{m^2_{d_j}}{m^2_W}\biggr)
\biggr] ~,
\end{align}
for model I and 
\begin{align}
{d^{WW}_e}
&=
-\frac{g^2 e}{(4\pi)^4}\frac{m_e}{m^4_W}
\nonumber
\sum_{i}{m_{\nu}m_{e_i}}
\left[
Q_e\text{Im}[C^{i*}_{LWL}C^{i}_{LWR}] \,f_{\rm BZ}\left(0,\frac{m^2_{\nu}}{m^2_W},\frac{m^2_{e_i}}{m^2_W}\right)
\right] ~,
%~~~\mbox{(for model II),}
\end{align}
for model II, where the mass function is given by
\begin{align}
%f_{h\gamma}(r)
%&=
%\int^1_0 dx\frac{1}{1-x}j\left(0,\frac{r}{x(1-x)}\right) ~,\\[3pt]
%%
%f_{hZ}(r,r_1,r_2)
%&=
%\int^1_0 dx\frac{1}{(1-x)}j\left(0,\frac{xr_1+(1-x)r_2}{x(1-x)}\right)~,\\[3pt]
%
%%
%f_{hZ}(r,r_1,r_2)
%&=
%\frac{1}{2}\int^1_0 dx\frac{1}{x(1-x)}j\left(0,\frac{xr_1+(1-x)r_2}{x(1-%x)}\right)~,\\[3pt]
%%
%\tilde{f}_{hZ}(r,r_1,r_2)
%&=
%\frac{1}{2}\int^1_0
% dx\frac{2x-1}{x(1-x)}j\left(0,\frac{xr_1+(1-x)r_2}{x(1-x)}\right) ~,\\[3pt]
% 
%
%f_{\rm BZ}(r_1,r_2)
%&=
%\int^1_0 dx\frac{1}{1-x}j\left(0,\frac{xr_1+(1-x)r_2}{x(1-x)}\right) ~,\\[3pt]
%
f_{\rm BZ}(r_,r_1,r_2)
&=
\int^1_0 dx\frac{1}{1-x}j\left(r,\frac{xr_1+(1-x)r_2}{x(1-x)}\right) ~,\\[3pt]
j(r,s)
&=\frac{1}{r-s}\left(\frac{r\ln r}{r-1}-\frac{s\ln s}{s-1}\right) ~.
\end{align}
%and $m^{(\pm)ij}_{fhZ}=(m^{ij}_{fhZ}\pm m^{ji}_{fhZ})/2$ with
%$m^{ij}_{fhZ}={m_{f_i}}(C^{ji}_{fZR}C^{ij}_{fhR}-C^{ji}_{fZL}C^{ij}_{fhL})$. 
Here, $m_h$, $m_Z$, and $m_W$ are the Higgs, $Z$, and $W$ boson masses, respectively, and $m_e$ ($\equiv y_e v/\sqrt{2}$) is the electron mass. 

In Model I, the CEDM for quark $q$
\begin{align}
\mathcal{L}
=
-\frac{i}{2}g_s\tilde{d}_q\bar{q}G^A_{\mu\nu}\sigma^{\mu\nu}T^A\gamma_5 q ,
\end{align}
is also generated from the upper left diagram in Fig.~\ref{fig:BZ} with
the $\gamma/Z$ lines replaced with gluon lines. The
resultant expression is given by
\begin{align}
\tilde{d}_q
=
-\frac{32g_s^2y_q}{3(4\pi)^4 m^2_h}
\sum_{f,i}
{m_{f_i}}\text{Im}[C^{ii}_{fhR}]f_{\rm BZ}\left(0,\frac{m^2_{f_i}}{m^2_h},\frac{m^2_{f_i}}{m^2_h}\right)
 ~.
\end{align}
where the quark Yukawa coupling constant $y_q$ is given as $m_q= y_q v/\sqrt{2}$ ($m_q$: quark mass).

%%%%%%%%%%%%%%%%%%%%%%%%%%%%%%%%%%%%%%%
\newpage
{\small
\bibliographystyle{aps}
\bibliography{ref}
}
%%%%%%%%%%%%%%%%%%%%%%%%%%%%%%%%%%%%%%%

\end{document}